\documentclass[letter,fleqn,usenatbib]{mnras}
\usepackage[T1]{fontenc}
\usepackage{ae,aecompl}

\usepackage{graphicx}	
\usepackage{amsmath}	
\usepackage{amssymb}	

\usepackage{amsmath, mathtools}

\newcommand\plotone[1]{\includegraphics[width=\columnwidth]{#1}}
\newcommand\plottwo[2]{\includegraphics[width=\columnwidth]{#1}\includegraphics[width=\columnwidth]{#2}}

\newcommand{\be}{\begin{equation}}
\newcommand{\ee}{\end{equation}}

\newcommand{\pc}{\,\rm pc}
\newcommand{\kpc}{\,\rm kpc}
\newcommand{\kms}{\,\rm km\,s^{-1}}
\newcommand{\cm}{\,\rm cm}

\newcommand{\g}{\,\rm g}
\newcommand{\yrs}{\,\rm yrs}
\newcommand{\yr}{\,\rm yr}
\newcommand{\AU}{\,\rm AU}
\newcommand{\ergs}{\,\rm erg\, s^{-1}}
\newcommand{\Myrs}{\,\rm Myrs}

\newcommand{\partA}{particle A }
\newcommand{\PartA}{Particle A }
\newcommand{\partB}{particle B }
\newcommand{\PartB}{Particle B }
\newcommand\plotonesmall[1]{\includegraphics[width=0.4\textwidth]{#1}}

\title{Collapse in Self-gravitating Turbulent Fluids}

\author[Murray, Chang, Murray, \& Pittman]{Daniel W. Murray$^1$\thanks{DM: dwmurray@uwm.edu; PC: chang65@uwm.edu; NM: murray@cita.utoronto.ca; JP: jbpittmanjr@gmail.com}, Philip Chang$^1$, Norman W. Murray$^{2,3}$, \&  John Pittman$^{1}$\\
  $^{1}$Department of Physics, University of
  Wisconsin-Milwaukee, 3135 North Maryland Ave., Milwaukee, WI 53211, USA\\
$^{2}$Canadian Institute for Theoretical Astrophysics, 60
  St.~George Street, University of Toronto, Toronto ON M5S 3H8, Canada\\
  $^3$Canada Research Chair in Astrophysics
}

\pubyear{2016}

\begin{document}
\label{firstpage}
\pagerange{\pageref{firstpage}--\pageref{lastpage}}
\maketitle


\begin{abstract}

Motivated by the nonlinear star formation efficiency found in recent numerical simulations by a number of workers, we perform high-resolution adaptive mesh refinement simulations of star formation in self-gravitating turbulently driven gas. As we follow the collapse of this gas, we find that the character of the flow changes at two radii, the disk radius $r_d$, and the radius $r_*$ where the enclosed gas mass exceeds the stellar mass. Accretion starts at large scales and works inwards. 
In line with recent analytical work, we find that the density evolves to a fixed attractor, $\rho(r,t ) \rightarrow \rho(r)$, for $r_d<r<r_*$; mass flows through this structure onto a sporadically gravitationally unstable disk, and from thence onto the star. 
In the bulk of the simulation box we find that the random motions $v_T \sim r^p$ with $p \sim 0.5$, in agreement with Larson's size-linewidth relation. In the vicinity of massive star forming regions we find $ p \sim 0.2-0.3$, as seen in observations. For $r<r_*$, $v_T$ increases inward, with  $p=-1/2$.  
Finally, we find that the total stellar mass $M_*(t)\sim t^2$ in line with previous numerical and analytic work that suggests a nonlinear rate of star formation.

\end{abstract}

\begin{keywords}
  galaxies: star clusters: general -- galaxies: star formation -- stars: formation -- turbulence
\end{keywords}

\section{Introduction}

The star formation time on galactic scales is long when compared to the dynamical time. 
\citet{1998ApJ...498..541K} expressed this in the form
\begin{equation}
\dot{\Sigma}_* = \epsilon\Sigma_g \tau_{\rm DYN}^{-1}
\end{equation}
where $\dot{\Sigma}_*$ is the star formation rate per unit area, $\Sigma_g$ is the 
gas surface density, $\tau_{\rm DYN}$ is the local dynamical time, and $\epsilon = 0.017$ 
is the efficiency factor.  Naively, if the gas self-gravity dominates the dynamics, $\epsilon \sim 1$, so 
the low efficiency of star formation is surprising.  More recent work has refined this and similar relations in regard to its dependence on molecular gas \citep{2008AJ....136.2846B} and by taking into account the error distributions of both $\dot\Sigma_*$ and $\Sigma_g$ \citep{2013MNRAS.430..288S}, but the best current estimates of the efficiency of star formation on galactic scales remains low. 

Whether this low efficiency applies to scales comparable to giant molecular clouds, with radii of
order $100\pc$, is debated in the literature.
\citet{2010ApJ...723.1019H}, \citet{2010ApJ...724..687L}, \citet{2010ApJS..188..313W}, and \citet{2011ApJ...729..133M} 
find efficiencies a factor of ten or more larger, while \citet{2007ApJ...654..304K} and \citet{2012ApJ...745...69K} 
find $\epsilon\approx0.01$.
On these small scales, observations also suggest that the efficiency is not universal, but instead 
varies over two to three orders of magnitude (e.g.  
\citealt{1988ApJ...334L..51M,2016arXiv160805415L}).

There are a number of explanations for the low star formation rate, on 
either small or large scales (although they may not be necessary for the former!). These include 
turbulent pressure support \citep{1992ApJ...396..631M}, support from magnetic fields
\citep{1966MNRAS.132..359S,1976ApJ...207..141M}, and stellar feedback (e.g.
\citealt{1986ApJ...303...39D}).  Numerical experiments investigating the first two 
effects suggest that neither turbulence nor magnetic support is sufficient 
to reduce the rate of star formation to $\epsilon\approx 0.02$ on small scales
\citep{2010ApJ...709...27W,2011MNRAS.410L...8C,2011ApJ...730...40P,2012ApJ...754...71K,2014MNRAS.439.3420M}.  Feedback from radiative effects and protostellar jets and winds may be able to explain the low star formation rate, but the impact of these forms of stellar feedback remains uncertain despite recent progress
\citep{2010ApJ...709...27W,2014MNRAS.439.3420M,2015MNRAS.450.4035F}.

Until very recently, galaxy-scale or larger (cosmological) simulations 
were not able to reproduce the Kennicutt-Schmidt relation.  Nor 
did the cosmological runs reproduce correctly the mass of stars in galaxies of a given 
halo mass, despite including supernova and other forms of feedback, 
e.g., \citet{2010MNRAS.404.1111G,2010Natur.463..203G,2011MNRAS.410.2625P}.
To overcome this low resolution driven problem, \citet{2011MNRAS.417..950H,2012MNRAS.421.3522H}
performed high resolution (few parsec spatial, few hundred solar mass particle masses) simulations 
of isolated galaxies, modeling both radiative and supernovae feedback (among other forms). 
They recovered the Kennicutt-Schmidt relation, a result that they 
showed was independent of the small-scale star formation
law that they employed. The simulations in the second paper also generated galaxy 
scale outflows or winds, removing gas from the disk, thus making it unavailable for star 
formation. When the feedback was turned off, the star formation rate soared, demonstrating that
in the simulations at least, feedback was crucial to explaining the Kennicutt-Schmidt relation, and
the outflows. Simulations including supernovae but lacking the radiative component of the feedback
did not exhibit strong winds and so overproduced stars.

Cosmological simulations employing unresolved (or ``sub-grid") models for both 
radiative and supernovae feedback are now able to reproduce the halo-mass/stellar mass relation (e.g.,
\citealt{2013MNRAS.434.3142A,2014MNRAS.445..581H,2015ApJ...804...18A}). Again, 
these simulations {\em require} 
stellar feedback to drive the winds that remove gas from the disk, so as to 
leave the observed mass of stars behind. 


\citet{2015ApJ...800...49L} emphasized that the star formation efficiency on 
parsec scales is nonlinear 
in time, i.e., $\epsilon \propto t \rightarrow M_* \propto t^2$, on small scales, where $M_*$ is the total stellar mass. Magnetic fields slowed the initial star
formation rate somewhat, but did not change the $M_*(t)\sim t^2$ scaling. 
Using a detailed numerical simulation, they showed that this nonlinear star formation 
rate is driven by the properties of collapsing regions.
In particular, they demonstrated that the turbulent velocity near or in
collapsing regions follows 
different scaling relations than does turbulence in the global environment, which follows
Larson's law, $v_T(r)\sim r^{1/2}$ \citep{1981MNRAS.194..809L}. They also showed that the density PDF 
is not log-normal, but rather develops a power law to high density.
This latter result was hinted at by \citet{2000ApJ...535..869K} and shown convincingly, 
as well as explained, by \citet{2011ApJ...727L..20K}. 

The increasing rate of star formation found by \citet{2015ApJ...800...49L} is important 
in that it may provide an explanation for the observed range in star formation rates on 
small scales.  It suggests that the star formation rates on small scales 
vary in part because of the age of the star forming region; slow star forming 
regions, with very low instantaneous efficiencies, will ramp up their stellar production over time. If this result
can be firmly established, it will highlight the need for a form of very rapid 
feedback. In particular, since the dynamical time in massive star forming regions 
is much smaller than the time delay of $\sim 4{\,\rm Myrs}$ between the start of star 
formation and the first supernovae, rapid star formation on small scales would 
have to be halted by some form of feedback other than supernovae. 

The simulations of  \citet{2015ApJ...800...49L} explicate
the link between the rate of star formation with the gravitational collapse 
of high density regions, which  is 
an analytically well studied problem. An early model of \citet{1977ApJ...214..488S} estimated 
the accretion rate onto stars by assuming that stars form from hydrostatic cores 
supported by thermal gas pressure. The accretion rate in his model was independent
of time, given by $\dot{M} = m_0c_s^3/G$, where $c_s = (k_b T / \mu)^{1/2}$  is the sound
speed in molecular gas, and $m_0 = 0.975$. \citet{1977ApJ...214..488S} predicted a 
maximum accretion rate of $\sim 2\times10^{-6} M_\odot\,\yr^{-1}$, which is too small to explain 
the origin of massive ($M_*\sim50-100M_\odot$) O stars, which have lifetimes $\lesssim 4\times10^6\yrs$

\citet{1992ApJ...396..631M} overcame the difficulty with slow accretion rates by adopting 
the turbulent speed in lieu of the sound speed (see also \citet{1997ApJ...476..750M} and
\citet{2003ApJ...585..850M}). In doing so they were able to replace the slower signal 
speed of sound with the faster turbulent speed. However, they continued to assume 
the initial condition was that of a hydrostatic core that is supported by
turbulent pressure. They also assumed that the turbulence is static and unaffected by the collapse.  

Collectively, these models, \citep{1977ApJ...214..488S,1992ApJ...396..631M,1997ApJ...476..750M,2003ApJ...585..850M}, are referred to as inside-out collapse models; the collapse starts at small radii (formally at $r=0$ in the analytic models) and works its way outward, at the assumed propagation speed ($c_s$ or $v_T(r)$).  At any given time, the infall velocity and mass accretion rate both decrease with increasing radius $r$.  The analytic models assume the existence of a self-similarity variable $x = r/vt$, where $v=c_s$ in \citet{1977ApJ...214..488S} or the turbulent velocity $v_T(r)$ in \citet{1992ApJ...396..631M,1997ApJ...476..750M,2003ApJ...585..850M}.  
These models predict velocity and mass accretion profiles very different than those seen 
in the simulations of \citet{2015ApJ...800...49L}.

Motivated by this discrepancy,
\citet{2015ApJ...804...44M}, hereafter MC15, developed a 1-D model of
spherical collapse that treats the turbulent velocity, $v_T$, as a dynamical variable and
does not assume that the initial condition is a hydrostatically
supported region. They used the results of \citet{2012ApJ...750L..31R} on
compressible turbulence; the evolution
of the turbulent velocity in a collapsing (or expanding) region is
described well by the following equation:
\be
\frac{\partial v_{T}}{\partial t} + u_{r} \frac{\partial v_{T}}{\partial r} 
+ \left( 1 + \eta \frac{v_{T}}{u_{r}} \right) \frac{v_T u_r}{r} = 0
\label{eq: Robertson}
\ee
The first two terms are the Lagrangian derivative, and $u_r$ is the radial infall velocity. The first term in the brackets 
describes the turbulent driving produced by the infall, while the second 
is the standard expression for the turbulent decay rate;  $\eta$ is a dimensionless constant of order unity.

MC15 used this in place of an energy equation. Together with the
equations for mass continuity and momentum, equation (\ref{eq: Robertson}) gives a
closed set of equations that can be solved in spherical symmetry
numerically. In addition, they were able to analytically show that the
results of their calculations gave density and velocity profiles that
appear to be in line with both recent numerical calculations
\citep{2015ApJ...800...49L} and observations (e.g.,
\citealt{1995ApJ...446..665C,1997ApJ...476..730P}).

To summarize, MC15's major results were:
\begin{itemize}
\item The gravity of the newly formed star introduces a physical scale into the problem, which MC15 called the stellar sphere of influence, $r_*$. This is an idea familiar from galactic dynamics. 
The radius $r_*$ is where the local dynamics transitions from being 
dominated by the mass of the gas to being dominated 
by the mass of the star. As a result,the character of the solution, in particular that of the velocity, differs dramatically between $r<r_*$ and $r>r_*$.  The existence of this physical scale modifies the form of the self-similarity on which inside-out theories rely.

\item The small scale density profile is an attractor solution.  MC15 showed numerically and argued analytically that at small scales, the density profile is an attractor solution. 
In particular, MC15 showed the density profile asymptotes to: 
\be
\rho(r,t)=
\begin{dcases}
\rho(r_0)\left({\frac{r}{r_0}}\right)^{-3/2}, & r<r_*(t)\\
\rho(r_0,t)\left({\frac{r}{r_0}}\right)^{-k_\rho}, \ k_\rho\approx1.6-1.8 & r>r_*(t).
\end{dcases}
\label{eq:density}
\ee
where $r_0$ is some fiducial radius. 
\item The existence of $r_*$ implies that the infall and turbulent velocities have different scaling for $r<r_*$ and $r>r_*$.  In particular, MC15 showed
\be
u_r(r,t), v_T(r,t) \propto
\begin{dcases}
\sqrt{\frac{GM_*(t)}{r}} \sim r^{-1/2} & r<r_*(t)\\
\sqrt{\frac{GM(r,t)}{r}} \sim r^{0.2} & r>r_*(t),
\end{dcases}
\label{eq:infall_behavior}
\ee
Thus the scaling of the turbulent velocity differs from that predicted by Larson's law ($\propto r^{1/2}$) inside the sphere of influence. In other words, the turbulent velocity in  massive star forming regions will deviate from Larson's law, which has long been observed, but without theoretical explanation. 
\item The stellar mass increases quadratically with time.  This result  arises naturally from the attractor solution nature of the density profile at small $r$, Equation (\ref{eq:density}), and the scaling with Keplerian velocity for the turbulent and infall velocities at small $r$, Equation (\ref{eq:infall_behavior}).

The mass accretion rate: 
\be
\dot M(r,t)=
\begin{dcases}
4\pi R^2\rho(R)u_r(r,t), \sim t\,r^{0} & r<r_*\\
4\pi R^2\rho(R)u_r(r,t) \sim t^0\,r^{0.2} & r>r_*.
\end{dcases}
\label{eq:Mdot_behavior}
\ee
\end{itemize}

MC15's predictions for $r<r_*$ could not be checked using the simulations
of \citet{2015ApJ...800...49L} as those fixed grid simulations were
too coarse.  In this paper, we study the collapse of gas and formation
of stars in a turbulent GMC using roughly a dozen high resolution
adaptive mesh refinement (AMR) simulations in this paper.

We employ large-scale (16 pc) hydrodynamic AMR simulations of star-forming clouds with continuously driven supersonic
turbulence. The initial conditions for our simulations are exactly
the same as the FLASH simulations in \citet{2015ApJ...800...49L}.

If the equations are non-dimensionalized, two dimensionless variables appear, the Mach number ${\cal M}$ and the virial parameter $\alpha_{vir} \equiv (5/3) v_T^2 / GM_{box}$, e.g., \citet{1984oup..book.....M}.
We want to model massive star forming regions in the Milky Way, so we choose the Mach Number ${\cal M} = 9$ 
and the virial parameter $\alpha_{vir} = 1.9$ respectively.
In addition, we choose the size of the box $L = 16 \rm pc$, and the sound speed $c_s = 0.264\,{\rm km\,s^{-1}}$, 
so that the turbulent velocity lies approximately on the observed size-line width relation, Larson's Law.
These choices fix both the density and the mass scale.

The simulations described in this paper disregard several physical effects.
We do not include radiative, stellar wind, or proto-stellar jet feedback.
While the feedback physics we neglect can have significant effects on both the rate of star formation and the initial mass function (IMF), we aim to address the role the random motions captured by the Reynolds stress play in the dynamics of gravitational collapse in turbulent fluids.

Our equation of state is that of an isothermal gas. It is possible, and even likely, that thermal effects play a role in setting the initial mass function of stars, e.g \citet{2005MNRAS.359..211L}.
With this in mind, we relegate the discussion of the IMF to an appendix, as the details are unlikely to be reliable.

This paper is organized as follows. In Section \ref{sec:simulation setup}
we describe our numerical methods and simulation setup. In Section
\ref{sec:results} we present and analyze the results of our simulations. 
In particular, we make detailed comparisons with the results of MC15.  
We discuss our results and compare
them to previous work in Section \ref{sec:discussion}

\section{Detailed Simulations of Turbulent Collapse} \label{sec:simulation setup}

Most of the simulations described here use the adaptive mesh refinement code FLASH ver. 4.0.1 
\citep{2000ApJS..131..273F, 2008ASPC..385..145D} to model self-gravitating, hydrodynamic turbulence in isothermal gas with three-dimensional (3D) periodic grids and a minimum of 8 levels of refinement on a root grid of $128^3$, 
giving an effective resolution of $32,768^3$.  
Following \citet{2015ApJ...800...49L} our FLASH runs use the Harten-Lax-van Leer-Contact 
Riemann solver and an unsplit solver \citep{2009ASPC..406..243L}. We have also used the RAMSES code \citep{2002A&A...385..337T}, but unless explicitly stated otherwise, the results below come from FLASH simulations.

As just mentioned, we start with a box with the physical length set to $L = 16$ pc using 
periodic boundary conditions. The initial mass density is 
$\rho = 3\times10^{-22}\,{\rm g\,cm}^{-3}$ (number density $n \approx 100\,{\rm cm}^{-3}$), 
corresponding to a mean 
free-fall time $\bar\tau_{\rm ff}\approx3.8\,{\rm Myrs}$; the total mass in the
box is $M\approx 18,000\,M_\odot$.
The sound speed is set to $c_s = 0.264\,{\rm km\,s^{-1}}$.
We use pure molecular hydrogen in this simulation so the ambient 
temperature $T \approx 17 {\rm K}$.

To initialize our simulations, we drive turbulence by applying a large scale 
($1 \le kL \le 2$, corresponding to $1.3 - 2.7 \, {\rm pc}$) fixed solenoidal acceleration field as a momentum
source term. 
We use solenoidal driving because it is known that compressive turbulence increases the star formation rate compared to solenoidal driving \citep{2008ApJ...688L..79F}.
We apply this field in the absence of gravity and star particle formation for 3 dynamical times until a statistical steady state is reached.
The resulting Mach number is ${\cal M} = 9$, i.e a turbulent velocity of $v_T = 2.37 {\rm km \, s^{-1}}$.

Stirring the initial turbulence using a fixed driving field is a
technique used by a number of workers in the field
\citep{2011ApJ...730...40P,2011ApJ...731...59C}.  Other groups
initialize the turbulence by initializing the velocity field with
Gaussian random perturbations having some assumed power spectrum
\citep{2014MNRAS.439.3420M,2015ApJ...809..187S}.  While neither of the
resulting velocity fields are generated the way the turbulence in the
interstellar medium (ISM) of our Galaxy is, the stirring allows one to
perform simulations which have nontrivial initial density structures
and velocity fields that are at least reminiscent of those inferred
from observations of the interstellar medium of our Galaxy.

\citet{2012ApJ...761..156F} use a time varying driving field to produce random motions.  They argue that a time-varying driving field allows one to avoid large spatial correlations that would result from a fixed driving field acting for a time longer than the dynamical time of the simulation box. 
In our simulations we do not run for longer than a box dynamical time after turning on star formation. 
We run for $600,000  \, {\rm yrs}$, about $0.16$ dynamical times, after the first star forms.
The limiting factor on the length of the runs was our available compute time.
Hence, we do not expect the large scale turbulent flow to vary much over such a short time.
In addition, there is some evidence \citep{2010A&A...512A..81F} that the results of turbulent driving are not sensitive to the exact large-scale mechanism.

This fully developed turbulent state is the initial condition to 
which we add self-gravity and star particle formation for our star 
formation experiments. We enable AMR to follow the collapse of overdense regions.
Even after turning on star formation, we continue to drive the large scale fixed solenoidal acceleration field.

To follow these collapsing regions, we have implemented an algorithm 
for mesh refinement in these simulations, similar to that of \citet{2010ApJ...713..269F}.
In supersonically turbulent flows, certain regions rapidly increase in density.
For a given density and temperature, or sound speed, regions larger than the 
Jeans length
\be
\lambda_J \equiv \sqrt{\frac{\pi c_{s}^2}{G \bar{\rho}}} \approx 3.5 \text{pc}
\label{eq:Jeans_length}
\ee
are prone to gravitational collapse. Our base grid's resolution of $N_{\rm root}^3 = 128^3$ 
gives a cell length of 
$1.25 \text{x} 10^{-1}$pc which is sufficient to resolve the Jeans 
length for the mean density.

In most of our simulations, the AMR grid is refined when the \citet{1997ApJ...489L.179T} criterion 
%
\be
\lambda_J \le N_J \Delta x, 
\ee
is met.
In this expression $\Delta x$ is the cell length, and $N_J$ is an integer; \citet{1997ApJ...489L.179T} found that in order to avoid artificial fragmentation, one requires $N_J\gtrsim 4$.

This corresponds to a condition on the density 
\begin{eqnarray}
\frac{\rho}{\rho_0} &=& 45 \cdot 4^l
\left( \frac{N_{\rm root}}{128} \right)^2 
\left( \frac{N_{\rm J}}{4} \right)^{-2} 
\left( \frac{16 \rm{pc}}{L} \right)^2 \nonumber\\
&&\times\left( \frac{c_s}{0.265\, {\rm km\, s}^{-1}} \right)^2
\left( \frac{3 \times10^{-22} {\rm g\, cm}^{-3}}{\rho_0} \right)
\label{eq:refinement_criteria}
\end{eqnarray}
where $l$ is the refinement level, with $l = 0$ corresponding to the root grid.
When this density condition is met the local grid is refined by a factor of 2, 
provided that the maximum refinement level has not been reached. 
When the transisition to the maximum refinement level is triggered i.e. when $l$ goes from $7$ to $8$ (the maximum refinement level),
the density contrast is $\rho / \rho_0 \approx 10^6$. 

In the Appendix we describe a number of test simulations in which we refined the grid when 
$N_J=4, 8, 16$ or $32$ \citep{2011ApJ...731...62F}. We show that many of the quantities in our runs, including the density and the mass accretion rates, are converged for $N_J=4$.

The maximum dynamic range is a little larger than 6 orders of magnitude, because we allow the density to increase further before forming star particles.
When the Truelove criterion is exceeded by a factor of three at the highest refinement level, the excess 
mass in a cell is transferred either to a newly created star particle or to a 
star particle whose accretion radius includes the cell. The factor of three allows only the highest density regions to form star particles.  It is inspired by the work of \citet{2011ApJ...730...40P} whose sink particle formation criteria of 8000$\times$ mean density is a factor of 3-4 above the Truelove criteria at their highest resolution of $1000^3$. 
Additionally, the 3 cells immediately around a star particle can rise above this density criterion. 
This is done so that we do not form star particles within 2 cells of each other. 
Instead these close surrounding cells can only accrete onto the previously formed star particle.
We should also note that like our previous work in \citet{2015ApJ...800...49L}, our star particle 
creation prescription is different from the prescription of 
 \citet{2010ApJ...713..269F} where additional checks are performed; in the appendix
we present the results of runs in which we used these additional checks, finding that they
do not affect the $t^2$ scaling of the stellar mass, or the dynamics of the infall.

To calculate the gravity, we use the same algorithm as described in 
\citet{2015ApJ...800...49L}, which we now briefly describe. 
To compute the self gravity on gas, we first map star particles 
to the grid and then use a multi-grid Poisson solver 
(see \citealt{2008ApJS..176..293R}), coupled with a fast-Fourier 
transform (FFT) solution on the root grid, to solve for gravity.  
To compute the gravitational acceleration on the star particles, 
we first compute the particle-particle forces using a direct N-body calculation.  
To compute the particle-gas forces, we use the same multigrid solver 
(with root grid FFT) on the grid, but with the star particle {\it unmapped}.   
As a result, two large scale gravity solutions (one with and one without 
mapped star particles) must be found per timestep as opposed to one.
This allows us to avoid the computationally expensive task of computing 
gas-star particle forces via direct summation. As discussed in \citet{2015ApJ...800...49L}, this splitting of particle-particle and particle-gas/gas-particle forces does not strictly obey Newton's second law, breaking down on order the size of the smallest grid cell.  As a result, errors in the orbits of particles may result.   However, we believe that our runs are short enough to avoid buildup of significant errors. 

In the FLASH runs, to obtain a useful number of star particles with long accretion histories, we have 
taken the initial turbulent box and have only run our refinement algorithm 
(and hence, star particle algorithm) on only one octant at a time.  
This forces us to run eight high resolution simulations, each on a 
difference octant and so allows us to treat each octant as a separate 
distinct simulation.  This is necessary as FLASH does not have individual 
timesteps, which results in the code grinding to a halt once a single 
region collapses.

\section{Results}\label{sec:results}
\begin{figure} 
\plotone{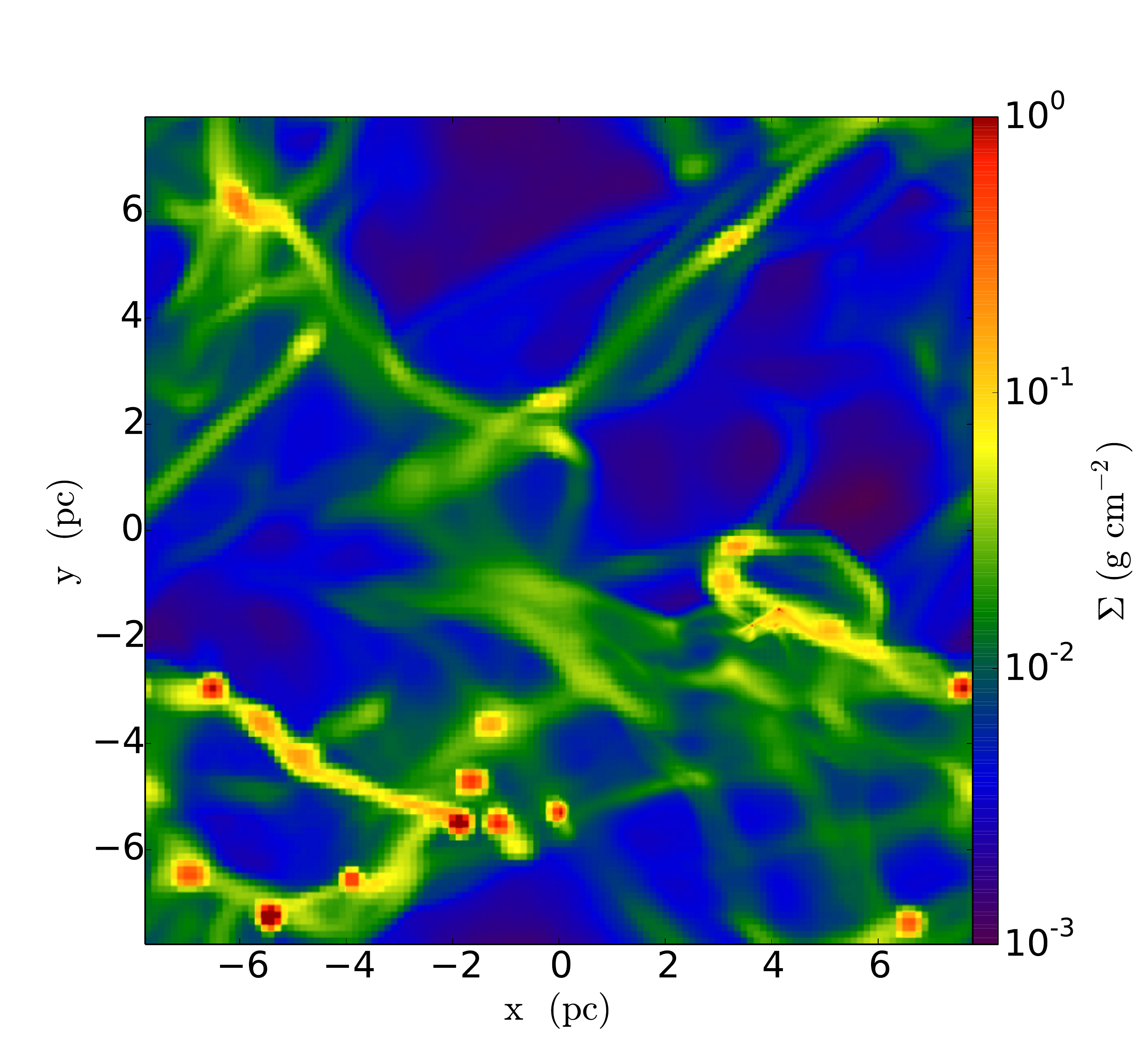}
\caption{Projection of the density along the z-axis of the entire simulation volume. 
The root grid is $128^3$ with up to 8 levels of refinement, giving an effective 
resolution of $32\,{\rm K}^3$. This snapshot is taken $2.8$ Myr after star 
formation was turned on. \label{fig:entire projection}}
\end{figure}

In Figure \ref{fig:entire projection} we show a projection along the z-axis 
 of the entire simulation volume 
for one of the high resolution octant simulations, 2.8 Myr after gravity has been turned on. 
The image shows up to 8 levels of refinement, giving an effective resolution 
of $32768^3$, or a minimum cell size of $5\times 10^{-4} \pc$.
Regions that are highly refined are the densest regions, for which the image 
is smoother than the low-density more pixelated regions.  Note that the highly 
refined regions are limited to the lower right, which is the octant that
this particular simulation focused on. The other seven simulations refine the other octants.

The high density regions are organized into filaments.
These filaments span most of the simulation box, with lengths up to 
several parsecs and widths of order a few tenths of a parsec. Some filaments appear 
to flow into large clumps. This is in accord with many previous simulations, e.g., 
\citep{1998ApJ...504..300P, 2015ApJ...800...49L}. These clumpy regions 
have the highest densities and, hence, are the first to fulfill 
the criterion for star particle formation.

In this section we focus on the regions around  two individual star 
particles, which we refer to as \partA\ and \partB.

\PartA\ formed about a quarter of a parsec away from its nearest neighbor star particle.
At the end of the run it was $\sim 736,000$ years old and had 
a mass of $\sim 17.5 \, M_\odot$, although it was still accreting rapidly.

\PartB\ formed and remained in isolation.
At the end of the run, the particle was $\sim 512,000$ years 
old and had a mass of $\sim 10.7 \, M_\odot$.
Throughout the simulation \partB had a steady supply of gas.

\subsection{The Run of Infall ($\vec{v}_r$), Circular ($\vec{v}_\phi$), and 
 Random Motion ($\vec{v}_T$) velocities with Radius ($r$) Before Star Particle Formation.}

\begin{figure} 
\plotonesmall{f2a.pdf}
\plotonesmall{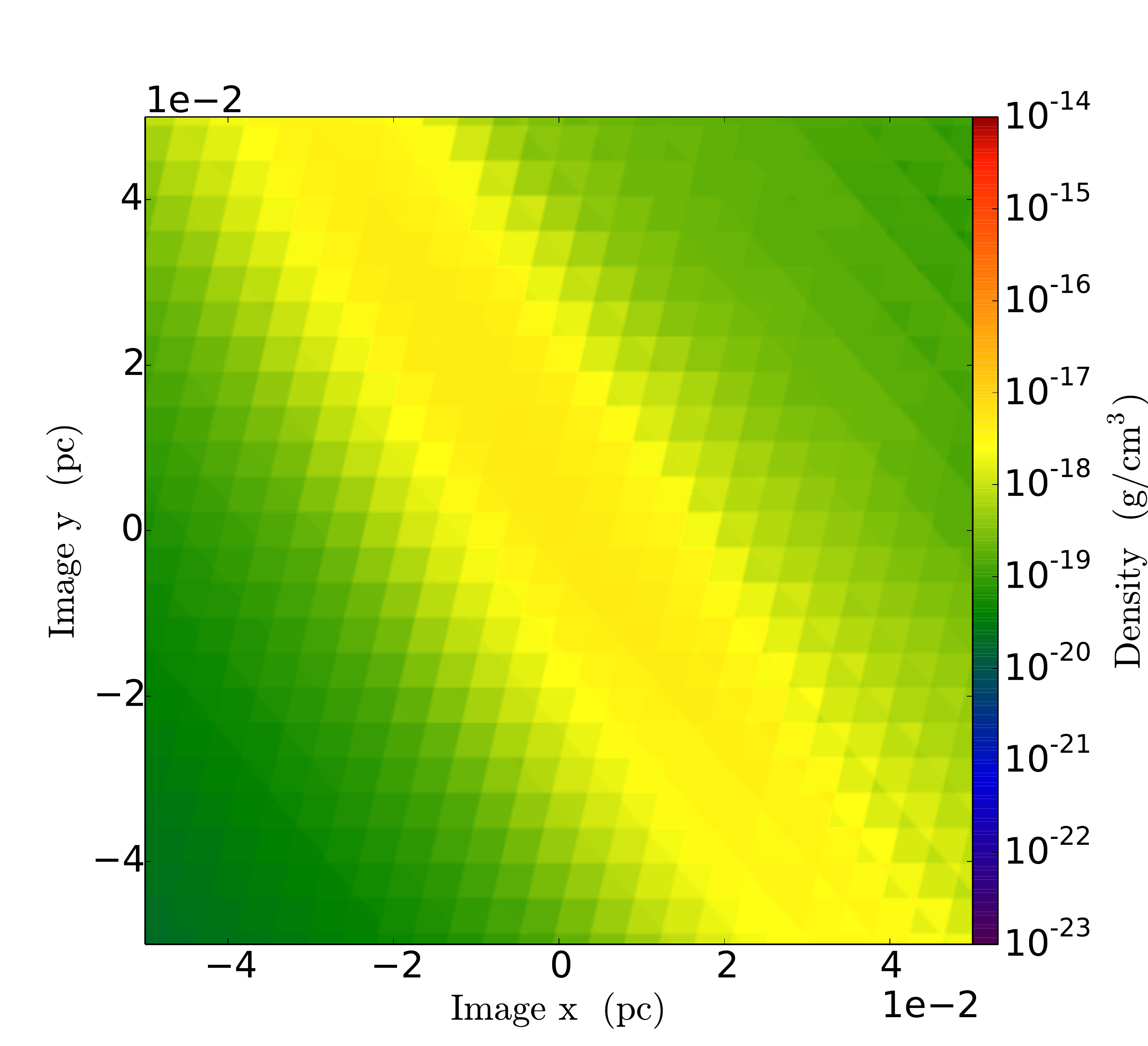}
\caption{The top plot shows the run of velocity with radius measured from the density peak; 
this density peak will develop into \partA in 100,000 years.
The sound speed is the black horizontal line while the infall velocity 
$|u_r|$ is given by the blue triangles, connected by a solid blue line.
The green circles connected by a solid green line show 
$v_T$ while the black crosses show the rotational velocity $v_\phi$.
The red dashed line is the Keplerian velocity $v_K\equiv\sqrt{GM(r)/r}$.
Even at this early stage the structure is far from hydrostatic equilibrium, 
as the infall velocity is $\sim 25\%$ of the free-fall velocity.
The refinement level is $l = 6$, which corresponds to a cell size of $\sim2\times10^{-3}\pc$.
The bottom plot shows the density in a slice along the direction of the 
angular momentum vector centered on that peak.}
\label{fig:Sphere_of_influence_quad2_well_before}
\end{figure}

Figure \ref{fig:Sphere_of_influence_quad2_well_before} shows the infall 
velocity, $u_r$, circular, $v_{\phi}$, and random motion, $v_T$, velocities as a function of  radius (top panel) and the density in a slice of the local volume 
(bottom panel) around the density peak that will form \partA\ 100,000 years in the future.   In Appendix \ref{appendix:velocity}, we describe how we calculate 
each of these velocities.

We will compare $v_T$ to what  MC15 referred to as a turbulent velocity. Our current definition of $v_T$ is simply that of a random velocity. 
We are agnostic about whether or not $v_T$ characterizes an isotropic turbulent pressure; close examination of the velocity field indicates that the random motions are not isotropic on the scale of their distance from the density peak. It is also clear, however, that $v_T$ characterizes a Reynolds stress that does provide a net outward support against gravitational collapse. This follows from a simple energy argument; the infall velocity in the vicinity of the density peak is well below the local free-fall velocity, and remains so throughout the simulation, even after a star particle forms. Thus, some of the potential energy released by the infall goes into some channel other than inward motion. A fraction of the potential energy release goes into shocks, and in our code is effectively removed immediately. At this early stage, the rotational motion represents a small fraction ($\lesssim10\%$) of the energy at all but the smallest radii. But the inward flattening of the green line in Figure
\ref{fig:Sphere_of_influence_quad2_well_before}, and the inward increase seen in later figures, shows that a substantial fraction of the potential energy released by the inflow goes into random motions. By energy conservation, this fraction is not available to the inflow, so that $|u_r|$ is smaller than it would be if the random motions were not absorbing some of the energy. This shows that there is an effective outward force on the infalling gas. 
\begin{figure} 
\plotonesmall{f3a.pdf}
\plotone{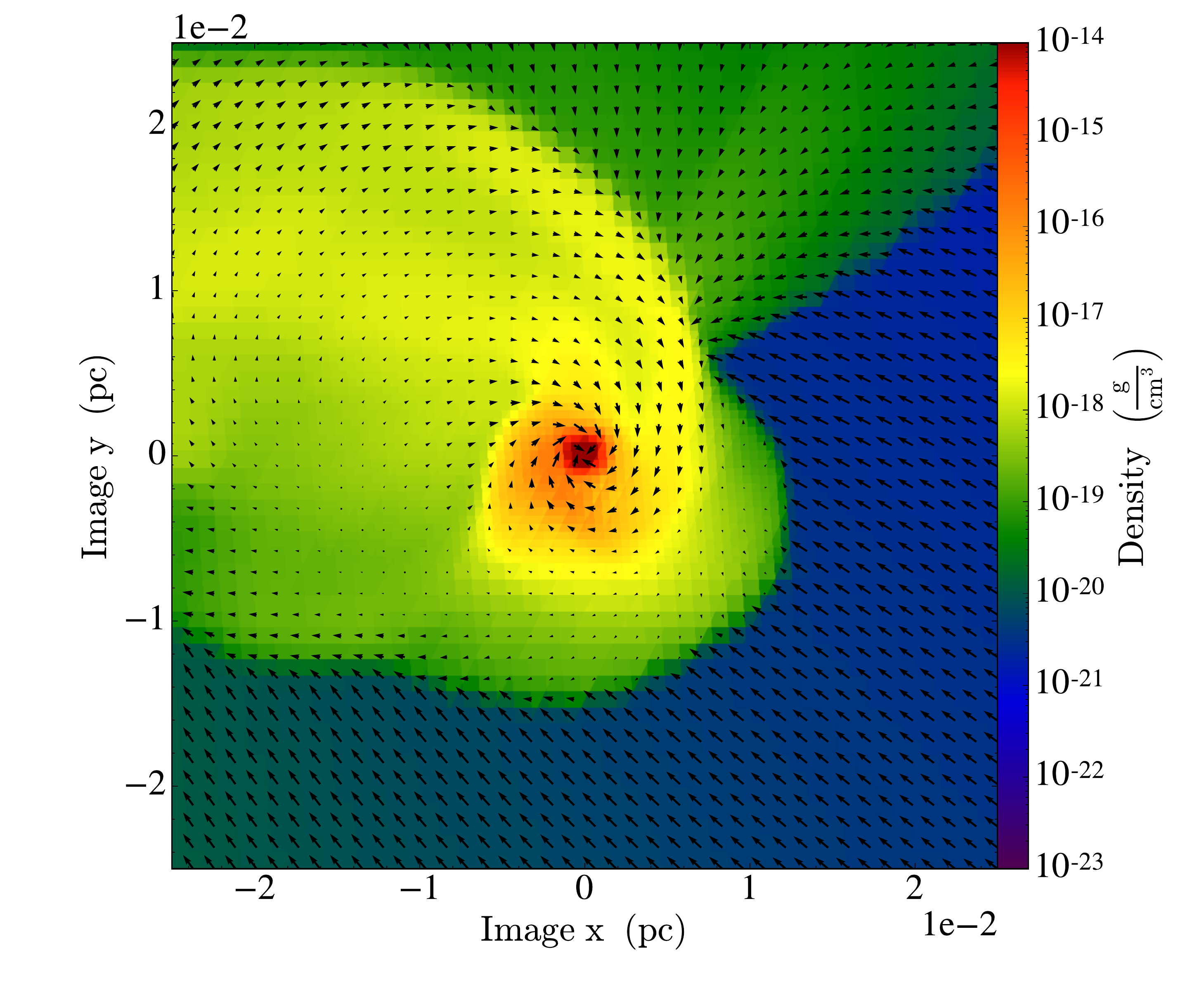}
\caption{The top panel shows the run of velocity around the same density peak 
as that shown in Figure \ref{fig:Sphere_of_influence_quad2_well_before}, 
but now only $\sim 30,000$ years before the formation of \partA .
The color and linestyles are the same as in the top panel of Figure \ref{fig:Sphere_of_influence_quad2_well_before}.
The bottom panel again shows the density in a slice centered on the density peak.
The plotted arrows show the velocity in the plane of the slice.
The longest arrows correspond to roughly $2 {\rm km \, s^{-1}}$.
In the intervening $\sim 70,000$ years since the time shown in Figure \ref{fig:Sphere_of_influence_quad2_well_before}, an accretion disk-like structure has formed, which has a 
mass of $\sim 0.7 \, M_\odot$.
The radius of the sphere of influence (of the disk) is $\sim$0.02 pc. All three velocities, 
$|u_r|$, $v_T$, and $v_\phi$, increase inward of $r_*$; the inflow is 
disrupted at $r \sim0.015\pc$ a feature that we interpret as a shock, where 
the flow meets the nascent 
accretion disk, at which point $v_T$ also drops in magnitude. 
At yet smaller radii the infall resumes, because at this early time the disk is not 
yet fully rotationally supported. The resolution at the location of the star 
particle has reached the refinement limit $\Delta x=5\times10^{-4}\pc$; the 
errors in the calculation of the velocities that are associated with the 
finite resolution are substantial inside $r\approx 0.002\pc$, so features 
inside this radius are not reliable, and thus not plotted.}
\label{fig:Sphere_influence_quad2_prior_to_particle}
\end{figure}

The infall velocity, $u_r$, and random motion ($v_T$) 
 velocity are similar in magnitude, and 
somewhat smaller than the Keplerian velocity, $v_K = \sqrt{GM(<r) /r}$.
Note that $u_r$ is roughly equal to the sound speed while $v_T$ is supersonic.
The fact that the infall velocity is $\sim 25\%$ of the free-fall velocity 
over all radii less than a parsec shows that this system is not in hydrostatic equilibrium.
The density distribution is smooth and filamentary.
The run of density versus radius, not shown, is a simple power law with a small inner core.

Figure \ref{fig:Sphere_influence_quad2_prior_to_particle} shows the region 
around the same density maximum some 70,000 years later, 30,000 years before 
star \partA\ forms.
Once again the infall velocity is a substantial fraction of the Keplerian velocity, 
showing that the core remains far from hydrostatic equilibrium.
However, $v_{\phi}$ in the innermost regions (inside $0.01\pc$) is 
comparable to both $v_K$ and $v_T$, showing that the innermost 
region is partially rotationally supported. The density slice, shown in the bottom 
panel of Figure \ref{fig:Sphere_influence_quad2_prior_to_particle}, confirms this  
interpretation, showing a disk-like structure with a radius of order $\sim 0.01 \pc$. 
The mass inside this radius is $\sim 0.7\,M_\odot$.
We note that the particle forms near the tip of a filament (not shown).

\subsection{The Stellar Sphere of Influence} \label{sec:r_star}
We begin by developing an operational definition of $r_*$.  We choose to define $r_*$ 
as the radius where the enclosed mass, $M(<r, t)$ is three times the mass 
of the star, i.e., 
\be
3M_*(t) = M(<r_*(t), t)
\label{eq:rstar def}
\ee
similar to \citet{2015ApJ...804...44M}. 
We use the factor of 3 to ensure that the gravity of the gas dominates the gravity 
of the star.\footnote{
The gas in the disk around the protostar is rotationally supported, so 
it essentially acts as a part of the star. We include the mass of the disk 
when calculating $r_*$ and discuss how we define the disk mass in 
\ref{sec:Rotationally_supported_disks}. \label{footnote:disk_mass}}
In particular, the factor of 3 essentially means that the mass in gas is twice the mass of the central mass (star and disk) and implies that the gravitational acceleration of the gas is twice that of the central mass, which is where the dynamical effects of the gas begins to dominate the dynamical effects of the central mass.

\begin{figure} 
\plotonesmall{f4a.pdf}
\plotonesmall{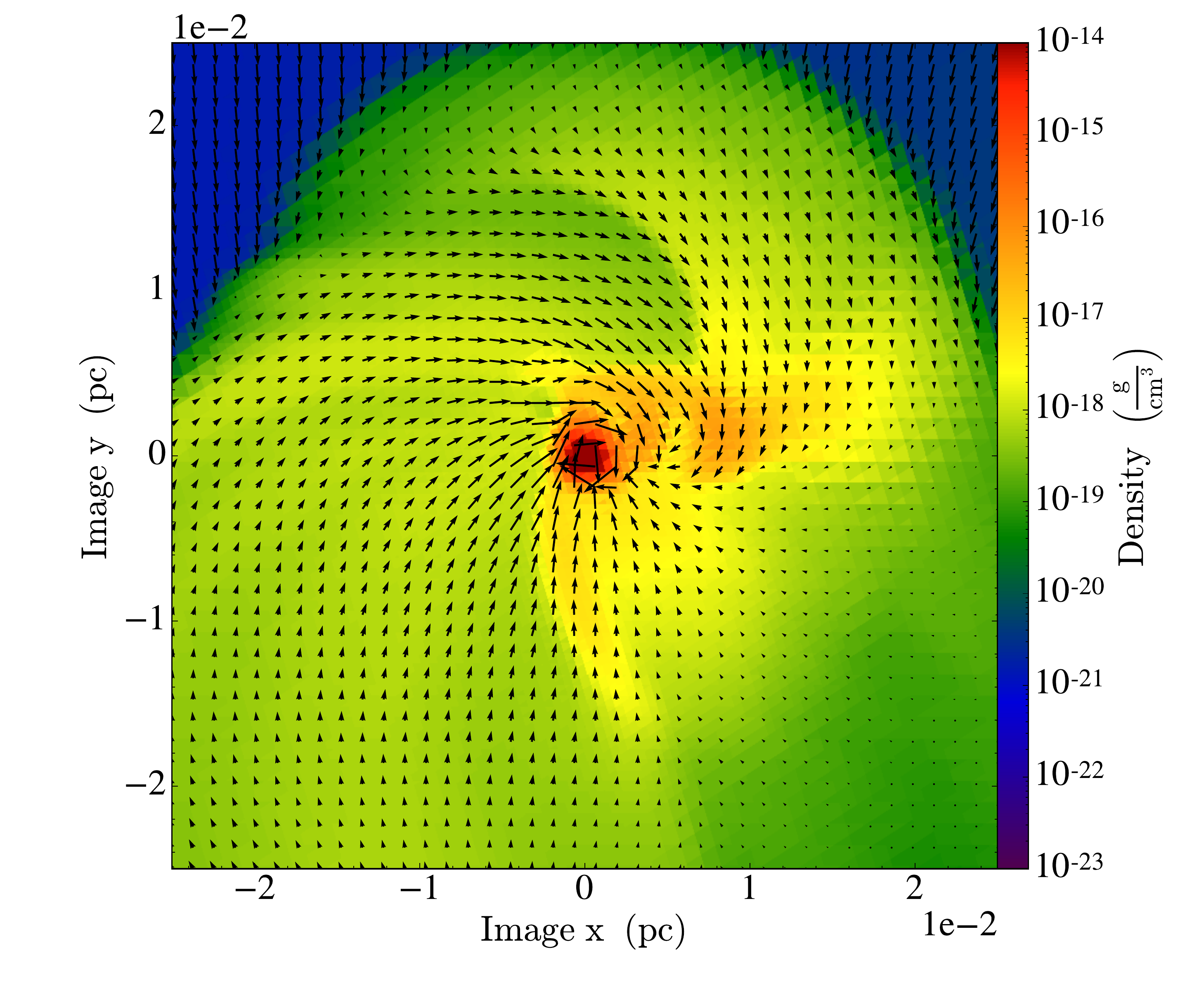}
\caption{The run of velocity (top panel) of \partA\ 24,000 years after 
star particle formation.
The color and linestyles are the same as in the top panel of Figure \ref{fig:Sphere_of_influence_quad2_well_before}.
This panel shows that the disk around the particle is rotationally supported for $r\lesssim 5\times10^{-3}\pc$; inside that radius, 
the black pluses are higher than either the green or blue points, i.e., the rotational velocity is larger than either the random motion
or infall velocity.
The radius of the sphere of influence is $r_*\sim 0.06 \pc$.
The star mass is $2.47 M_\odot$ and the disk mass is $0.58 M_\odot$, 
thus the disk is about 23\% the mass of the star.
The bottom panel is a density slice centered on particle A, in the center of mass frame, face-on to the rotationally supported disk. 
The arrows show the velocity in the plane of the slice. 
The longest arrows correspond to nearly $3 {\rm km \, s^{-1}}$.
With a cell size of $\Delta x=5\times10^{-4}\pc$, we have $\sim 20$ cells across the diameter of the disk.
}
\label{fig:Sphere_influence_quad2_just_after}
\end{figure}

\begin{figure} 
\plotonesmall{f5a.pdf}
\plotonesmall{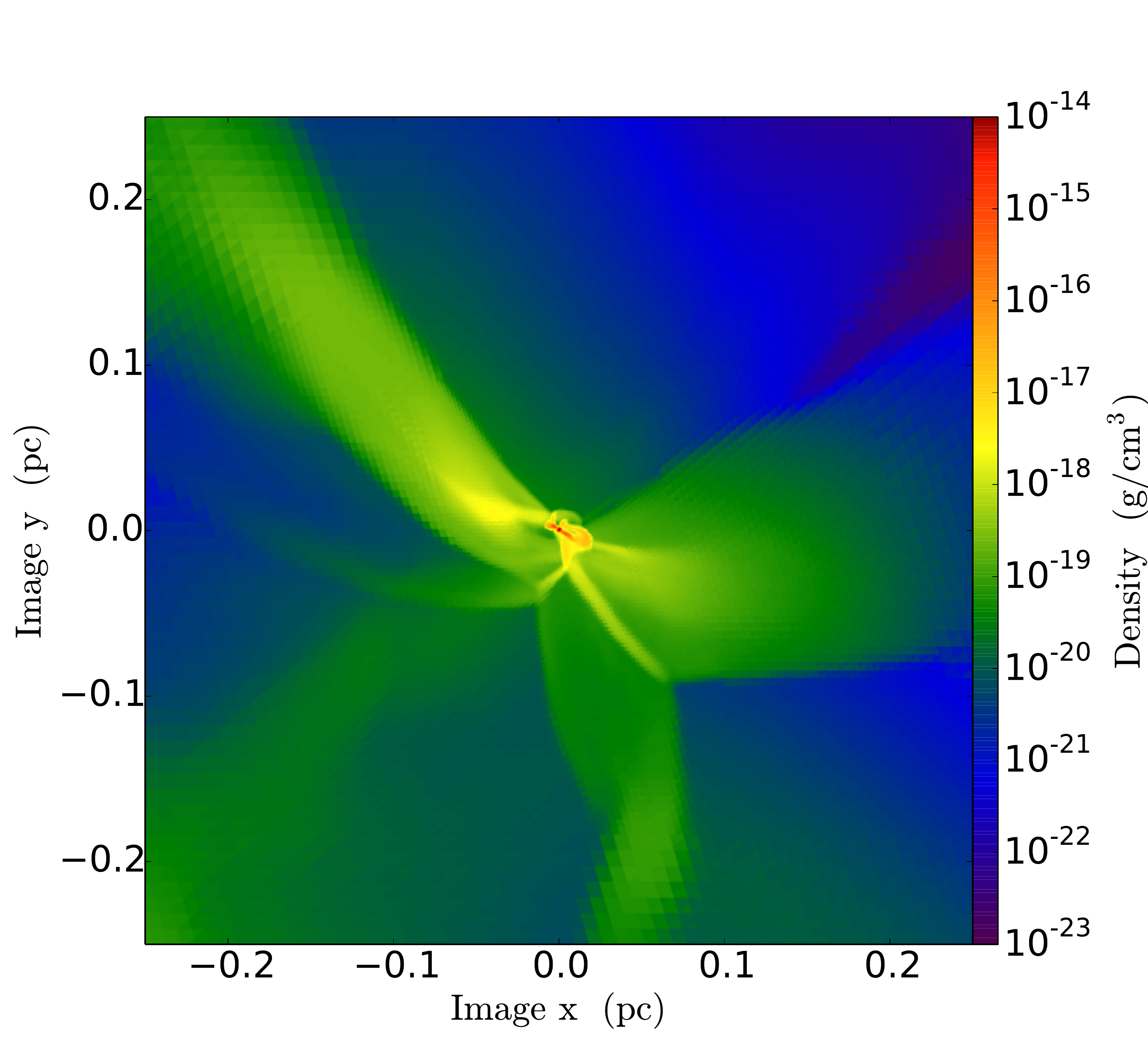}
\caption{The top panel shows the run of velocity for \partB $\sim 100,000$ years after formation.  The radius of the sphere of influence is $~0.18 \pc$.  The stellar mass is $4.5 M_\odot$ and the disk mass is $2.47 M_\odot$ so the disk is $\sim 55\%$ of the
stellar mass. The bottom panel shows the density in a slice centered on \partB.}
\label{fig:Sphere_influence_quad2_end_time}
\end{figure}

Equations (\ref{eq:density}), (\ref{eq:infall_behavior}), and (\ref{eq:rstar def})
predict that the 
character of the solution should change at $r_*$ and that $r_*$ 
increases with time. Our numerical results support this prediction.
Figure \ref{fig:Sphere_influence_quad2_just_after} shows that 
$v_T$ decreases with decreasing radius down to $r_*$ 
and then  increases with decreasing radius inside the sphere of influence. We see that $v_T$ reaches a minimum near $r=r_*$.  
The inward decrease in $v_T(r)$ is not monotonic near $0.2\pc$, probably due to a shock, 
as suggested by jumps in both the infall and random 
 velocities, and 
in the density, at $ r \sim 0.02 \pc$.
This trend of increasing $v_T$ with decreasing radius inside $r_*$ is repeated in Figure 
 \ref{fig:Sphere_influence_quad2_end_time}.

We don't see an increase in the infall velocity for $r > r_*$ for 
this object because the star particles are forming about $1 \pc$ from 
the end of a filament, but we do see an increase in $|u_r|$ in other 
particles, see below.

Comparing Figure \ref{fig:Sphere_influence_quad2_just_after}, which shows 
the velocity and density of the same region 24,000 years after star 
\partA forms, with Figure \ref{fig:Sphere_influence_quad2_prior_to_particle} 
demonstrates that the radius of the change in character of the flow 
associated with $r_*$ increases over time. In particular, the global minimum 
of the random motion 
velocity is now at $0.06 \pc$ rather than somewhere 
between $0.01-0.02 \pc$.

The drop in $u_r$ at large radii in Figure 
\ref{fig:Sphere_influence_quad2_just_after} reflects the vagaries 
of the large scale Reynolds stress pressure gradient
; we already mentioned that this particle is forming near the end of a filament.

Figure \ref{fig:Sphere_influence_quad2_end_time} shows the velocities and the density in a slice centered on particle B, 100,000 years after that star particle forms. 
This star is more isolated than particle A, and as a result $|u_r|$
increases from $r\gtrsim r_*$ out beyond $r\approx 3\pc$. This is in
accord with equations (\ref{eq:infall_behavior}) and (\ref{eq:Mdot_behavior}), but
it contrasts with the result in Figure \ref{fig:Sphere_influence_quad2_just_after}. 

The behavior of $|u_r(r)|$ at large radii is not set by the collapse dynamics, but rather by
the properties of the random motions, 
 most importantly the outer scale of the Reynolds stress gradient
In particular, we do not expect $|u_r(r)|$ to be significant on scales larger than some moderate 
fraction, say 1/4, of the outer scale. In our simulations, the outer scale is given by $k=2$, 
or $L/2$, and we use solenoidal stirring, so that the cascade starts out with no compressive
component, although one develops as the cascade proceeds. In fact 
we will show in \S \ref{sec: average profiles} that the typical radius of a converging region 
is more like $r\approx 1\pc$ in our simulations.

\subsection{A Fixed Point Attractor for $\rho(r, t)$ Inside $r_*$}\label{sec:attractor}

\begin{figure}
\plotone{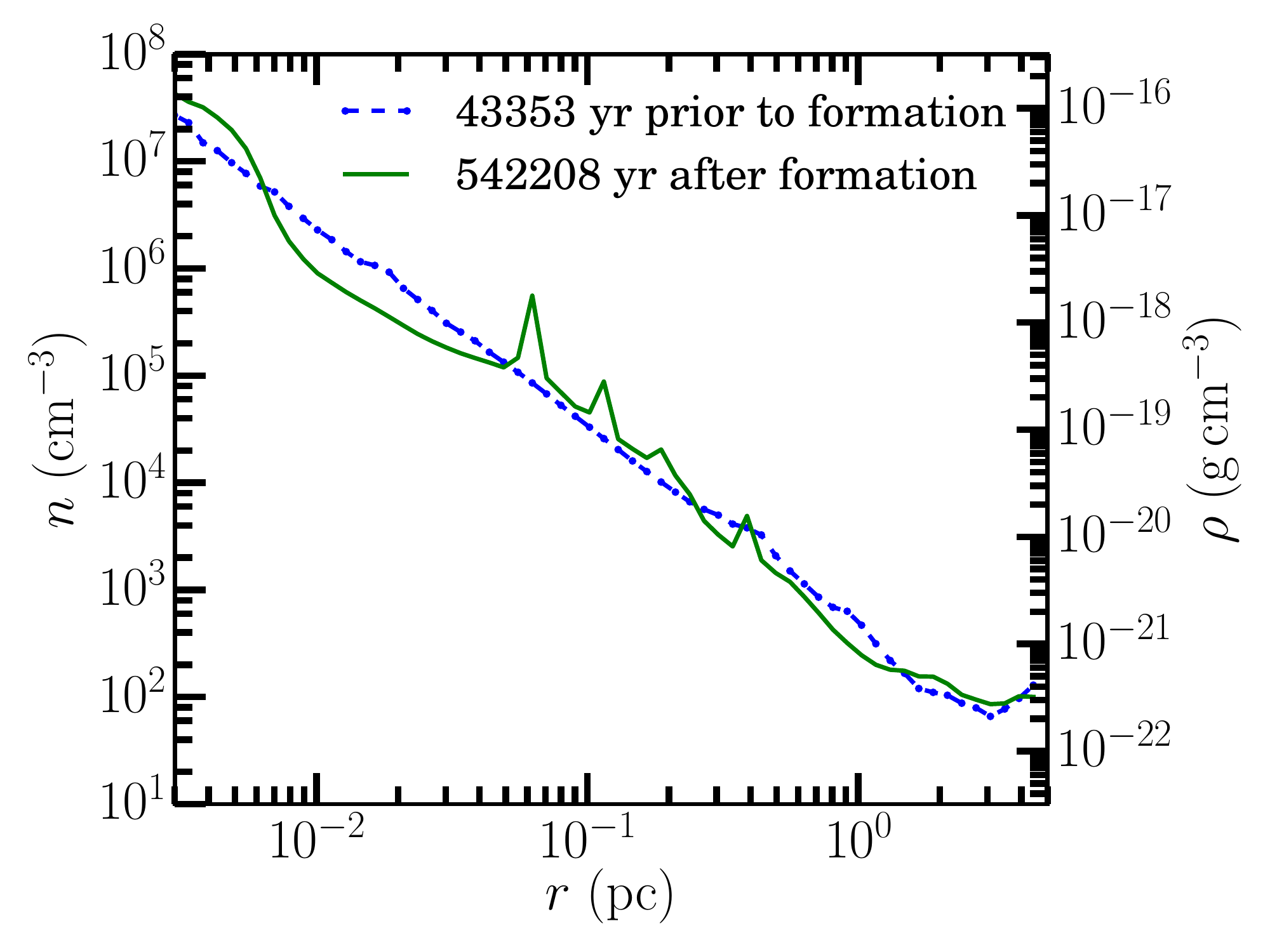}
\caption{The run of density for \partA. The dotted blue line is the 
density $\sim 40,000$ years before the star forms.
The solid green line is the run of density $\approx 540,000$ years after formation. 
The gap in time corresponds to the free-fall time at a radius of $0.24\pc$.
For $r < 0.24 {\rm pc}$ the range of time spanned in the plot is more than a local free-fall time, yet the density does not vary significantly.
There are fluctuation in the density, e.g. the spikes around $r \sim 0.1\pc$
In Figure \ref{fig:density_avg} we average over a number of objects to remove these fluctuations: we also demonstrate that the density in the disc does increase.}
\label{fig:quad2_run_of_density_double}
\end{figure}

One of the most striking findings of MC15 was that the run of density is 
independent of time for $r<r_*$. Our simulations confirm that finding, 
as illustrated in Figure \ref{fig:quad2_run_of_density_double}.
The plot shows the run of density for two separate times.
The dotted blue line shows the run of density $\sim 40,000$ years 
before \partA forms, while the solid green line is the run of 
density $\sim 540,000$ years after the star particle forms.
The elapsed time corresponds to nearly two tenths of the mean free-fall 
time of the box, and to many free-fall times at radii less than a tenth of a parsec.
We emphasize that the density can change on the local free-fall time, which is much smaller than the global free-fall time (by a factor of 10 or more for $r<0.1\pc$). We will show that in fact the density inside $r_d$ does change rather rapidly, after the star particle forms, but that for $r_d<r<r_*$ the density does not change; see \S \ref{sec: average profiles}

The mean power law slope of the density before the star forms (the blue dashed line in the figure) is $k_\rho \approx 1.9$, consistent within the star-to-star variations we see  with the range $k_\rho \approx 1.6 - 1.8$ from equation (\ref{eq:density}) for $r>r_*$ (since in this case $r_*=0$).

%
\subsection{Mass accretion rate}\label{sec:Mass_accretion}

\begin{figure}
\plotone{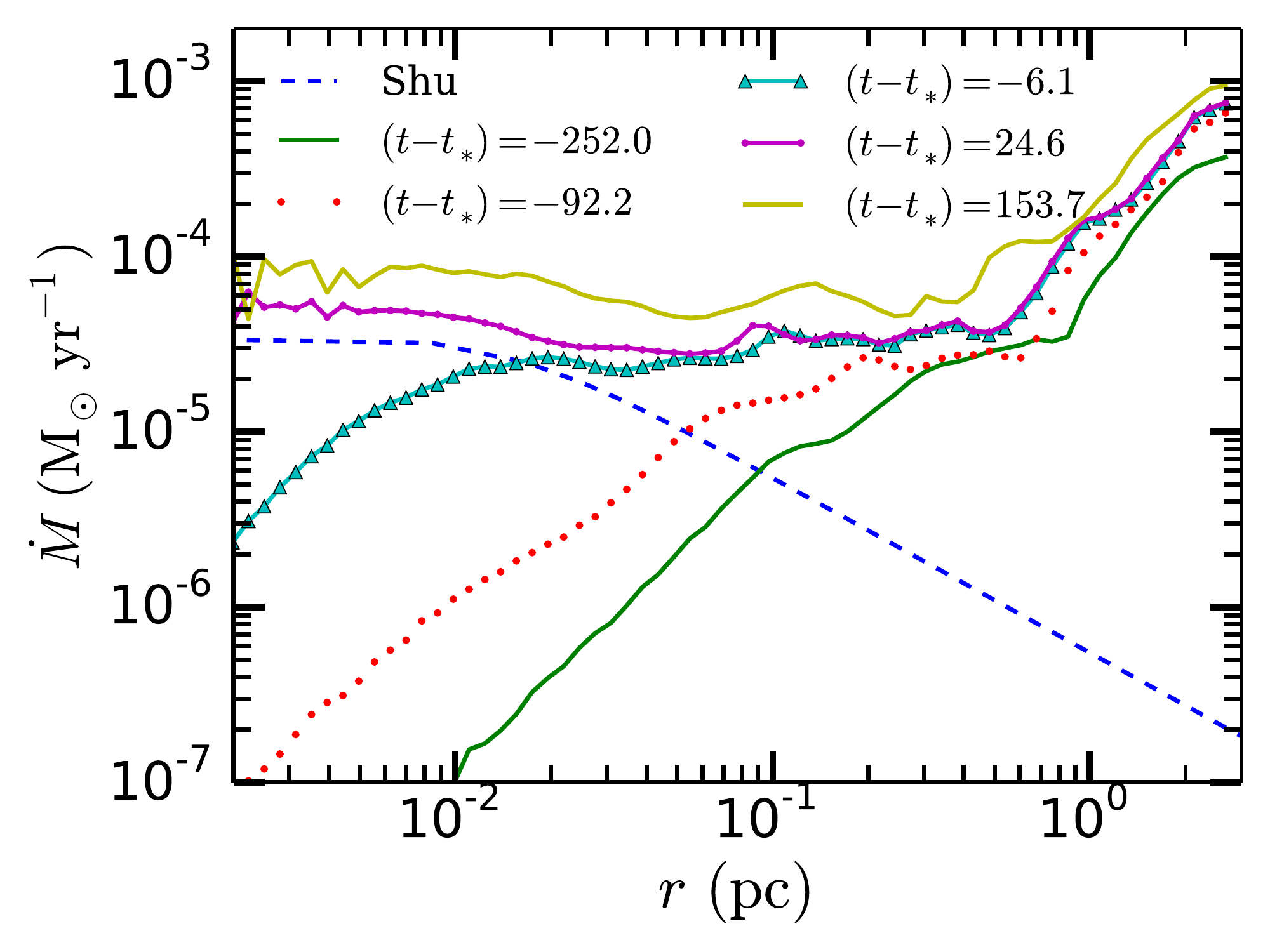}
\caption{The run of $\dot{M}$ for a star particle in a simulation with $N_J=32$ cells per jeans length resolution, at a maximum effective resolution of $16384^3$. The lowest solid (green) line is the run of $\dot{M}$  for the density peak that will form the star particle 252,000 years after the time plotted. The accretion rate is about an order of magnitude lower at small radii (say $10^{-2}\pc$) than at $1\pc$. The purple line connecting the dots is the run of accretion soon after the star particle forms, when the stellar mass is about a solar mass. The accretion rate at $1\pc$ still exceeds that at all smaller radii, showing that the collapse is outside-in, not inside-out. As an example of an inside-out collapse, we show the accretion rate for the \citet{1977ApJ...214..488S} model (the dashed line) for a star of a solar mass with $A = 5.5$. 
\label{fig:Mdot_vs_shu}}
\end{figure}

In Figure \ref{fig:Mdot_vs_shu} we show the mass accretion rate $\dot{M}$ as a function of r around a star particle ($t-t_* > 0$) and from the corresponding density peak in which the star particle eventually formed ($t-t_*<0$). This plot is taken from a RAMSES simulation. Before the star particle forms, $\dot{M}$ decreases inward at all radii. 

Following the establishment of the power law solution for the density, at $t=t_*$, a star particle forms and the $\dot{M}$ profiles flatten at small radii. An examination of the density profile (not plotted) reveals that $\rho\propto r^{-3/2}$, while for $t-t_*=24$kyrs, the gravitational force (and hence $u_r$) is dominated by the central mass for $r\lesssim0.1\pc$, so that $v\propto r^{-1/2}$ out to that radius. We also note that while the $\dot{M}$ profile is flat, it does increase in time as shown by the difference between the $t-t_* = 24.6$ kyrs and $154$ kyrs curves. All this behavior agrees well with the prediction of Equation (\ref{eq:Mdot_behavior}).

At all times, the accretion rate is either nearly flat or increasing with radius, which is a natural result of the near balance between gravity and Reynold's stress support, as posited in the theory of MC15. We contrast this with an inside-out collapse model, which we exemplify using a \citet{1977ApJ...214..488S} solution (blue dashed line) obtained by directly integrating equations (11) and (12) of \citet{1977ApJ...214..488S} at a fixed time. The asymptotic behavior of $\dot{M}$ follows from Shu's equations (15) and (17);
recall that $x = r/ (c_s  t)$ is a function of the radius. 
In the limit of small $x$, $\dot{M}$ approaches a constant.
However, for large values of x,  
$\dot M(r,t)=-A(A-2) c_s^3/Gx$ (Equation [15] of \citealt{1977ApJ...214..488S}), i.e., the mass accretion rate falls like 1/r at a fixed time at large r as seen in Figure 7. 

In other words, for inside-out collapse models, the accretion rate is monotonically decreasing with increasing radius. This is qualitatively different from the prediction of MC15 or the results of this work. We note that while we have chosen to plot the Shu solution, other collapse solutions \citep{1997ApJ...476..750M,2002Natur.416...59M,2003ApJ...585..850M} have the same general profile: the mass accretion rate is roughly independent of $r$ at small radii, and decreases with increasing $r$ at large radii.  

In summary, at no time do we see any indication of an inside-out collapse in our simulated massive star forming regions.   

\subsection{Rotationally Supported Disks}
\label{sec:Rotationally_supported_disks}
Many of the qualitative and even quantitative features predicted by MC15 are 
found in our simulations as discussed above, including the approach of the 
density profile inside $r_*$ to an attractor 
solution, the minimum in the velocity profile around the sphere of influence, 
and the expansion of the sphere of influence with time.
However, our simulations display additional dynamics that were not modeled by MC15.  

A particularly interesting bit of dynamics neglected by MC15 is the 
development of a rotationally supported disk, which we alluded to above.
This development is evident in the velocity plots, starting from the absence of a disk in 
Figure \ref{fig:Sphere_of_influence_quad2_well_before} to a proto-disk 
with no central star particle in Figure 
\ref{fig:Sphere_influence_quad2_prior_to_particle}, to a fairly well 
developed rotationally dominated disk, at $r \lesssim 0.05 \pc$ in 
Figure \ref{fig:Sphere_influence_quad2_just_after}. 

We define the outer edge of the accretion disk $r_d$ as the largest radius 
where $v_\phi$ exceeds both $|u_r|$ and $v_T$, that is, where the 
disk is rotationally dominated. The development of the disk is best followed 
by examining the rotational velocity seen in 
Figures \ref{fig:Sphere_of_influence_quad2_well_before},
\ref{fig:Sphere_influence_quad2_prior_to_particle} and
\ref{fig:Sphere_influence_quad2_just_after}. In the last 
figure, $r_d\approx7\times10^{-3}\pc$. 
We have also used a second definition for the disk radius, i.e., 
where the derivative of the density has a sharp drop, see footnote \ref{footnote:disk_mass}. The two definitions of the disk radii agree 
well with each other.

We note that the disks in our simulation have $r_d\sim 1,000$ AU. 
This is somewhat larger than the radii of the largest observed disks, e.g., \citet{1999AJ....117.1490P} find
$500\AU\lesssim r_d\lesssim 1000\AU$. Of course we are simulating massive
star formation, and most observations of disks are of nearby, low mass stars. Another
factor to keep in mind is that we are doing hydrodynamic simulations, so there are
no magnetic fields, which are believed to be effective at transporting angular momentum; 
the inclusion of magnetic fields might therefore tend to reduce the sizes of the accretion
disks in our simulations. 

\subsection{Gravitationally Unstable Disks}

The plot of $\dot{M}$ in Figure \ref{fig:Mdot_vs_shu} shows that the accretion rate varies little across the transition from the rotationally supported disk to the 
radial infall dominated part of the flow at slightly larger radii. In other 
words, the disk is transporting angular momentum efficiently enough so that 
the disk accretion rate matches the rate at larger radii. Since our simulations 
do not include magnetic fields, this efficient disk accretion is not due to the magneto-rotational instability \citep{1991ApJ...376..214B,1998RvMP...70....1B}.

\begin{figure} 
\plotone{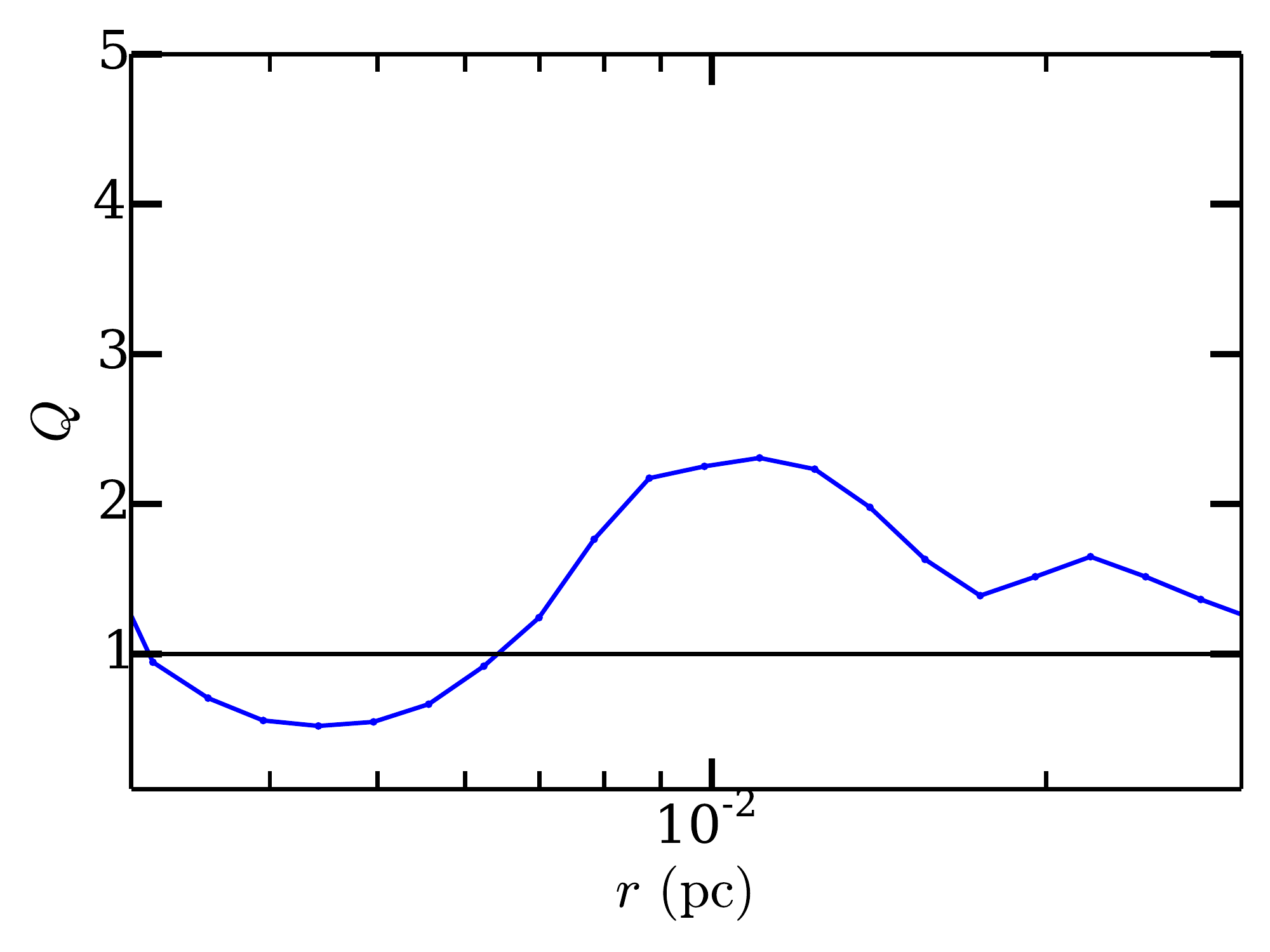}
\caption{The Toomre Q parameter for \partB, at the same time as shown in
Figure \ref{fig:Sphere_influence_quad2_end_time}, $\sim 100,000$ years 
after star particle formation. From Figure 
(\ref{fig:Sphere_influence_quad2_end_time}) we see that the disk is rotationally
dominated for $r<2\times10^{-2}\pc$. 
For $3\times10^{-3} \lesssim r \lesssim 6\times10^{-3} \pc$, $Q\lesssim 1$. 
This indicates that the disk is gravitationally unstable at these radii, 
while it is marginally stable at larger radii.
Figure \ref{fig:Q_avg} provides a more representative view of disk stability.}
\label{fig:Toomre_Q_2B}
\end{figure}

Following \citet{2010ApJ...708.1585K}, we suggest that angular momentum is transported via gravitational torques. We have not yet
tried to calculate these torques, but as a first check, we have calculated the 
Toomre Q parameter, as shown in Figure \ref{fig:Toomre_Q_2B}; recall that 
\be 
Q = \frac{v_\phi \sqrt{v_T^2 + c_s^2}}{\pi G r \Sigma}.
\label{eq:Toomre_Q}
\ee
In this expression $\Sigma$ is the gas surface density of the disk.
The figure shows $Q$ for the disk around \partB at the time shown in Figure
\ref{fig:Sphere_influence_quad2_end_time}.
For the region $3\times10^{-3} \lesssim r \lesssim 6\times10^{-3} \pc$, $Q$ 
is below one, which supports the notion that the efficient accretion 
is due to gravitational torques resulting from a gravitational instability 
in the disk. However, in the next section, we find results suggesting that
the accretion disks in our simulation are not gravitationally unstable at all
times.

\subsection{Average Profiles}\label{sec: average profiles}
Thus far, we have focused our attention on two of our stars and shown
that their $v_T$, $u_r$ and $\rho$ profiles are qualitatively similar
to the profiles predicted in the analytic work of MC15. Now we will
show that this behavior is generic, in the sense that this is true on
average over all the star particles in our simulations.

\begin{figure*}
\plottwo{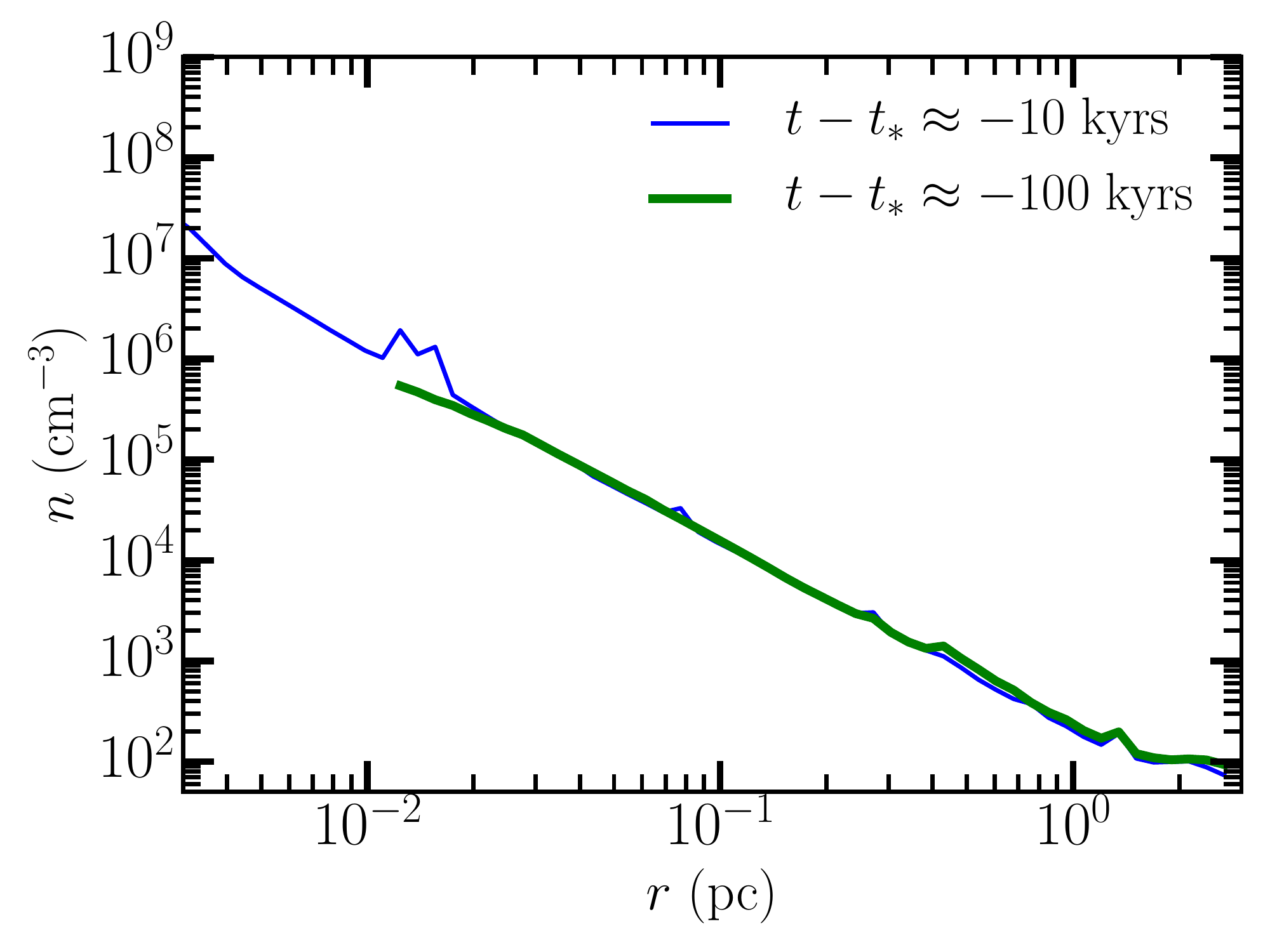}{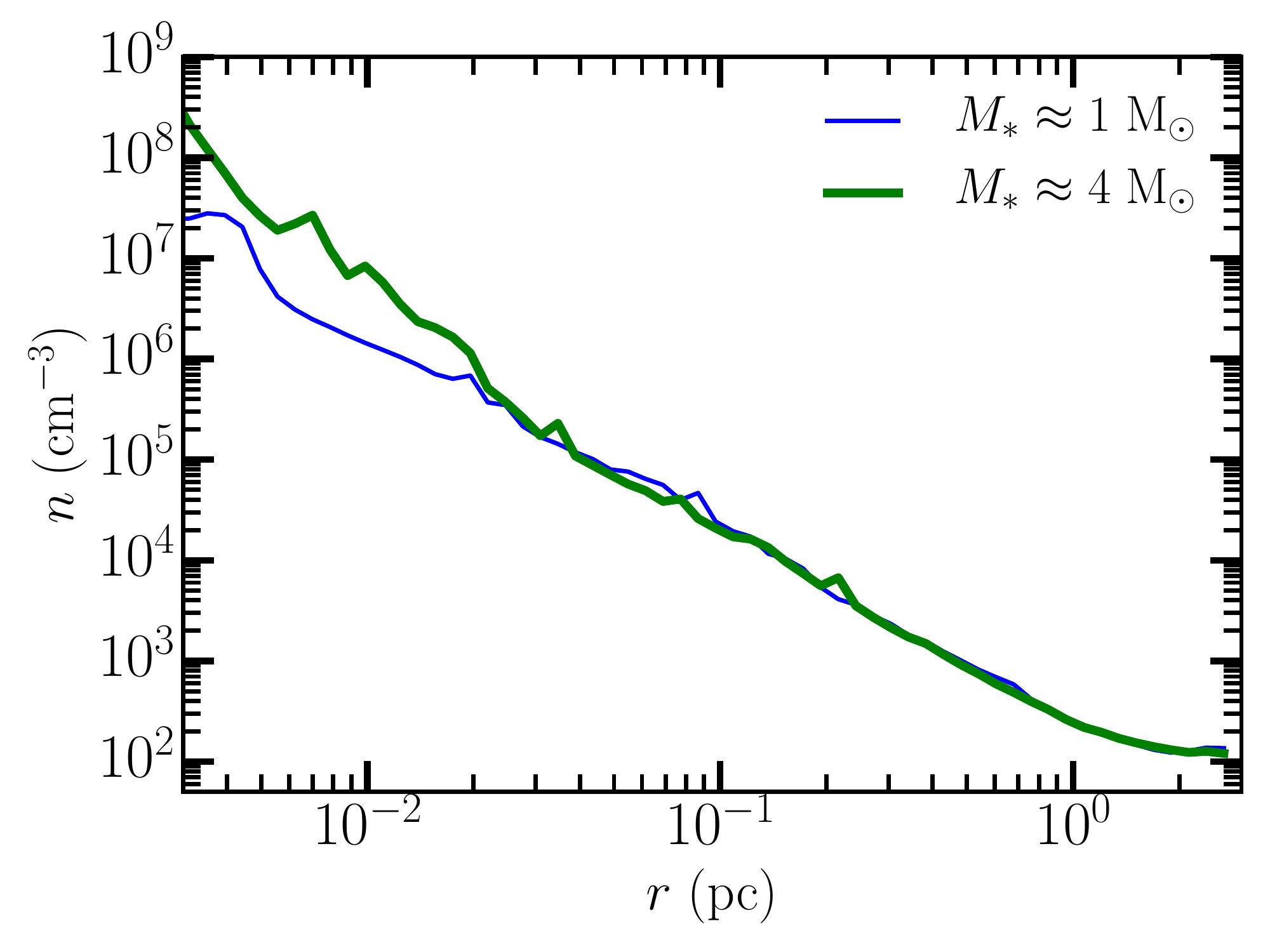}
\caption{Number density as a function of radius at 10,000 and 100,000 
years before the star particle forms (left plot) and when the star 
reaches 1 and 4 $M_{\odot}$ (right plot).  For the left plot, we average 
over 6 and 7 regions that are within 25\% of 10,000 and 100,000 years 
prior to star particle formation. The line corresponding to 
$t-t_*=-100,000\yrs$ terminates at $r=0.02\pc$ because that corresponds 
to the level of refinement for the local density peak 
($n\approx 10^6\,\cm^{-3}$) at that time. For the right plot, we 
average over 14 and 6 regions that contain 1 or 4 $M_{\odot}$ 
star particles (within 0.5 $M_{\odot}$).  A power law fit to either curve between 0.02 pc and 1 pc gives a power law of $n \propto r^{-\kappa_{\rho}}$ with $\kappa_{\rho} \approx 1.8$.
At $r=0.1\pc$ the free fall time is $\approx 250,000$ years, roughly the span of time show across the two panels of the plot.
For radii between $r_d\approx 0.02 \pc$ and $1 \pc$ the density profile does not 
change appreciably between the left and right plots. The lack of change for 
$0.1\pc \lesssim r\lesssim 1\pc$ is unsurprising, since the elapsed time is 
less than the local dynamical time at these radii. However, the same cannot be 
said about the lack of evolution of $\rho(r,t)$ in the range $r_d<r\lesssim 0.1\pc$. 
The collapse theories of \citep{1977ApJ...214..488S,1997ApJ...476..750M,2003ApJ...585..850M} 
predict that the density should vary with time, but this is not what we see.  
Note that the density profile does increase for $r<r_d$; compare the green line 
(the $4 M_\odot$ profile) versus the blue line (the $1 M_\odot$ profile) in the 
right hand panel. This is consistent with mass accreting onto the disk faster 
than it is transported in towards the star, in such a way that $Q\approx1$ at 
all times. The change in the density for $r<r_d$ demonstrates that the density 
{\em can} evolve on the time scale of our simulations. Thus the fact that the 
density does not change for $r_d<r<r_*$ between the two plots is striking. 
It is, however, what is predicted by Equation (\ref{eq:density}).
}
\label{fig:density_avg}
\end{figure*}
At the end of our base FLASH simulations, we have found roughly 60 star particles.  
To study these systems in a generic way, we look at the 
average velocity and density profiles. Motivated by the results of MC15, 
we average the profiles at fixed stellar mass;  
by fixing the stellar mass, 
we fix $r_*$, and hence the velocity, $\rho$, and $\dot{M}$ profiles.  

For epochs before a star particle forms, it is less clear how these profiles 
should be averaged.  However, equation (\ref{eq:density}) predicts that 
$\rho(r,t)$ approaches a time independent function as soon as any 
non-pressure supported structure, such as a disk, forms.
As a result, we elect to follow the methodology in
\citet{2015ApJ...800...49L} and average profiles at fixed times 
(10 and 100 kyrs) before the formation of a star particle.  The 
choice of these two times allows us to study the conditions in the 
collapsing region immediately before and well before the formation of 
the star particle, while retaining several (six to seven) density peaks and hence
reasonable statistics.  

In Figure \ref{fig:density_avg}, we plot $n$ as a function of 
$r$, 10,000 and 100,000 years before star particle formation (left plot),
and for stellar mass $M_* = 1$ 
and 4 $M_{\odot}$ (right plot). The plots show 
that  $\rho(r,t) \to \rho(r)$ for $r_d< r<r_*$, i.e., the density approaches the attractor solution,
early in the collapse, and this profile persists through formation 
of the star particle and well after. This generalizes what we found for 
our two example star particles in \S \ref{sec:attractor}.  

It is important to note that the lack of change in the run of density is not due to the fact that we integrate for only a few tenths of a global free fall time. To emphasize this point, observe that the density at $r<r_d$ {\em does} increase, while the density for $r_d< r<r_*$ 
{\em does not} increase with time.

The reason for the increase of $\rho(r,t)$ with time for $r<r_d$ is easy to understand: the gas is in a rotationally supported disk, which (as Figure \ref{fig:Toomre_Q_2B} shows) is marginally unstable. As the central stellar mass grows, the mass of the disk surrounding it will grow as well, in such a way that $Q\sim M_d/M_*$, where $M_d$ is the disk mass, is roughly constant. 

The fact that the density inside $r_d$ increases illustrates the general point that the relevant time scale for the run of density to change can be much shorter than the global dynamical time scale. If one had to wait for a global dynamical time, the density in the disk would not change over the entire course of our simulation, but Figure \ref{fig:density_avg} shows that the density in the disk does change over a tenth of the global dynamical (or free-fall) time. Thus the result that $\rho(r,t)\to\rho(r)$ for $r_d< r < r_*$ is not a result of our short (relative to the global dynamical time) integration.

The one dimensional numerical models in MC15 also showed that
$\rho(r,t)\to \rho(r)$; MC15 find that the fixed point solution is
approached from outside-in (see their Figure 1). We see the
same behavior in the simulations we have run with $N_J = 16$ and
with $N_J=32$. In those runs we see the flattening of the density at
small radii and early times, before the star particle forms. 
 
\begin{figure} 
\plotone{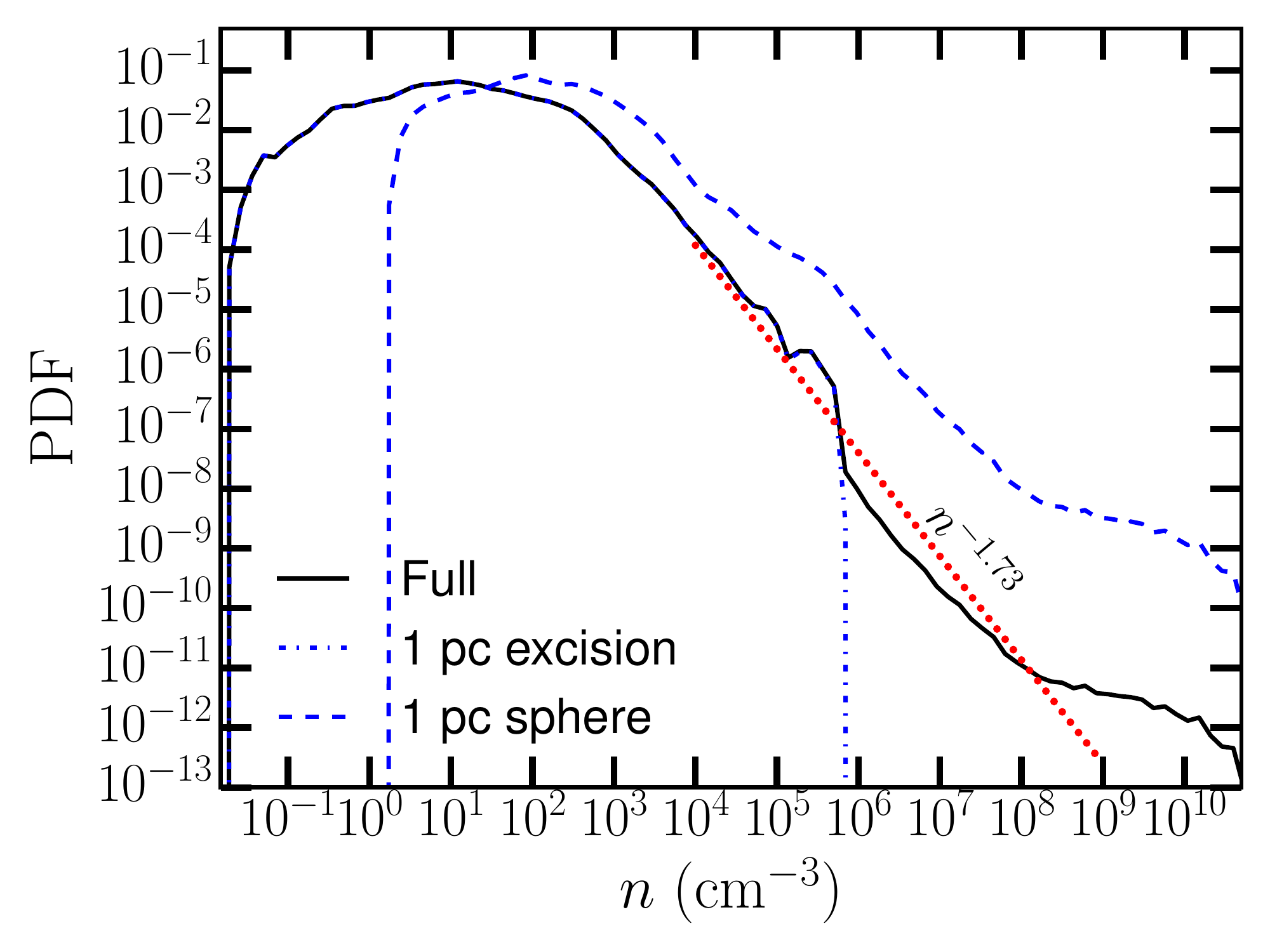}
\caption{The probability distribution function of $n$. The black thin solid line
shows the full PDF, the blue thin dot-dashed line shows the PDF with 1 pc spheres 
cut out around star particles, and the blue thin dash line shows 
the PDF within those spheres. 
The red dotted line show the power law $\propto n^{-1.73}$.  }
\label{fig:PDF_quad1}
\end{figure}

 In Figure \ref{fig:PDF_quad1}, we show the density probability 
 distribution function (PDF) of one of our simulations. 
The black line shows the result for the full box.
The blue thin dot-dash line shows the result when we excise a 1 pc sphere around each star particle.
Finally, the thin blue dashed line shows the PDF of all the 1 pc spheres around each star particle.
At high densities, the PDF exhibits power 
 law behavior, as found by previous workers
 \citep{2000ApJ...535..869K,2011ApJ...727L..20K,2015ApJ...800...49L}.  
Moreover these high density regions are localized around star particles, 
 as the PDF with 1 pc spheres excised around star particles shows (blue dot-dashed line). 
We also note that the regions around star particles are not devoid of low density regions, 
as the PDF of the 1 pc spheres around star particles (blue dashed line) shows.
 \citet{2011ApJ...727L..20K} first argued that the power law tail of the 
 density PDF at high densities is related to the scaling of the density with radius; 
 for $\rho \propto r^{-\alpha}$, the density PDF $\propto \rho^{-3/\alpha}$.  
 For the values of $\alpha$ ($\alpha \approx 1.5-2$) that we expect from 
 analytic theory (MC15) and from previous numerical calculations
 \citep{2015ApJ...800...49L}, we expect the density pdf 
 to scale like $\rho^{-2}$ to $\rho^{-3/2}$. We fit a power law between 
 $n=10^4-10^9\,{\rm cm}^{-3}$ (red dotted line) and find a scaling like 
 $n^{-1.7}$, in line with these expectations.

The power law shows a break to a flatter slope at $n\approx 10^8\cm^{-3}$. A similar 
break was seen by 
\citet{2011ApJ...727L..20K}, who argued that at very high densities, the density PDF 
flattens due to the presence of disks, which they also found. 
In our simulations,  we have found that 
material with $n>10^9\,{\rm cm}^{-3}$ always resides within 
$0.01 \pc$ of a star particle. Since $0.01\pc$ is the typical outer 
radius of our simulated disks, this suggests that the highest density material is strongly associated with the disk.
 
\begin{figure*}
\plottwo{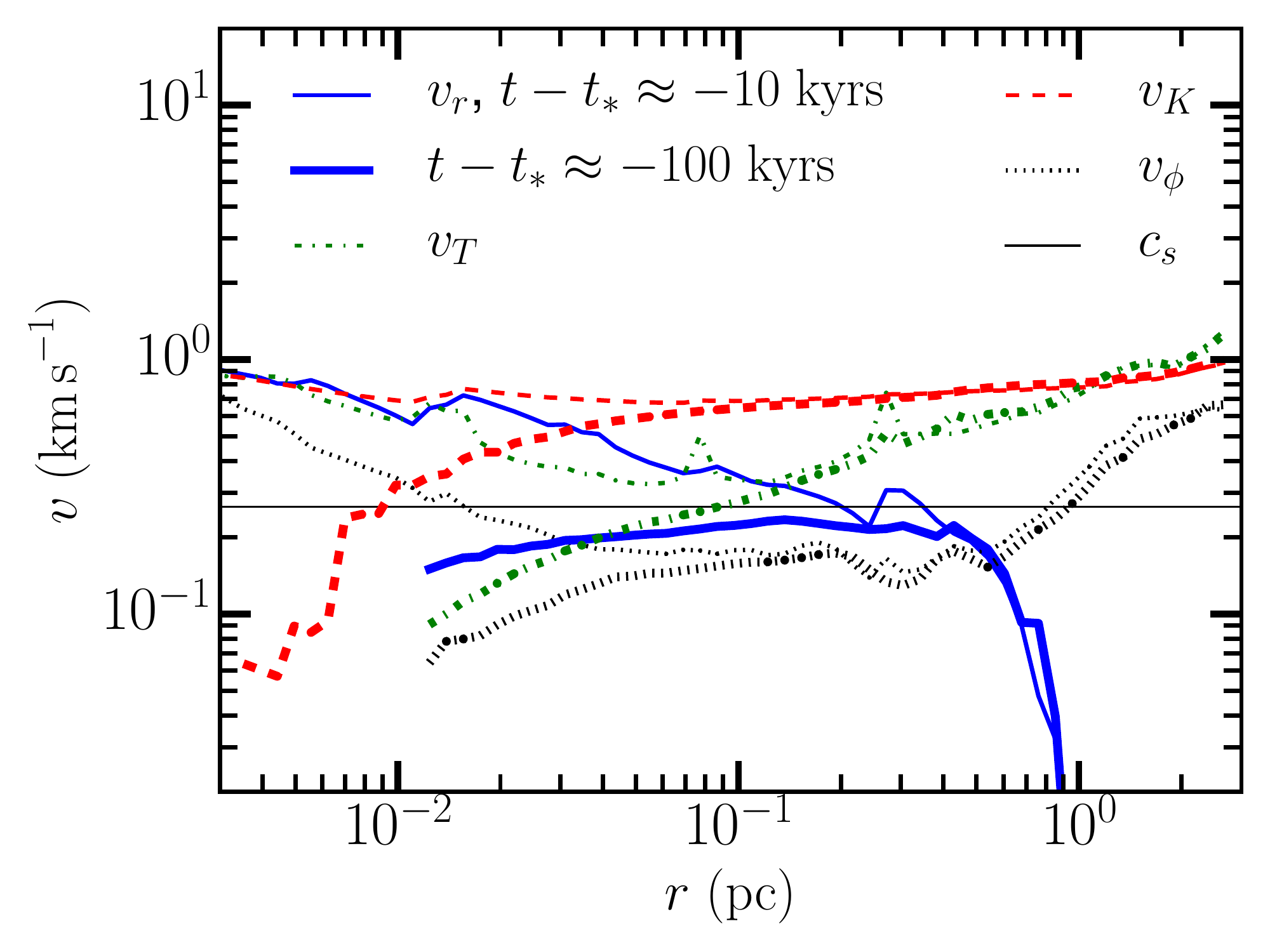}{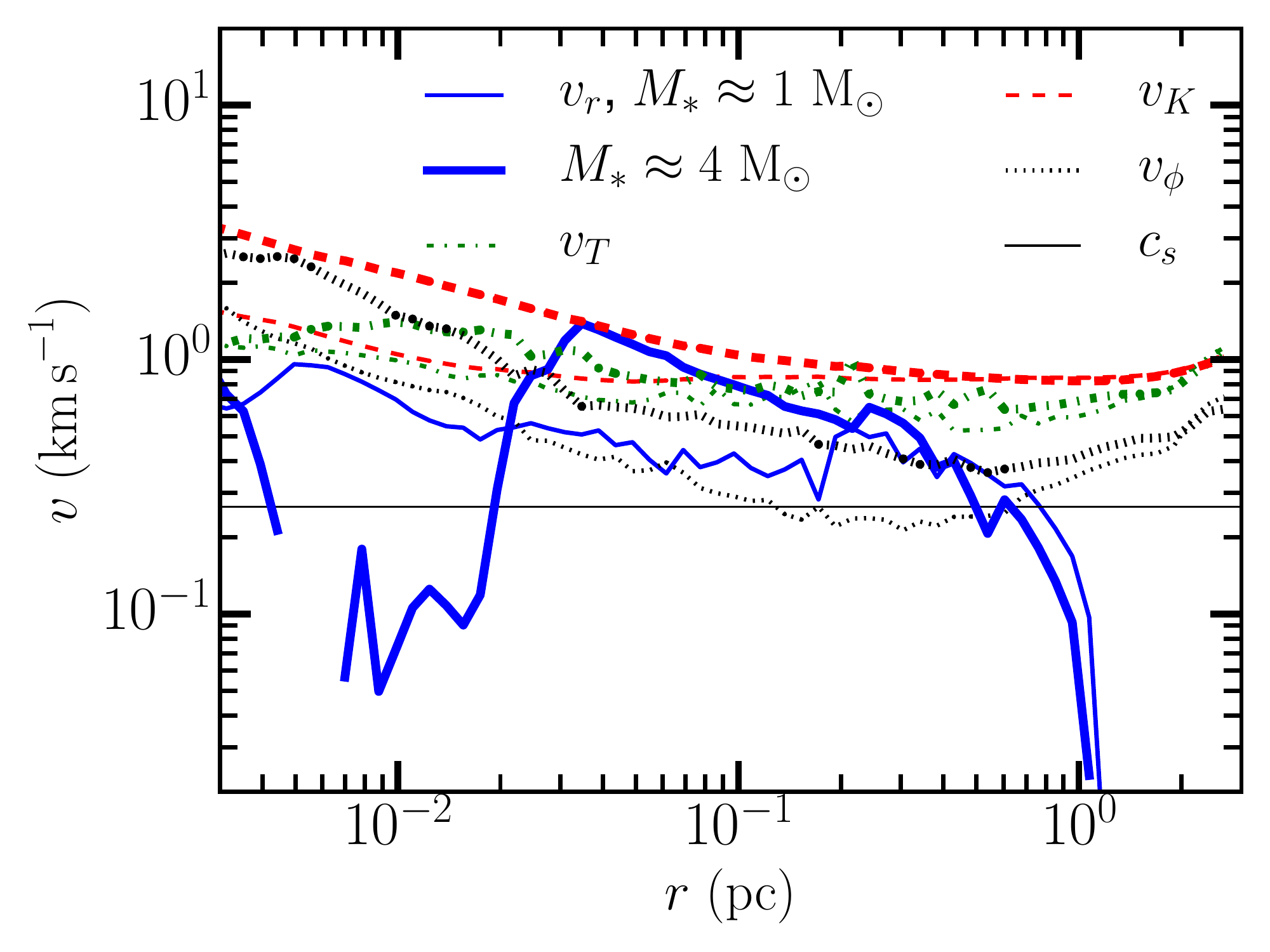}
\caption{$v_r$, $v_T$, $v_{\phi}$ and $v_K$ as a function of $r$ at 10,000 
(thin lines) and 100,000 (thick lines) years before the star particle forms 
(left plot) and when the star reaches 1 and 4 $M_{\odot}$ (right plot).  
The averages are over the same regions as those used in producing 
Figure \ref{fig:density_avg}. The infall $|u_r|$ and random $v_T$ velocities show 
the behavior predicted by the theory of adiabatic turbulent heating for times 
later than $-10,000$ years: 
at large radii, where $|u_r|$ is small, $v_T>|u_r|$ and $v_T$ decreases 
inward, but more slowly than in non-collapsing regions; $p\approx0.2$ 
rather than $p=0.5$. Inside $r_*$, 
where $|u_r|>v_T$ (or $|H|>v_T$ in the notation of 
\citet{2012ApJ...750L..31R}) $v_T$ {\em increases} towards $|u_r|$ 
as $r$ decreases, with both increasing inward.  \label{fig:velocity_avg}}
\end{figure*}

Figure \ref{fig:velocity_avg} shows the averaged $v_r$, $v_T$,  
$v_{\phi}$, $v_K$ as a function of $r$ before the star particle forms 
(left) and after (right panel).  As in Figure \ref{fig:density_avg}, 
we have selected the same fixed times (10 and 100 kyrs) before the star 
particle forms and the same fixed masses ($1$ and $4\,{\rm M}_{\odot}$) 
after star particle formation.  Here the dynamics of $v_T$ follow 
quantitatively the behavior of $v_T$ found in MC15. In MC15, $v_T$ 
scales with radius as $r^{1/2}$ at very large $r$, where self-gravity 
is not important. For example, at $t=100$ kyrs before 
the star particle forms, we find that $v_T(r)\sim r^{0.48}$, in line with
Larson's law, i.e., $\propto r^{1/2}$. 


However, at $t=10 $ kyrs before star particle formation, one can see
that the $v_T$ scaling has reversed itself at small 
radii ($r\lesssim0.1\pc$) due to the accumulation of mass in a 
proto-disk; the gas in the disk deepens the potential well, but does 
not provide radial pressure support. The figure also shows that $|u_r|$ 
increases inward from $r\approx0.1\pc$.

The reversal of the power-law form of both $|u_r|$ and 
$v_T$ as a function of radius tracks 
the position of $r_*$, as can be seen
comparing the lines for $t=-10$ kyrs in the left
plot with $M=1$ and $M=4$ $M\odot$ in the right plot. This confirms another
aspect of the MC15 solution -- that as $r_*$ moves outward with time the 
inflection point in $|u_r|$ and $v_T$ moves outward as well, as we found
earlier in \S \ref{sec:r_star}.

\begin{figure*} 
\plottwo{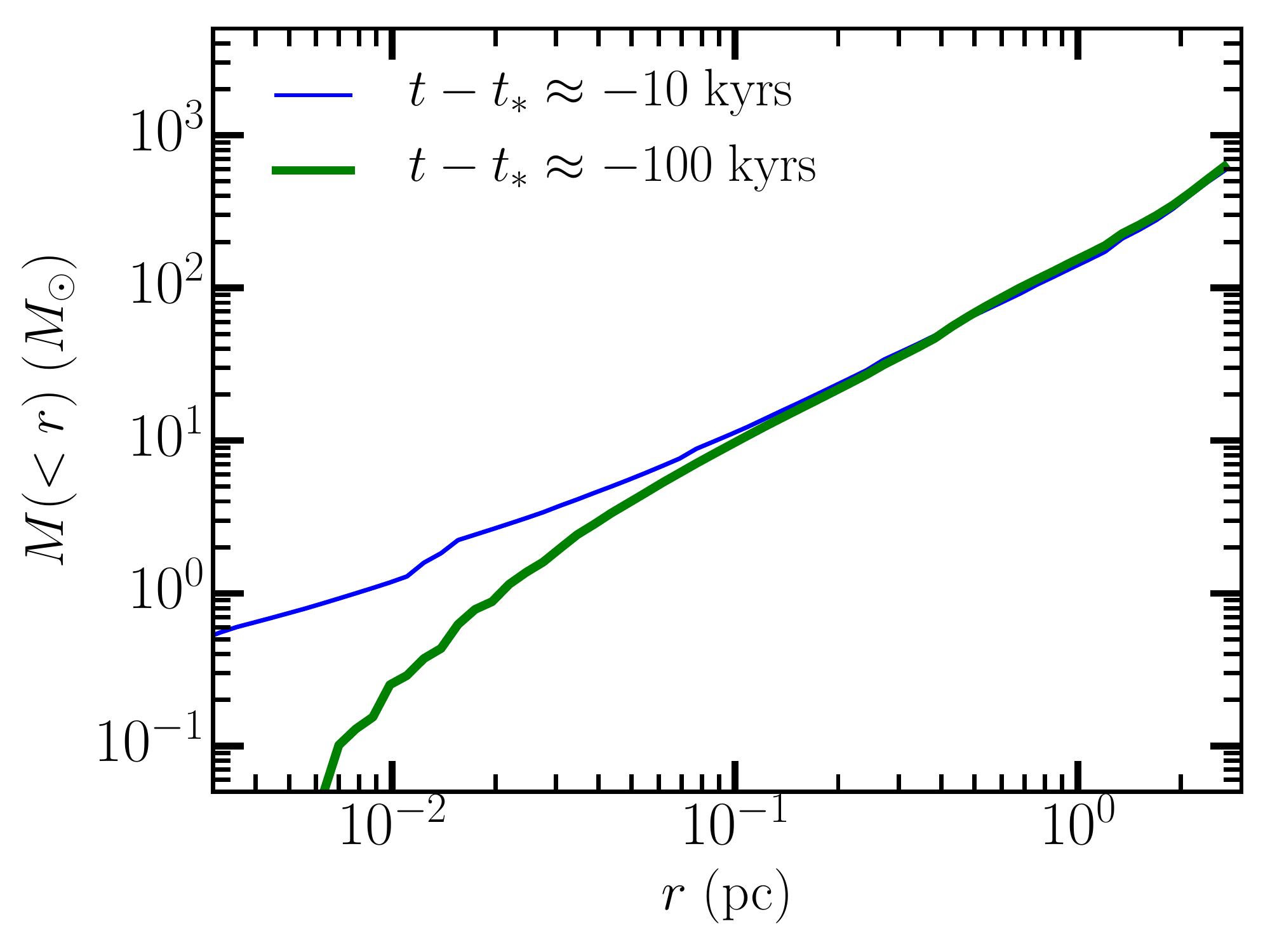}{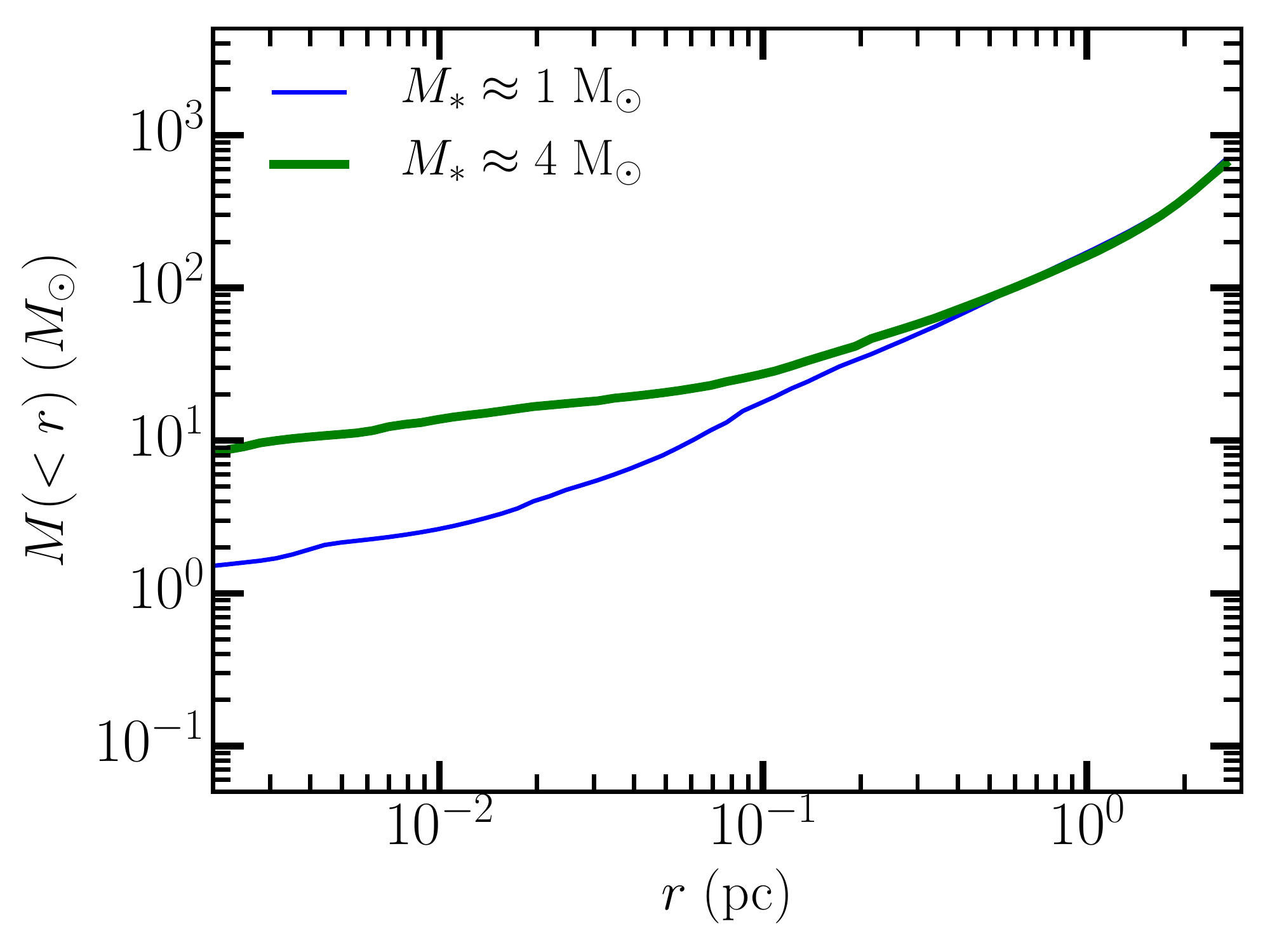}
\caption{Mass of gas and stars as a function of $r$ at 10,000 (thin lines) 
and 100,000 (thick lines) years before the star particle forms (left plot), 
and when the star reaches 1 and 4 $M_{\odot}$ (right plot). Averages 
are as described in Figure \ref{fig:density_avg}. \label{fig:mass_avg}}
\end{figure*}

The steady outward march of the sphere of influence 
is demonstrated in Figure \ref{fig:mass_avg}, 
which shows the run of enclosed mass at four different times.  
At $t=-100,000$ years, in Figure \ref{fig:density_avg} the density cusp is not yet in place; 
correspondingly, at small radii
the enclosed mass is convex, curving down as $r$ decreases. By $t=-10,000$
years cusp formation is complete, and a small disk has formed, evidenced by a slight
upward concavity in the enclosed mass profile inside $0.02\pc$.  
The radial extent of this upward concavity is increased to $r\approx 0.1\pc$ 
for $M_*=1$, and further to $r\approx0.3\pc$ by the time the stellar mass reaches 
$M_*=4\,{\rm M}_{\odot}$. The position of $r_*$ can be inferred by the position 
in the curves where the concave portion of the curve meets the linear portion.  
The concave regions are dominated by a central mass and hence are inside of $r_*$.  

The growth of the central mass forces $r_*(t)$ outward because $\rho(r)$ 
is independent of time, and hence the gas mass at small radii remains fixed, while
$M_*(t)$ grows. 

Returning to the velocities, the fact that $v_T(r) \sim r^{0.48}$ for $t=-100$ kyrs in the left plot of Figure \ref{fig:velocity_avg} shows that the 
turbulence in the 
initial collapse obeys the same scaling law found in non-collapsing
regions in the molecular cloud \citep{2015ApJ...800...49L}.  
This suggests that the turbulence in incipient collapsing 
regions is governed by the same large scale turbulent cascade as in 
non-collapse regions. 

However, the flattening and reversal of $v_T(r)$ 
at small radii and late times shows that some mechanism other than a 
turbulent cascade is at work at these radii and times. We interpret the 
behavior of $v_T(r)$ as the combined result of compressional heating and 
turbulent decay, as suggested by MC15 and by \citet{2012ApJ...750L..31R}.  

The relatively large infall velocity demonstrates that, even $100,000$ years before 
the proto-disk or star particle forms, these regions 
are not in hydrostatic equilibrium, in which Reynold's stress or turbulent pressure balances the force of gravity.  This calls into question
the assumption made by previous analytic models of massive star formation, 
such as the turbulent core model.
At early times, $|u_r|$ is between $v_K/3$ and $v_K$, except 
at $r\gtrsim 1$ pc, where the clump fades into the ambient 
molecular cloud.  These high ratios of $|u_r|/v_K$ show that 
hydrostatic equilibrium is not a valid description of the star forming 
regions at any time. 

In fact these plots show that $|u_r|$ is of order $v_K/3$ or larger at all 
times for $r_d\lesssim r\lesssim 1\pc$.


At small radii, the fact that $\rho(r,t)=\rho(r)\sim r^{-3/2}$ for $r<r_*$, 
combined with the fact that $|u_r(r,t)|\sim r^{-1/2}$, ensures that 
$\dot M(r,t)=\dot M(t)$, i.e., the mass accretion rate is independent of 
radius for $r < r_*$. 

This result for the accretion rate was shown previously in Figure \ref{fig:Mdot_vs_shu}. At early times, $(t-t_*) \sim -100$ Kyrs (the red dotted curve),
$\dot{M}$ decreases by a factor of 20 between $r=0.5\pc$ and $r=0.01\pc$
because the density profile is still
evolving toward the attractor solution. But for later times $\dot M(r)$ 
is flat at small radii. This demonstrates that the attractor solution, once established, imposes a 
major effect on the accretion profile. 

\begin{figure}
\plotone{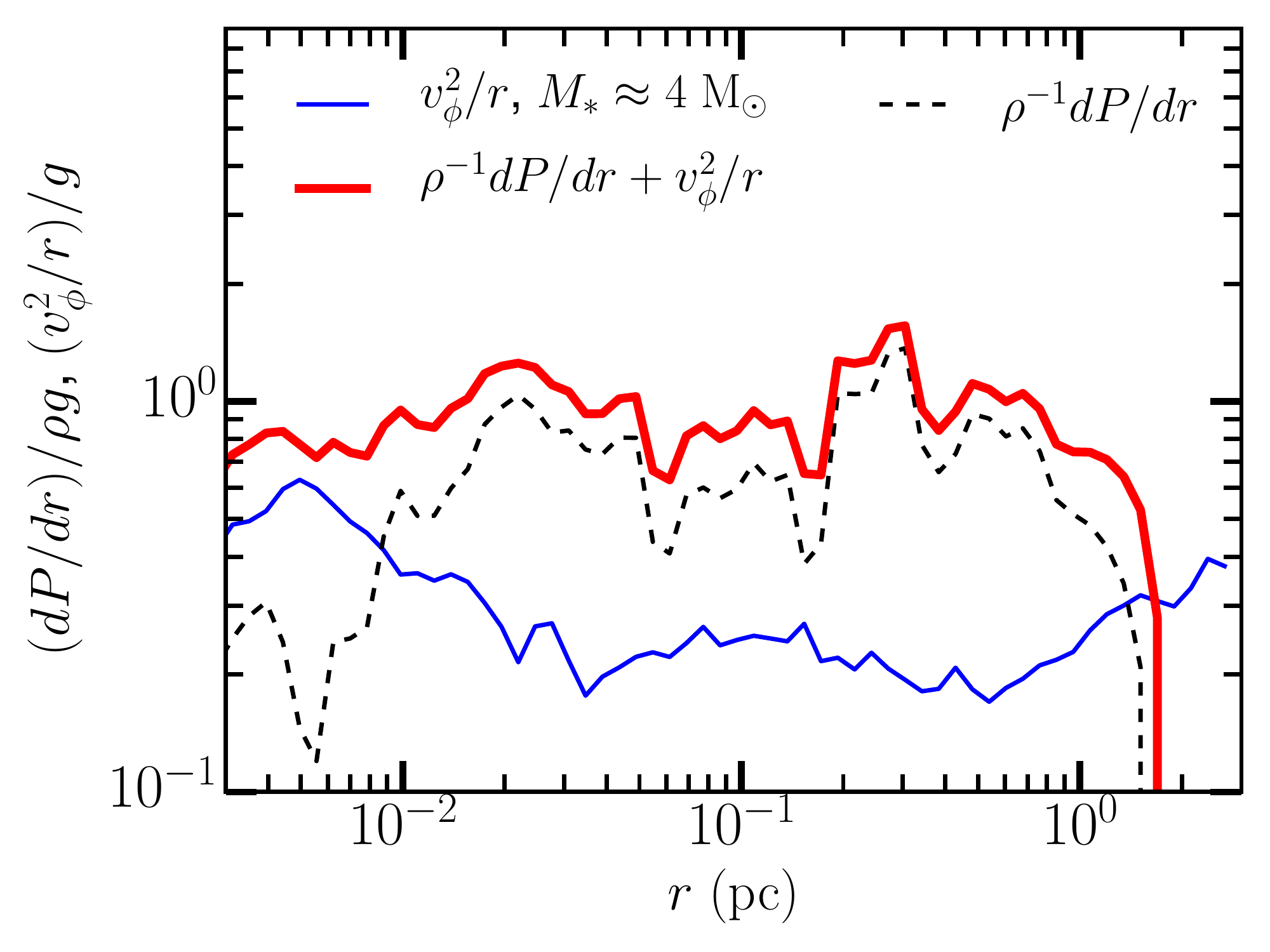}
\caption{The ratio of rotational (solid line) support, Reynolds stress and gas 
(dashed line) pressure support to the local gravitational acceleration 
$g = GM(<r)/r^2$, as a function of $r$ when the star reaches 4 
$M_{\odot}$. \label{fig:dPdr_avg}}
\end{figure}

While the gas is never hydrostatic, the gradient of the Reynolds
stress does roughly balance gravity as can be seen in Figure
\ref{fig:dPdr_avg}.  The figure shows the rotational support,
which we define as $v_{\phi}^2/r$ (solid blue line), Reynolds stress
plus thermal pressure support $\rho^{-1}dP/dr = \rho^{-1}d\rho
(v_T^2/2+c_s^2)/dr$ (dashed line), and total pressure support
$\rho^{-1}dP/dr + v_\phi^2/r$ (thick red line).  We have scaled these
quantities to the local gravitational acceleration, $g=GM(<r)/r^2$.

Inside of $r\approx 0.01$ pc, the gas settles into a rotationally supported 
disk and the support from other sources drops. However,  
the sum of the Reynold stress and rotational support (thick red line) nearly balances 
the local gravity. 

\begin{figure}
\plotone{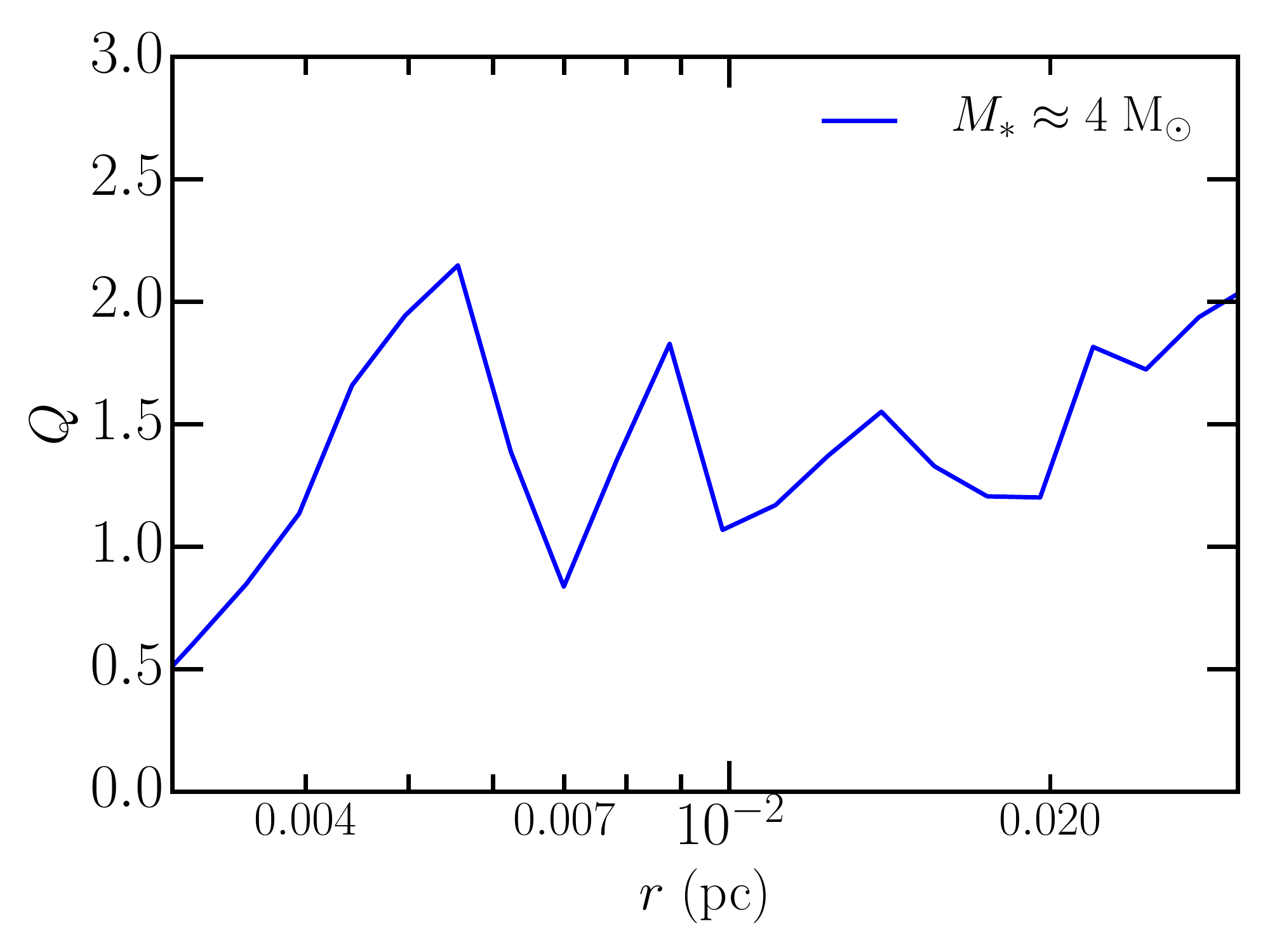}
\caption{The average (over many disks) of Toomre $Q$ as a function of $r$ when  $M_*\approx 4M_{\odot}$. \label{fig:Q_avg}}
\end{figure}

The rotationally supported disks in our simulations are, on average, 
roughly marginally stable,  
as seen in Figure \ref{fig:Q_avg}, which should be compared with Fig. 3 in \citet{2010ApJ...708.1585K}.  For $0.007\pc\lesssim r \lesssim 0.02\pc$, 
we find  $1\lesssim Q\lesssim 1.6$. Examining individual disks, some of 
the time the disk is unstable and rapidly dumps material toward the central 
star, while at other times the disk is stable, building up material to approach marginal stability. For $r\gtrsim 0.02$ pc,  $Q$  rises, though the interpretation of 
$Q$ as a measure of stability is questionable, as the gas is no longer 
rotationally supported, nor is it in a flattened or disk-like configuration. 

\subsection{Mass Accretion Rates}

Finally, we discuss the mass accretion rates in our simulation.  Previously,
\citet{2015ApJ...800...49L} (see also \citealt{2014MNRAS.439.3420M}) found 
that the star formation efficiency is nonlinear in time, with $M_* \propto t^2$.  
This nonlinear rate is evident in the work of previous workers, but was 
often interpreted as an initial transient \citep{2011ApJ...730...40P}.  MC15 
showed that $M_*(t)\sim t^2$  is a natural consequence of the density approaching an 
attractor solution and the scaling of the infall velocity with the Keplerian 
velocity at small radius, as we have clearly demonstrated in this work.  

First we address the question of whether the $M_*\propto t^2$ phase is an initial ``transient".
\citet{2012A&A...540A.113T} and \citet{2015MNRAS.451.3089T} estimate the lifetimes of massive 
star-forming clumps found using the Apex telescope and the Herschel telescope, respectively.
\citet{2012A&A...540A.113T} identify clumps with  column densities $\Sigma_g> 0.1\g\cm^{-2}$,
masses up to $10^5M_\odot$. Since they have a fairly complete catalog of such clumps, they can
estimate the typical lifetime of a clump by comparing to the number of massive stars formed in 
the Milky Way every year. They find a mean clump lifetime of $6\times10^4\yr$, and a clump 
free-fall time of $\approx1.5\times10^5\yr$; the clumps live 1 free-fall time.

Similarly, \citet{2015MNRAS.451.3089T} identify clumps with sizes ranging from $0.1-1\pc$, masses
ranging up to $10^4M_\odot$. They estimate an upper limit lifetime for the starless phase of $10^5\yr$
for clumps with $M>500M_\odot$, and a ratio of starless to total clumps (the rest of the clumps host
protostars) of 39\%. Thus the total lifetime of clumps in their mass range (above $500 M_\odot$) is 
$\sim 2-4\times10^5\yr$. The clumps in their sample have 
$10^4\cm^{-3}\lesssim n_H\lesssim10^5\cm^{-3}$, so free-fall times
$1.6\times10^5\yr\lesssim\tau_{ff}\lesssim5\times10^5\yr$. The clumps live $0.2-2.5$ free-fall times, 
similar to the estimate of \citet{2012A&A...540A.113T}.

Thus the lifetimes of massive star forming clumps, when measured in units of free-fall times, is
similar to the lifetimes of GMCs, again measured in free-fall times, e.g.,
\citet{2007prpl.conf...81B}, who find that GMCs live 2-3 free-fall times.

Our simulations run for only a fraction of a free-fall time, but those of other workers 
have often run for two to three  \citep{2010ApJ...709...27W,2011ApJ...730...40P}, or, in 
some cases, up to five free-fall times. The simulations are often
halted when $\sim10-30\%$ of the mass has been converted into stars, since that is a rough 
observational estimate of the maximum fraction of clump gas turned into stars, e.g.,
 \citet{2003ARA&A..41...57L}. In most cases this star formation efficiency is reached in
one or two free-fall times, while the $M_*(t)\sim t^2$ behavior is still apparent from the
plots. In the case of \citet{2015MNRAS.450.4035F}, in the MHD run with stellar wind/jet feedback, the 
$M(t)\sim t^2$ behavior ceases after 4 free-fall times, when the star formation efficiency is 
about 15\%. Since it is unlikely that star forming clumps live so long, the simulated star 
formation rate is probably too low. 

We conclude that the time scale over which the $t^2$ scaling is seen in simulations is similar to the 
lifetimes of massive star forming regions, so that, while the behavior we focus on may be of short duration, it is not ``transient".

\begin{figure}
\plotone{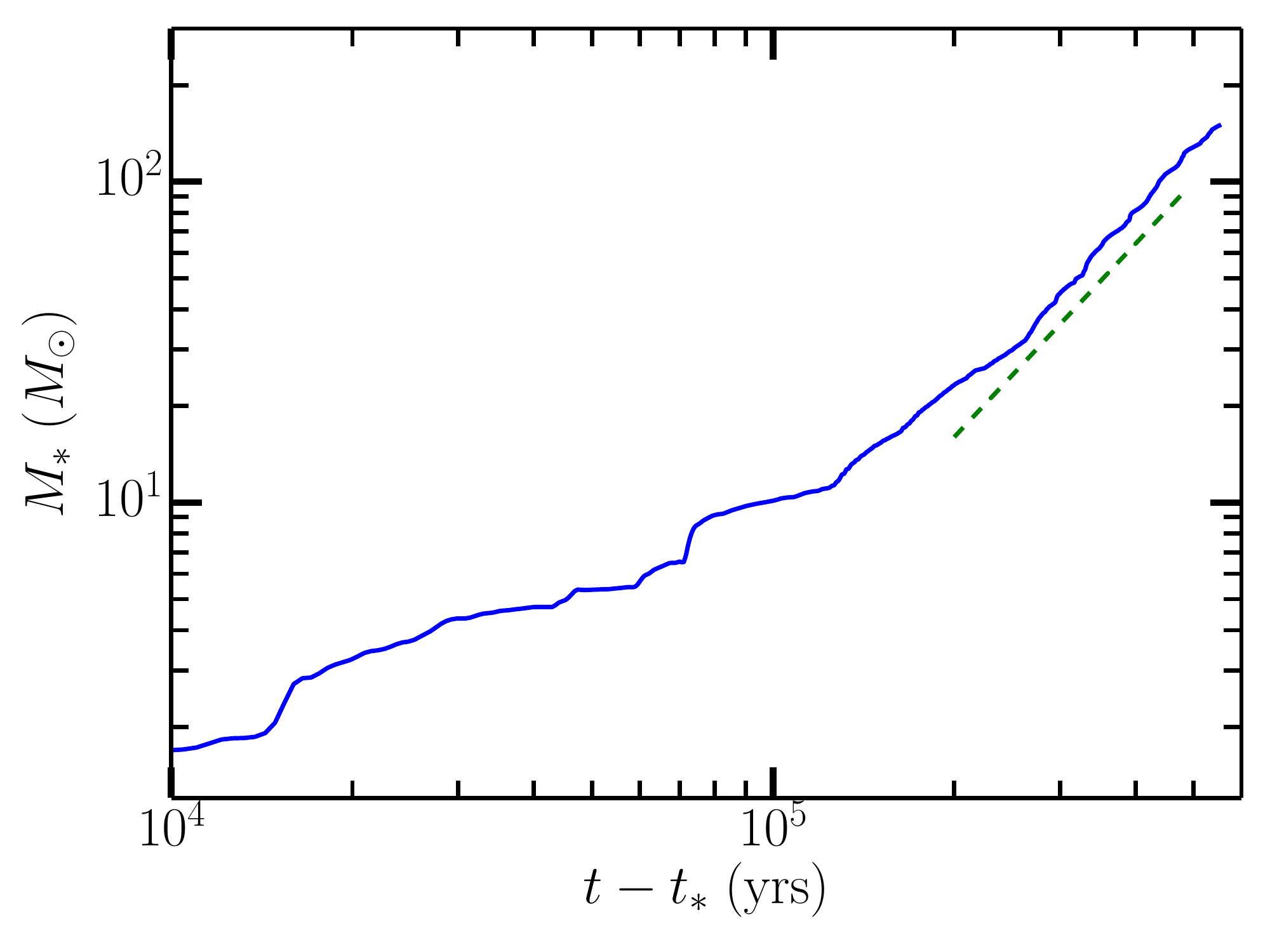}
\caption{The total mass in stars plotted as a function of time since the first star 
particle formed, $t_*$. The final value of the total stellar mass $M_*\approx180M_\odot$ 
is about $1\%$ of the total gas mass in the box, while the final $t-t_*\approx 0.6\Myrs$, 
about 15\% of the free fall time $3.8\Myrs$ for the mean density of the box. The dashed green line
shows a slope of $2$. }
\label{fig:M_t2}
\end{figure}

In Figure \ref{fig:M_t2} we show the total $M_*$ as a function of time since 
the first star particle was formed, $t_*$.  This is exactly the same analysis 
as \citet{2015ApJ...800...49L}.  However, because the simulations are distributed 
among the eight different octants, each with a different star formation time, we 
produce the total SFE history as follows.  First, we analyze the simulations to 
find the earliest time at which a star particle formed, which we define as $t_*$.  
We then look at all the simulation to find the earliest time at which a simulation 
ended or $t_{\rm end}$, which defines the time over which all our simulations 
have data.  Because each snapshot for each simulations are taken at different 
times,  we define a number of times at fixed intervals between $t_*$ and 
$t_{\rm end}$ and interpolate the total stellar masses for each simulation on 
those times.  These masses are then summed to produced $M_*(t)$, which we plot 
in Figure \ref{fig:M_t2}

As shown in Figure \ref{fig:M_t2}, $M_*(t)$ grows roughly linearly for 
$\approx100,000\yrs$ after the first star particle is formed.  However 
once the total stellar mass reaches about $M_* \gtrsim 10\,M_{\odot}$, at a time 
$t-t_*\gtrsim $100 kyrs, $M_* \propto t^2$.  At this stage, 
$M_*/M_{\rm GMC} \sim 0.001$. This agrees well with the results of
\citet{2015ApJ...800...49L}, who found $M_*\sim t^2$ for stellar masses between 
$M_*/M_{\rm GMC} \approx 0.015$ and $0.3$.  Due to the computationally expensive 
nature of our much higher resolution simulations, even given the use of AMR,
we are not able push our simulation to the same total $M_*/M_{\rm GMC}$ 
as \citet{2015ApJ...800...49L} were able to in their fixed grid, but much lower resolution,
simulations. However, our simulations do show that their simulations were already at 
sufficiently high spatial resolution to recover the scaling relation. 

The reason for this is not hard to find. Figure \ref{fig:density_avg} shows 
that at $r\approx1\pc$ the density has already settled onto its time-independent form, while 
Figure \ref{fig:velocity_avg} shows that the infall velocity is scaling as $r^{-1/2}$ for 
$r\lesssim 0.3-0.7\pc$, with the smaller value corresponding to the time of star particle 
formation, and the larger value to times for which $M_*\approx 1M_\odot$. As long as
a simulation resolves this radius (which corresponds roughly to $r_*$), it will recover the
$M_*\sim t^2$ scaling. 

The slower growth at earlier times is due to fact that, at the time 
of star formation, the infall velocity is non-zero, despite there being no star to 
attract the gas; see Figure \ref{fig:velocity_avg}. In other words, the initial 
infall velocity is larger than $\sqrt{GM_*/r}$; it takes time before the Reynold stress can slow the infall to the steady state value given by equation 
(\ref{eq:infall_behavior}), and hence before the mass accretion rate settles onto the steady 
state value given by equation (\ref{eq:Mdot_behavior}).

The same comments apply to the accretion rates of individual stars; individual stars start
out accreting mass at a roughly constant rate. At later times, when they have substantial
masses, the accretion rate grows linearly in time. This happens only with the most massive 
star particles in our simulation. 

The total number of star particles also grows roughly
  linearly with time; the combination of this linear growth in number
  of stars, together with the roughly linear mass growth of most of
  the stars, produces the over all $M_*\sim t^2$ scaling we see for
  the simulation region as a whole. In regions where the summed
  accretion rate is highest, individual star particles are not able to
  accrete all the collapsing mass, leading to the formation of new
  star particles in the immediate vicinity. In other words, our
  simulations produce clustered star formation. This is similar to
  what has been found in other recent simulations
  \citep{2015ApJ...800...49L,2015ApJ...806...31G}.

We interpret Equations (\ref{eq:density}-\ref{eq:Mdot_behavior}) as a description of star cluster formation; the total mass of the cluster will grow as $t^2$; the most massive stars may spend a significant fraction of their accretion history growing as $t^2$, but many of the less massive stars will not undergo such rapid growth in their accretion rate.  

\section{Discussion}\label{sec:discussion}
\subsection{Basic Results of this Work}
We begin with a summary of our results. 

\subsubsection{Collapse is not self-similar}
First, in our isothermal, 
driven turbulence simulations, star formation is not a self-similar process. 
Two length scales, in addition to the radius of the outer boundary condition or turbulence
outer scale,
and the stellar radius, enter the problem, the Keplerian or outer disk radius 
$r_d$, i.e., the radius at which the gas becomes rotationally supported, and 
the radius $r_*$ of the stellar sphere of influence, the smallest radius at 
which the gravity of the gas dominates the gravity from the star and disk. 
The existence and significance of $r_d$ has been known since the time of 
Kant and Laplace; the recognition that $r_*$ plays a role in star formation 
is recent, so we concentrate on the effect of $r_*$ on the dynamics in 
what follows. The value of $r_*$ increases monotonically with time, since 
the stellar mass increases monotonically, while $r_d$ may vary in or out 
with time, depending on the (turbulently determined) distribution of 
angular momentum of the accreting gas. 

The non-self-similar behavior of the collapse is most strongly reflected in the
variation of $v_T$ with radius; for $r_d<r<r_*$, the 
random motion velocity is a decreasing function of radius 
$v_T(r)\sim r^{p}$ with $p\approx -1/2$, while for $r_*<r$, it is
an increasing function of $r$, scaling like $r^{0.2}$.
Similarly, the infall velocity $|u_r|\sim r^{-1/2}$ for $r_d<r<r_*$,  
while for $r_*<r$ it is flat or even increasing outward (as in the top 
panel of Figure \ref{fig:Sphere_influence_quad2_end_time}), 
with substantial variations both from particle to particle and at different times
for the same particle, due to the vagaries of the turbulent flow at large radii. 

\subsubsection{Density approaches an attractor solution}
Second, we find that inside the sphere of influence of the star, the density 
remains constant over several to tens or even hundreds of (local) dynamical 
or infall times;  for $r_d<r<r_*$, $\rho(r,t)\to\rho(r)$. This is illustrated 
by Figures \ref{fig:quad2_run_of_density_double} and \ref{fig:density_avg}. 
One implication of this result is that one cannot use observations of the 
free-fall or crossing time of collapsing structures to infer either the age 
or lifetime of those structures.

The fact that $\rho(r,t)=\rho(r)\sim r^{-3/2}$ for $r<r_*$, combined
with the fact that $|u_r(r,t)|\sim r^{-1/2}$,  ensures that $\dot
M(r,t)=\dot M(t)$, i.e., the mass accretion rate is independent of
radius for $r<r_*$ (see Figure
\ref{fig:Mdot_vs_shu}). 

Since $|u_r|$ increases with stellar mass and hence with time where $r<r_*$ 
(see the second panel in Figure \ref{fig:velocity_avg}), while $\rho$ is fixed, 
$\dot M(t)$ increases with time, a result seen in many previous papers \citep{2011ApJ...730...40P,2012MNRAS.419.3115B,2012ApJ...754...71K,
2012ApJ...761..156F,2014MNRAS.439.3420M}, 
although this fact was usually not commented on. After some initial transient 
behavior, we find $M(t)\sim t^2$ (Figure \ref{fig:M_t2}) in line with the results of \citet{2015ApJ...800...49L}.

\subsubsection{Partitioning of the Collapsing Region's Potential Energy}
Our third result is to show how the potential energy released in collapse is partitioned.
In our simulations,
the support from random motions slows the rate of infall, so that $|u_r|$ is significantly smaller
then the free-fall velocity, but large enough to maintain $v_T$ at a sufficient level
that the acceleration due to the Reynolds stress is close to the acceleration of
gravity (Figure \ref{fig:dPdr_avg}).
In contrast, \citet{2010ApJ...721L.134S} and \citet{2011ApJ...731...62F} find an infall velocity which is equal to the free-fall velocity just outside their core. The dynamics in their simulation is very different to the dynamics in ours; in their case the acceleration due to the pressure gradient is negligible compared to the acceleration due to gravity. Their initial conditions incorporate transonic turbulence, but no driving. Since the collapse in their simulation does not take place for roughly four free-fall times, by the time the collapse starts, the turbulence is subsonic.

Because the infall in our simulations is typically supersonic, some of the kinetic energy can then 
be converted into thermal energy by shocks; of course, even in the absence 
of shocks, normal (molecular) adiabatic heating will convert a small fraction 
of the liberated potential energy into heat.

Figures \ref{fig:Sphere_influence_quad2_prior_to_particle},
\ref{fig:Sphere_influence_quad2_just_after}, and \ref{fig:Sphere_influence_quad2_end_time} 
show that for $r > r_*$ the bulk of the potential energy goes into random motion, 
and thence into shocks. Inside $r_*$, but outside of the disk, the potential energy 
that is not immediately radiated is shared roughly equally between the infall and 
random motions. Inside of the disk, the potential energy is converted to 
rotational and random motion, in roughly equal measure, and thence to thermal 
emission. At any radius, the ratio of kinetic energy to potential energy is
typically around a quarter to a half, although at large radii
($r\gtrsim 1\pc$) the turbulent kinetic energy can exceed the potential
energy: on the scale of our box, the ratio 
$v_T/\sqrt{GM_{\rm GMC}/r_{\rm GMC}}\approx \alpha^{1/2}_{\rm vir}\approx 1$,  
where $M_{\rm GMC}$ is the total mass contained in the simulation volume and 
$r_{\rm GMC} = 8\pc$ is the "radius" of the simulation volume, i.e., half of the side 
of the box.  This ratio scales as $\sim 1/\sqrt{r}$, so at 
$r\approx 1\pc$ the ratio is $\sim\sqrt{8}$.

We also found evidence that the ratio of radial to transverse
random motion depends on whether the collapse leads to radial
compression (for $r>r_*$) or dilation (for $r<r_*$), and on the
tendency for hydrodynamic turbulence to isotropize motions down the
cascade; see Figure \ref{fig:make_up_of_Turbulent} and the
discussion in \S \ref{sec:turbulent components}.

\subsubsection{Modification of Larson's Law}\label{sec:modification larson}
Our fourth result is the confirmation that the adiabatic heating of the turbulence 
alters Larson's law. On large scales or away from collapsing regions, Larson's 
law is $v_T\sim r^p$ with $p\approx 0.4-0.5$. It emerges naturally from the 
decay of supersonic turbulence that is driven on large scales.  We find, as did
\citet{2015ApJ...800...49L}, that in rapidly collapsing regions the decay of 
$v_T\sim r^p$ with decreasing radius is slowed for $r_*<r\lesssim 1\pc$. 
Least squares fits to $v_T(r)$ for over this range of radii result in 
exponents between $0.1\lesssim p\lesssim 0.4$, with an average around 
$p\approx 0.2$, in fair agreement with the prediction of equation (\ref{eq:infall_behavior}). Inside of $r_*$, we find $p=-1/2$, as predicted by equation (\ref{eq:infall_behavior}), representing a reversal of Larson's law. 

\subsubsection{Collapse does not proceed in an inside-out manner}

A fifth result is that the gathering and accretion of mass starts from large scales, 
and that, both before and after a star particle forms, $\dot M(r,t)$ is larger at 
large $r$ than it is at small $r$; in other words, the collapse proceeds in an outside-in manner.
The first point, that the accretion starts from large scales, is illustrated by 
Figures \ref{fig:Sphere_of_influence_quad2_well_before},
\ref{fig:Sphere_influence_quad2_prior_to_particle}, and the left panel of
\ref{fig:velocity_avg}, which show that $|u_r(r)|\sim (1/3) v_K(r)$ out to $r\sim 1\pc$ or
farther, and is trans- or supersonic tens or hundreds of thousands of years before the
density cusp forms, and hence before star or even disk formation 
starts\footnote{We remind the 
reader that the scale of the infall region in our simulations, and possibly in the ISM of
galaxies, is a simply a fraction of the driving scale of the turbulence.}.

Figure \ref{fig:Mdot_vs_shu} shows that just after the cusp/star particle
forms, the mass accretion rate is actually larger at larger radii. This behavior is 
the opposite of that predicted by  
inside-out collapse models, either that of Shu or of the turbulent core model. In Figure
\ref{fig:Mdot_vs_shu}, this inside-out behavior is illustrated with a Shu-type solution, 
shown by the dashed line; in that solution, the mass accretion rate decreases 
with increasing radius, in contrast to the results of our simulation.

Figures \ref{fig:Sphere_influence_quad2_end_time} and \ref{fig:velocity_avg} (right panel) show
that just after and well after the star forms, the surrounding region is also far from
hydrostatic equilibrium.

We conclude that there is no indication of gas in hydrostatic equilibrium 
prior to, during, or after star particle formation in our simulations. Nor is there
any indication of inside-out collapse. 

The violation of self-similarity and evidence against inside-out collapse shows that the assumptions made by previous analytic collapse models
\citep{1977ApJ...214..488S,1992ApJ...396..631M,1997ApJ...476..750M,2003ApJ...585..850M}, 
are not fulfilled in our simulations.  In addition, 
the collapsing regions in our simulations do not start from a hydrostatic equilibrium.

\subsubsection{The magnitude of the pressure gradient term is comparable to that 
of the gravity term for $r>r_d$}

Figure \ref{fig:dPdr_avg} shows that the 
acceleration due to the pressure gradient is comparable to the 
acceleration of gravity for $r_d<r<r_*$. 
Thus Reynold's stresses slow the infall compared to the free-fall rate, i.e., $|u_r(r,t)|< \sqrt{GM(r)/r}$.  
At small radii ($r<r_d$), the rotational support becomes important and 
the support from $v_T$ becomes much smaller than the radial component
of gravity, but comparable to the vertical component in the disk.

\subsubsection{The total stellar mass increases as $t^2$}

The total stellar mass in our simulation region, and in individual star forming
sub-regions, increases as the square of time after the first star (in the box,
or in the individual star forming region) forms.
Low mass (less than a few solar masses) stars have $M_*(t)\sim t^\gamma$, with
$0\lesssim\gamma\lesssim 2$, with a typical value $\gamma\approx1$, but the total 
number of low mass stars
 $N(t)\sim t$, so that the total mass in low mass stars grows as $t^2$. High 
mass stars, which tend to sit at density peaks (or at the bottom of potential
 wells) have $M_*(t)\sim t^2$.

\subsection{Comparisons to Observations}

\citet{1995ApJ...446..665C} showed that massive star forming regions have shallower 
line width-size relations than the classical Larson result, i.e., $v_T(r)\sim r^p$ 
with $p= 0.21\pm 0.03$, compared to $p\approx0.53\pm 0.07$ in low mass star 
forming regions. \citet{1997ApJ...476..730P} also found that Larson's law breaks 
down in massive star forming regions, i.e., their measured line widths are 
larger for a given source size than those found in low mass star forming regions.
As noted in \S \ref{sec:modification larson}, we find the same behavior in our
simulations, and we interpret this as the effect of adiabatic heating in a 
collapsing flow at $r>r_*$. 

In addition, \citet{1997ApJ...476..730P} plotted the mean velocity 
dispersion as a function of number density, which they derived 
from an excitation analysis of CO. 
They found that, contrary to expectations, the velocity dispersion increased with increasing density, 
which is opposite to the expectation based on Larson's law or supersonic turbulence driven from large scales. 
They concluded that the conditions in dense star forming cores 
are different from the rest of the cloud. The simulations presented here, and the 
analytic results of MC15, show the same behavior. In particular, the theory suggests 
that the enhanced turbulence or velocity dispersion at small radii in dense star 
forming regions is the result of gravitational collapse adiabatically heating the 
turbulence.

We find qualitative agreement between the observations of \citet{1997ApJ...476..730P} 
and our results, i.e., enhanced line-widths at high 
densities, which are associated with smaller radii.
Performing a more detailed 
comparison is more difficult as we have selected regions with the same stellar 
mass, whereas the stellar mass in \citet{1997ApJ...476..730P} is not well known.  
However, it is promising that the linewidths in our simulations are of similar 
magnitude and show the same trend with density as do the observations.

There are now numerous measurements of infall at large radii $\sim 0.1-1\pc$ 
in the literature. For example, \citet{2011A&A...527A.135C} Cygnus X, 
$D=1.7\kpc$ see infall $|u_r|=0.1-0.6\kms$, $\sigma_T\sim 0.6-2\kms$ at  
$r\approx 0.1\pc$, $n\sim 10^5-10^6$. Other examples include \citet{2012ApJ...746..174R,2015A&A...573A.119R} and 
\citet{2013A&A...555A.112P}. 

Infall is also seen on larger scales, $r\approx1\pc$, by \citet{2016A&A...585A.149W}, who observe $|u_r|$ in the range $0.3-3\kms$, corresponding to a fraction of the free-fall velocity ($1.4$ the Keplerian velocity) of $0.03-0.3$. In words, the gas at $r=1\pc$ is not in hydrostatic equilibrium, nor is it in free-fall. The turbulent velocity in the same clumps at the same radii is comparable or slightly in excess of the infall velocity, $v_{\rm T}\approx 1.0-2.3\kms$.

\citet{1986ApJ...304..501H}, \citet{2008ApJ...684.1273K}, and 
\citet{2011A&A...530A..53K} see infall at three different radii, 
$r\approx 0.5\pc$, $r\approx 0.3\pc$, and $r\approx 0.03\pc$ using 
different molecular tracers in the same object, G10.6-0.4. The infall velocity
is large at $r$, small at $r\approx 0.3\pc$, and large again at $r\approx 0.03\pc$. 
As noted by MC15, this is in qualitative agreement with the
picture of adiabatically heated turbulence.

\subsection{Missing physics}
Our current understanding of star formation suggests that the effects of magnetic 
fields, radiative and proto-stellar outflow feedback from stars, and the equation 
of state of the gas can 
all have significant effects on both the rate of star formation and the initial 
mass function (IMF) of the stars. We do not include any of this physics in the simulations
described in this paper. 

It is often argued that the turnover in the IMF, somewhere between $0.2$ and
 $0.6\,M_\odot$, is associated with the thermal state of the gas in the collapsing
region. If so, then our use of an isothermal equation of state suggests that the IMF found in our simulations is likely to be in error, so we have not discussed our computed IMF. However, as Figures \ref{fig:Sphere_of_influence_quad2_well_before}-\ref{fig:Sphere_influence_quad2_end_time} and \ref{fig:velocity_avg} show, both $|u_r|$ and $v_T$ exceed $c_s$, except at the earliest times ($\sim 100,000\yrs$ before a star forms), and then only for $r\lesssim0.1\pc$, so that the gas pressure does not dominate the dynamics in most regions and most of the time. Of course we do include the effects of gas pressure, so even in those regions and those times, our simulations capture the dynamical effects to lowest order, aside from, as we have just said, from fragmentation effects on the smallest scales. 

We have undertaken and made some preliminary analyses of magnetohydrodynamic simulations, which we will report on in future publications; as seen by other authors, we find that magnetic fields slow the star formation rate. But the runs of density and velocity have the same qualitative form in our MHD simulations as in the hydro runs presented here, and the MHD runs also give $M_*(t)\sim t^2$. 

Like magnetic fields, feedback from protostellar outflows are seen to slow the rate of star formation, e.g., \citet{2010ApJ...709...27W,2015MNRAS.450.4035F}. But those authors also find that $dM_*/dt$ increases with time even in runs that include outflows. 

Radiative feedback will also affect both the IMF and, for massive enough stars, the dynamics of the collapse at late times (after massive stars have formed). 

All the figures we show present results for stars with masses no larger than about $4 M_\odot$. To estimate the effects of radiation, we compare the force from the Reynold stress $F_T=4\pi r^2 \rho v_T^2$, to the radiation force $L/c$. 
From Figure \ref{fig:velocity_avg}, the (averaged over many stars) $v_T$ is slightly in excess of $1\kms$ at $r=0.01\pc$, while from any of the density figures the density is $\rho\approx5\times10^{-18} {\rm g/cm}^3$. The force from Reynold stress is then $F\approx4\times 10^{26}$ dynes. The luminosity of a 4 solar mass star on the zero age main sequence is $L\approx 2\times10^{36}\ergs$ \citep{1992A&AS...96..269S}, so the radiation force $L/c \approx 3\times10^{25}\,{\rm dynes}$, about a 10\% effect.  The force from Reynolds stress increases outward, see Figure \ref{fig:dPdr_avg}, so this statement holds at larger radii as well. 

Thus we expect that the effects of radiation pressure are not particularly significant in the situations we report;  the run of density and infall velocity, and hence the $M_*(t)\sim t^2$ scaling should not be affected, at least up to the times we are reporting on.  We note, however, that this estimate neglects the effect of radiative or ionization heating which is an important feedback mechanism.

Simulations including radiative feedback support this simple analysis. Figure 15 of  \citet{2014MNRAS.439.3420M} shows that in their simulations, which include feedback from both protostellar outflows and radiation (as well as magnetic fields), the stellar mass increases as the square of the time, up to masses of ~4.5 solar masses. Earlier work by the Berkeley group found similar results, forming stars with 10 solar masses, with $M_*(t)~ t^2$ even for such massive stars, see Figure 13 of \citet{2012ApJ...754...71K}. Their simulations included radiative effects, but no proto-stellar winds. 

\section{Conclusions}
\label{sec:conclusions}

Motivated by recent analytic (MC15) and numerical 
\citep{2014ApJ...797...32P,2015ApJ...800...49L} results, we perform deep AMR simulations 
of star formation in self-gravitating continuously driven hydrodynamic turbulence. We show 
that two length scales emerge from the process of star formation, $r_*$ and $r_d$, and demonstrate
that these length scales are clearly associated with physical effects. 
In particular, the character of the 
solution changes at $r_*(t)$, inside of which (but outside $r_d$) $|u_r|$ and  $v_T$ are both
$\propto r^{-1/2}$; outside of $r_*$, $v_T\sim r^{p}$ (with $p\approx0.2$), while $|u_r|$ 
is on average about constant. 
We emphasize that the length scales at which the character of the solution changes are time dependent. 
As the star grows in mass, the radius where the stars' gravity exceeds the gravity of the surrounding gas increases outwards away from the star, 
$r_*(t) \propto M_*^{2/3}(t)$.
The disk radius, $r_d$, also changes as a function of time as a result of the advection and transport of angular momentum from largescales to small scales (and vice versa).

We also found that the density profile evolves
to a fixed attractor, $\rho(r,t ) \rightarrow \rho(r)$ in line with the results of MC15 and 
the earlier numerical results of \citet{2015ApJ...800...49L}. 

Our results strongly support the basic premise of MC15, that turbulence is a dynamic variable 
which is driven by adiabatic compression \citep{2012ApJ...750L..31R}, and that the turbulence
in turn acts to slow the collapse. 
We note, as did MC15, that observations of massive star forming regions also find $v_T \propto r^p$ 
with $ p \sim 0.2-0.3$, and that at small radii or high density, $v_T$  
increases with increasing density, as seen in observations of massive star forming
regions \citep{1997ApJ...476..730P}. We find these departures from Larson's law only in
collapsing regions in our simulations. We also show that the acceleration 
due to the pressure gradient is comparable to that due to gravity at all $r>r_d$. 
As a result, the infall velocity is substantially smaller than the free fall velocity
even very close to the star or accretion disk. Inside $r_d$, rotational support takes over 
and as a result $u_r$ and $v_T$ both decrease.  

Our simulations capture rotational dynamics that MC15 did not capture in their 1-D model.  
In particular, we find the development of rotationally support disks at $r_d \sim 0.01$ pc.  
These disks have radii comparable to  or slightly larger than disks seen around 
young stars in Taurus \citep{1999AJ....117.1490P} in which stellar feedback effects are minimal,
and where the undisturbed disks are larger than in more active star forming regions such as Orion, 
where the disk radii are $\sim 100\AU$ \citep{2011ARA&A..49...67W}.  
This is despite the fact that we do not include magnetohydrodynamic effects 
in our numerical computations;  large scale magnetic fields may transfer angular 
momentum away from these disks, shrinking them.  

Like the disks modeled by \citet{2010ApJ...708.1585K},
our simulated disks are 
marginally gravitationally stable, suggesting that large scale gravitational torques are 
responsible for transport of material and angular momentum in our simulations; this may
also be true at early times in real protostellar disks.  

We have shown that the assumptions made by previous analytic collapse models
\citep{1977ApJ...214..488S,1992ApJ...396..631M,1997ApJ...476..750M,2003ApJ...585..850M}, 
are not fulfilled in our simulations.  In particular, 
the collapsing regions in our simulations do not start from a hydrostatic equilibrium, nor 
do they show any evidence of inside-out collapse.  The gathering of material before collapse, i.e.,
before the central cusp in the density power law is formed, involves transonic bulk motions 
and supersonic random motions (see Figure \ref{fig:velocity_avg}). The accretion of mass starts at large 
scales ($r\sim 1\pc$) with large initial infall velocities.  In addition, we find that $v_T$ 
scales differently in collapsing regions as opposed to the rest of the simulation box, 
whereas the turbulent collapse models \citep{1997ApJ...476..750M,2003ApJ...585..850M} assume 
that the scaling of $v_T$ with $r$ remains fixed.  

Finally, we close with a brief discussion of how our results relate to turbulence regulated theories of star formation. Here we find several points of disagreement. 
First, we find that the star 
particles accrete continuously from the surrounding large scale turbulent flow; there is 
no hydrostatic ``core'' that is cut-off from the turbulent medium. Second, 
the density distribution does not remain log-normal, but rather develops a power law tail that 
is directly related to the density profile \citep{2011ApJ...727L..20K,2015ApJ...800...49L}.  
Third, the fact that the density profile approaches an attractor solution that scales like 
$r^{-3/2}$ for $r<r_*$ and $u_r$ scales with the Keplerian velocity guarantees that $\dot{M}$ is 
constant with radius and $\dot{M}_*\propto t$ and hence a non-linear star formation efficiency, 
i.e., $M_* \propto t^2$ results. This is in contrast with turbulence regulated theories of star 
formation that predict a constant star formation rate, i.e., $\dot{M}_* = {\rm const}$ and, 
hence, a linear star formation efficiency $M_* \propto t$.

\section*{Acknowledgements}

We would like to thank the anonymous referee for useful comments.
DM, PC, and JP are supported in part by the NASA ATP
program through NASA grant NNX13AH43G, NSF grant AST-1255469, and the Research Growth 
Initiative at the University of Wisconsin-Milwaukee. NM is supported by the Canada 
Research Chair program and by NSERC of Canada.
Some of the computations were performed on
the gpc supercomputer at the SciNet HPC Consortium
\citep{2010JPhCS.256a2026L}. SciNet is funded by: the Canada
Foundation for Innovation under the auspices of Compute Canada; the
Government of Ontario; Ontario Research Fund - Research Excellence;
and the University of Toronto. 
The authors also acknowledge the Texas Advanced Computing Center (TACC) at The University of Texas 
at Austin for providing HPC resources that have contributed to the research results reported 
within this paper. URL: \url{http://www.tacc.utexas.edu}
This work was performed in part at the Aspen Center for Physics, which is supported 
by National Science Foundation grant PHY-1066293.

\bibliographystyle{mnras}
\bibliography{Bibliography}

\appendix
\section{Calculating the random motion and rotational Velocity}
\label{appendix:velocity}
In this appendix, we discuss how we calculate the random motion, $v_t$, infall, $u_r$, and rotational velocities, $v_{\phi}$ from the full three dimensional numerical solution.  To begin, we adopt a series of concentric, logarithmically spaced, spherical shells around either a star particle (if available) or around a density maximum in the case where the star particle has not yet formed. We then removed the bulk velocity from these shells by first calculating the enclosed mass, $M(<r)$, and momentum in each sphere ${\bf P}(<r)$, then dividing the two to find the bulk velocity, ${\bf V} = {\bf P}(<r)/M(<r)$ of the sphere of matter. We then subtract this bulk velocity from the corresponding {\em shell}. 

We also tried defining the bulk velocity using the total momentum in each of the spherical shells (rather than in the enclosed spheres), and found very similar results.

We then subtract the bulk velocity from the raw velocity of each cell in the spherical shell. We denote the result by ${\bf v}$.

Having removed the bulk velocity, we then calculate the radial infall velocity, $v_r = {\bf v}\cdot\hat{\bf r}$ per cell, where ${\bf v}$ is the velocity of the gas in a cell and $\hat{\bf r}$ is the radial unit vector (with the origin at the location of the star or local density peak).  Finally we find $u_r = <v_r>$ as the average over the spherical shell, where
\be 
\langle v_r\rangle \equiv \Sigma_i m_i v_{r,i}/ M_{\rm shell}
\ee
denotes a mass weighted average over each spherical shell, and $M_{\rm shell}$ is the mass of the shell. The sum is over all the cells in the thin spherical shell.

To calculate the velocity in the $\phi$ direction, where $\phi$ is defined by taking the $z$ axis along the angular momentum vector of the shell, we first calculate the angular momentum ${\bf L}_{\rm shell} = \int_{\rm shell} {\bf r} \times {\bf v} dm$, where $m$ is the mass in a cell. We next calculate the moment of inertia tensor ${\bf I}$ of each spherical shell.  In component form, ${\bf I}$ is
\be
I_{ij} = \int_{\rm shell} (\delta_{ij}r^2 - x_i x_j) dm
\label{eq:moment of inertia}
\ee
We then find the rotation vector ${\bf \Omega}$ by inverting
\be
{\bf L} = {\bf I}{\bf \Omega},
\ee
e.g., \citep{1992ApJ...399..551M}.
Next we calculate the rotational velocity in each cell from 
\be
{\bf v}_{\phi} = {\bf \Omega} \times {\bf r}.
\label{eq:calc v_phi}
\ee
This amounts to assuming that the gas in each spherical shell rotates rigidly; in other words we are averaging over the random motions in the shell. 
Finally, we calculate the spherical shell average as $v_{\phi} = <|{\bf v}_{\phi}|>$, i.e., the mass weighted average of the norm of ${\bf v}_{\phi}$.  

Armed with the coherent infall ($u_r$) and rotational ($v_{\phi}$) velocities, we define the remaining velocity as the random motion velocity (per cell) as 
\be
{\bf v}_T = {\bf v} - u_r\hat{\bf r} - {\bf v_{\phi}}
\ee
and the spherical average as $v_T = <|{\bf v}_T|>$. 
As a check that we were accounting for all of the velocities, we added the velocities in quadrature: $v_{\rm sum} = \sqrt{u_r^2 + v_T^2 + v_{\phi}^2}$ and verified that it traces the mass weighted average total velocity, $v_{\rm tot} = <|{\bf v}|>$ accurately.

\section{Filamentary or Spherical Accretion?}
\begin{figure*}
\plottwo{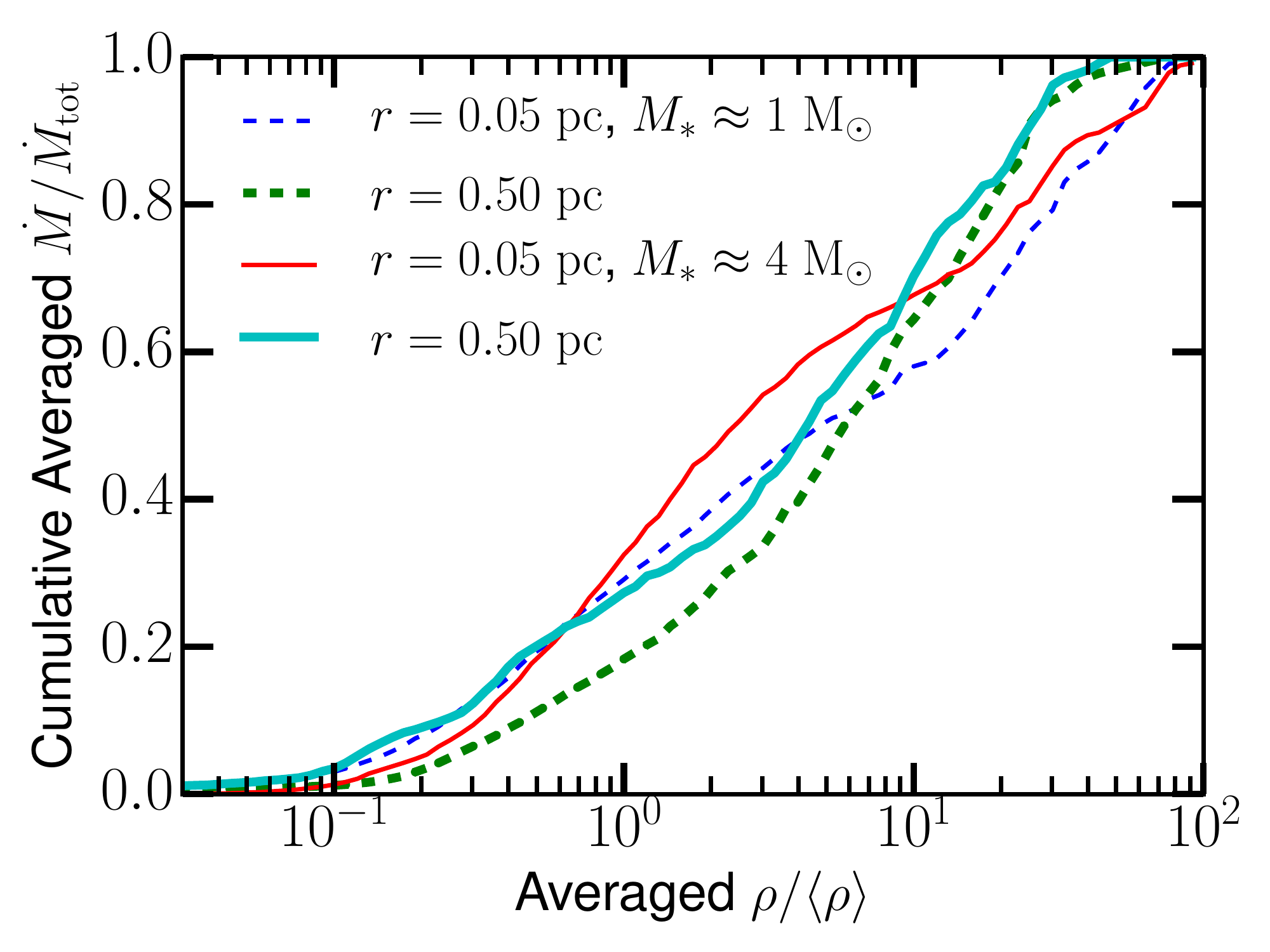}{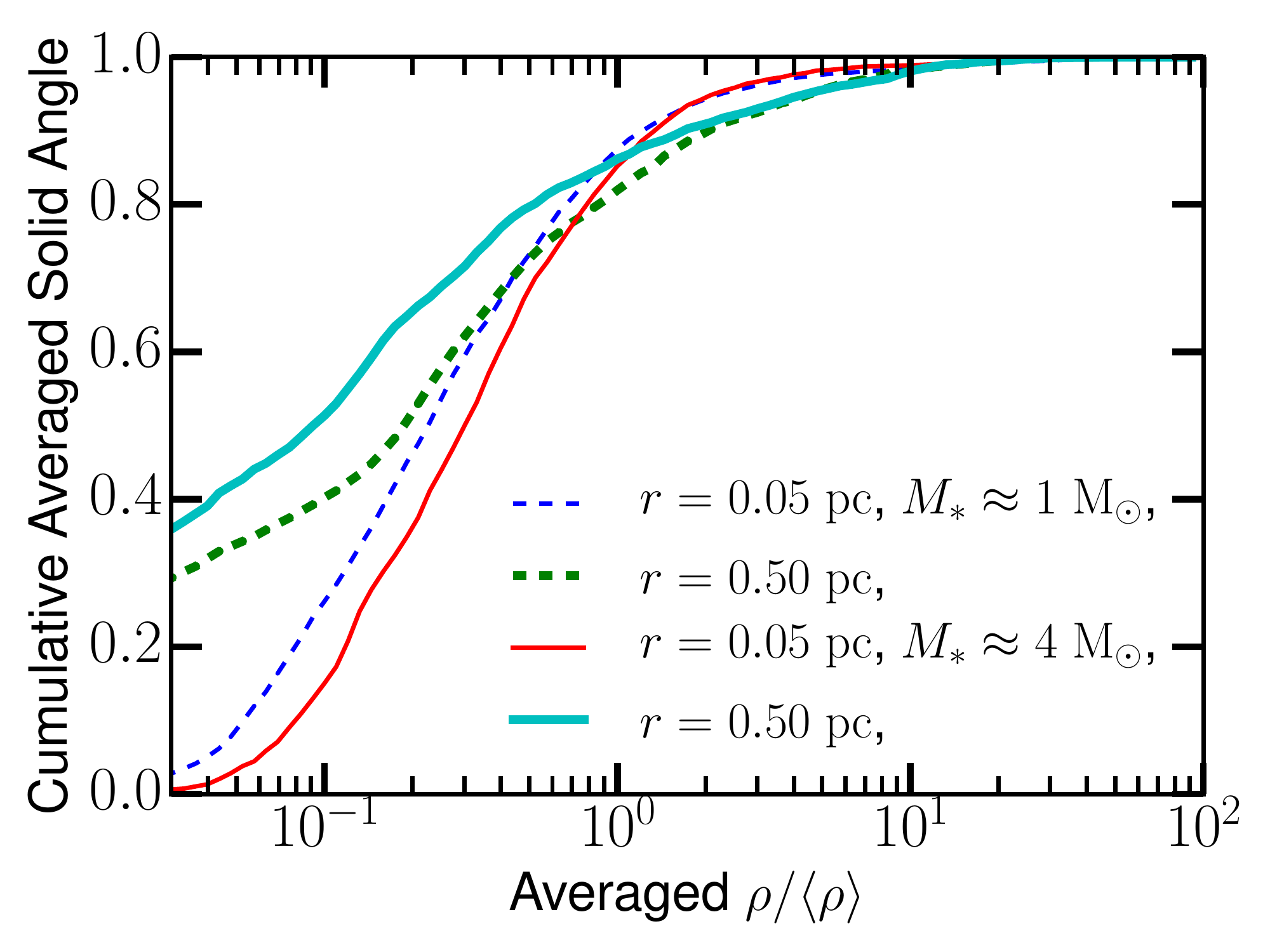}
\caption{Histogram of the cumulative normalized $\dot{M}$ (left) and cumulative normalized solid angle (right) as a function 
of normalized density for $r=0.05$ and $0.5\pc$ when the star reaches 1 and 4 
$M_{\odot}$. \label{fig:Accretion_avg}}
\end{figure*}

Figures 
\ref{fig:Sphere_of_influence_quad2_well_before}-\ref{fig:Sphere_influence_quad2_end_time} 
show that the density in the vicinity of collapsing regions
is decidedly non-spherical. Despite this, the results of MC15 appear to describe the 
accretion process well. For example, in those same Figures we have shown {\em mass-weighted} 
infall, random motion, and rotational velocities, the first two of which behave as predicted by
MC15 (they made no predictions for $v_\phi$). 

To understand this better, we examine how $\dot M$ depends on $\rho$, and how both
are distributed on the sky as seen by the accreting particles.
The left plot of Figure \ref{fig:Accretion_avg} shows a histogram of 
cumulative $\dot{M}(\rho)/\dot M_{\rm tot}$ through two spherical shells
at $r = 0.5 \pc$ and $0.05 \pc$, 
as a function of $\rho/\langle \rho\rangle$, where $\langle\rho\rangle$ 
denotes the density average over the (finite thickness) shell. 
We show average histograms when the central star has a mass $M_*=1M_\odot$ (dashed lines)
and $M_*=4M_\odot$ (solid lines).  The plot shows that $50\%$ of the accretion through the
sphere occurs via gas that has a density less than 2-5 times the average 
density of the shell,
where the low end of this range occurs at small radii at late times, with the high
end occurring at large radii and early times.

Since the mean density at $r=0.5\pc$ is 
$\langle\rho\rangle\approx 3\times10^{-21}\g\cm^{-3}$, see 
Figure \ref{fig:quad2_run_of_density_double} or Figure \ref{fig:density_avg}, an examination of 
Figure \ref{fig:Sphere_influence_quad2_end_time}, where gas with three times the 
mean density is depicted by dark green (and less dense gas is blue), shows that more than
half of the accretion is coming from gas that covers most of the sky as seen 
from each of those accreting particles; most of each slice is colored blue. If we take 
filaments to consist of gas that is
colored light green or yellow (with $\rho>10^{-20}$, or $\sim 3$ times 
the mean density $\langle\rho\rangle$ inside $r=0.5\pc$),
the filaments account for less than half the accretion.

A similar statement holds for the accretion inside $r=0.05\pc$, shown as the thin lines in Figure 
\ref{fig:Accretion_avg}.

To see more quantitatively how this gas is distributed on the sky,
we plot in the right panel of Figure \ref{fig:Accretion_avg} the cumulative solid 
angle as a function of $\rho/\langle \rho\rangle$, again for $r=0.5\pc$ (thick lines) 
and for $r=0.05\pc$ (thin lines).  Roughly 90\% of the sky is
covered by gas that is at three times the average shell density or lower, 
consistent with the qualitative analysis in the previous paragraph. 

\begin{figure}
\plotone{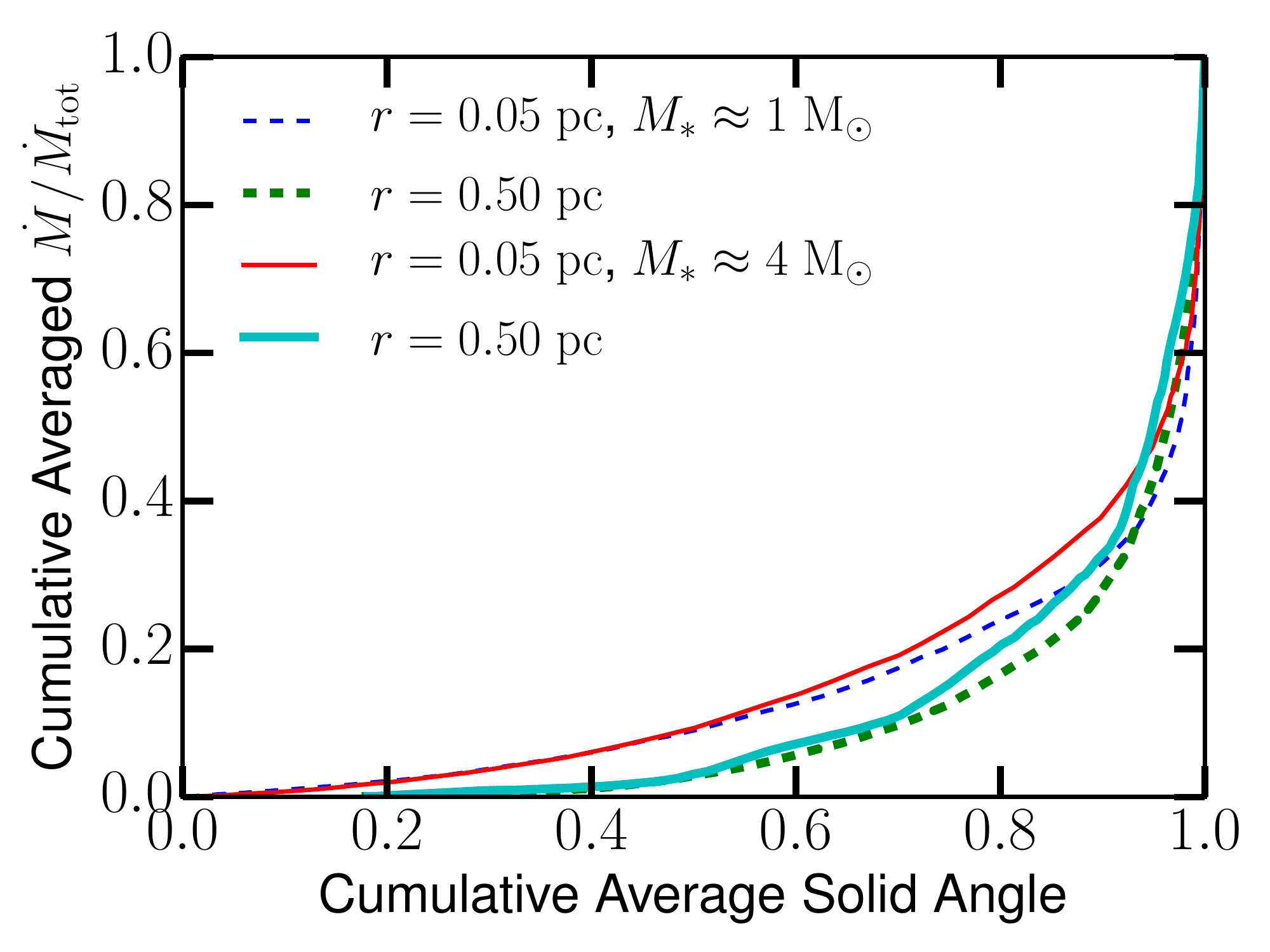}
\caption{Histogram of the cumulative normalized averaged $\dot{M}$ as a function of the cumulative normalized averaged solid angle for $r=0.05$ and $0.5\pc$ when the star reaches 1 and 4 
$M_{\odot}$. \label{fig:Accretion_angle_avg}}
\end{figure}

Figure 
\ref{fig:Accretion_angle_avg} shows a histogram of the cumulative normalized 
$\dot{M}$ as a function of the cumulative normalized solid angle.  
The plot shows that half the accretion occurs over about 10\% of the sky 
where the density is $\sim3$ or more times the mean density of the spherical shell.  
So, while about half the gas accretes from over most of the sky, and at about the mean density,
very dense gas entering the sphere from a very small covering fraction of the sky contributes the other half of the total accretion budget. 

Thus, while the filaments are readily identifiable by eye, and are important sources 
of accreting gas, much of the accretion (and much of the mass) lies in gas that 
is more nearly spherically distributed. 

\section{Star Formation Criteria}
\label{particle_criteria}
The majority of our simulations used a simple density condition of three times the Truelove condition (Equation \ref{eq:refinement_criteria}) at the maximum refinement level inspired by the sink particle formation criteria of \citet{2011ApJ...730...40P} as discussed in \S \ref{sec:simulation setup}.
\begin{figure}
\plotonesmall{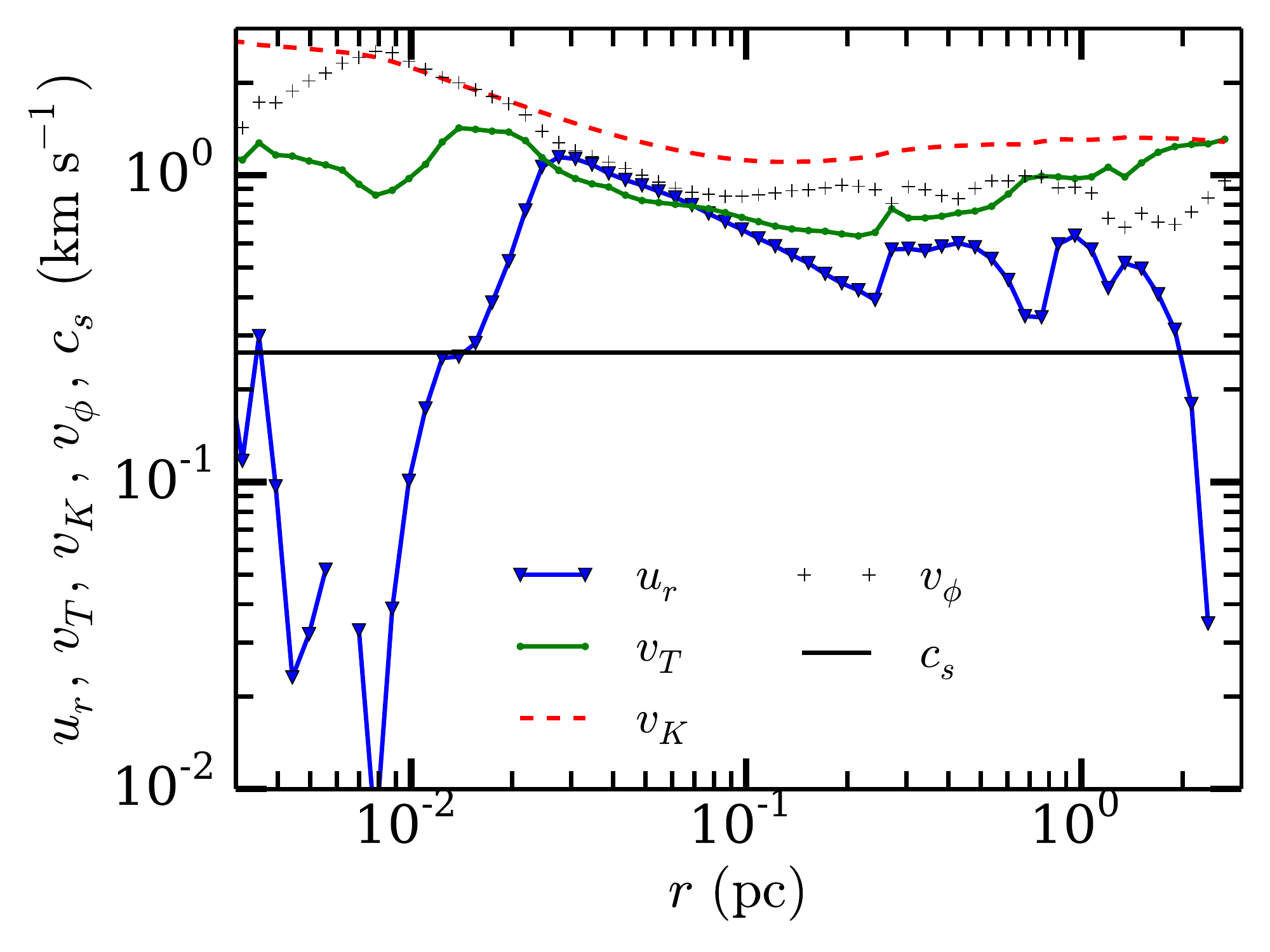}
\caption{The run of velocity for a particle formed in a simulation using FLASH's built in particle formation checks (as well as the density condition proposed by \citet{2011ApJ...730...40P}). 
This shows that the dynamics of 
collapse are not overly sensitive to the star particle creation algorithm.}
\label{fig:federrath_particle}
\end{figure}
We have experimented with additionally including the sink particle checks of \citet{2010ApJ...713..269F} to check the robustness of our results to these additional checks. In Figure \ref{fig:federrath_particle} we show the run of velocity in a simulation in which we included the star particle formation checks used in the default used in FLASH. The results do not differ significantly from runs lacking such checks. For example, both show that the stellar mass increases like $t-t_*$ squared, $M_*(t-t_*) \propto (t-t_*)^2$. There are however stochastic variations in the stellar mass ratio from runs with and without the extra checks. The mass ratio at a given $(t-t_*)$ can vary by a factor of roughly 2. For example, a hundred thousand years after the first star forms, in one run the total stellar mass is $10 M_\odot$ while in another it is $15 M_\odot$.

\subsection{Radial and Lateral Components of the Random Motion
  Velocity}\label{sec:turbulent components}
\begin{figure}
\plotone{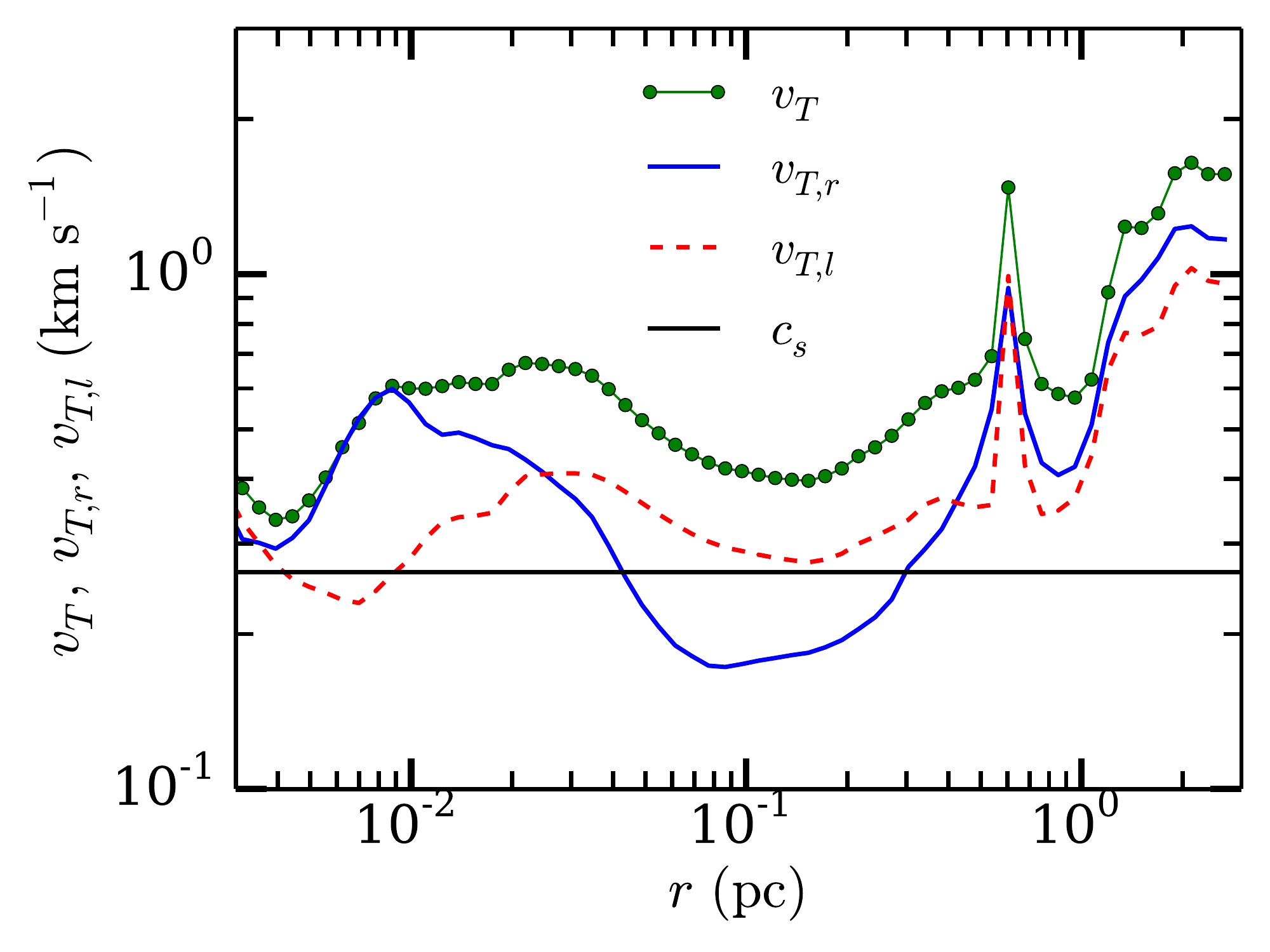}
\caption{The radial $v_{\rm T,\, r}$ (blue solid), lateral 
$v_{\rm T,\, l}=(v_{{\rm T},\,\theta}+v_{{\rm T},\,\phi})/2$ (red dashed) and total 
(green line over-plotted with dots) random motion velocities as a function of 
radius for \partB 
$100,000\yrs$ after the particle formed. The sound speed $c_s$ is shown for 
comparison (black horizontal line).
At large radii $v_{\rm T,\, r}\approx v_{\rm T,\, l}$; for 
$0.04\pc\lesssim r\lesssim 0.4\pc$
$v_{\rm T,\, r}< v_{\rm T,\, l}$, while inside of 
$0.03\pc$ $v_{\rm T,\, r}$ quickly recovers and
then exceeds $v_{\rm T,\, l}$.
The behavior of the lateral and radial velocities is dictated by the 
radial infall velocity in Figure \ref{fig:Sphere_influence_quad2_end_time} 
(see the main text).
}
\label{fig:make_up_of_Turbulent}
\end{figure}

Figure \ref{fig:make_up_of_Turbulent} shows the radial 
$v_{T,\,r}$ and lateral 
$v_{T,\,l}\equiv (v_{T,\,\theta}+v_{T,\,\phi})/2$ \footnote{Note that we define $v_{T,\,l}$ as an average so that we can compare it directly to $v_{T,\,r}$.}
components of the random motion velocity for the same collapsing region as 
shown in Figure (\ref{fig:Sphere_influence_quad2_end_time}), where 
$v_{T,\,\theta}$ and $v_{T,\,\phi}$ are the random motion velocities 
along the $\hat{\theta}$ and $\hat{\phi}$ directions defined 
from the z-axis. In the absence of self-gravity,
a turbulent hydrodynamic cascade to small scales tends towards
equipartition, ($v_{T,\,r}\approx v_{T,\,l}$), with a scaling behavior
$v_T\sim r^{1/2}$, similar to that seen in Larson's size-linewidth relation; this is
what we see in non-collapsing regions in our simulation. 

Figure \ref{fig:make_up_of_Turbulent} shows that both the radial and
transverse components of the random motion velocity decrease with
decreasing $r$ for $0.4\lesssim r\lesssim 3\pc$ (except for a spike
at $r\approx0.6\pc$). Furthermore, the ratio $v_{T,\, r}/v_{T,\, l}\approx 1$. 
We interpret the decrease as the decay of turbulence down a cascade.
However, the decrease in both the total and in the
longitudinal component, when fit with a simple power law, 
gives $v_T\sim r^p$ with $p=0.2$,
while the decrease in the radial component of the turbulence
corresponds to $p\sim0.35$. Since both exponents are less than the
value $p\approx0.5$ that we see on larger scales or away from collapsing
regions, we conclude that adiabatic heating is affecting both the
radial and transverse components of the turbulence.

At smaller radii, $0.04\pc\lesssim r\lesssim0.4\pc$, the inward decrease of both 
$v_{T,\, r}$ and $v_{T, \, l}$ slows and then reverses, as the flow passes $r_*$. 
However, the ratio $v_{T,\, r}/v_{T, \, l}$ is now only $1/2$. 
Finally, at and inside the disk radius $r_d\approx 0.02\pc$, the lateral 
turbulence once again decreases inward, while the radial component 
grows until much smaller radii, before decreasing again. 

If adiabatic heating is responsible both for the slower than normal decrease with
random motion velocity at $r>r_*$, and for the increase in random motion velocity 
inside $r_*$, why does the ratio of the radial and lateral components of the 
turbulence vary? 

In Figure \ref{fig:Sphere_influence_quad2_end_time}, $|u_r|$ is
decreasing with decreasing radius over the range $0.4\pc\lesssim r\lesssim 3\pc$.
What this decrease means physically is that as the
gas falls in towards the center, it is being compressed not just
in the $\theta$ and $\phi$ directions, but also radially. This
compression along the radial direction should drive radial turbulence,
while the lateral compression should drive lateral
turbulence. This is why the radial and lateral components of the random motion
velocity have the same magnitude. 

This physical reasoning also tells us that as the infall velocity
increases inward over the range $0.04\pc\lesssim r\lesssim 0.4\pc$
(see Figure \ref{fig:Sphere_influence_quad2_end_time}), 
the gas dilates in the radial direction even as it continues to compress 
in the transverse ($\theta$ and $\phi$) directions. Compression in the
$\theta$ and $\phi$ directions will tend to drive an increase in the
lateral components of the random motion velocity, but dilation
in the radial direction will tend to drive a decrease in the radial
component; of course both tendencies have to compete with (or add to,
in the case of radial motion) the usual tendency for turbulence to
decay, and the tendency, mentioned above, for hydrodynamic turbulence
to tend to equipartition as the motion cascades to small scales. 

We interpret the rapid inward decline of $v_{T,r}$
starting at $r\approx0.4\pc$ as the effect of adiabatic cooling. The result
is that the ratio $v_{T,\, r}/v_{T,\, l}\approx 1/2$ for $0.04\pc\lesssim r\lesssim 0.4\pc$. 

Between $r_d\approx0.02\pc$ and the local maximum of $|u_r|$ at
$r\approx0.04\pc$, the infall velocity is large but roughly constant,
meaning that the radial dilation ceases. We interpret the uptick in
$v_{T,\, r}$ toward small radii as the result of the cascade of
$v_{T,\,l}$ driving the radial component, as the turbulence strives
to reach equipartition, combined with the cessation of adiabatic cooling 
associated with the cessation of radial dilation.  

Inside $r\approx 0.02\pc$ the infall takes place primarily through a rotationally 
supported disk, in which both the vertical $v_{T,\, \theta}$ and 
azimuthal $v_{T,\, \phi}$ component of the turbulence is greatly reduced 
(although we do not show the separate components in the figure). At the 
outer edge of this 
disk we see a sharp rise in the radial component of the random motion velocity, 
followed at yet smaller radii by a decrease in the total turbulent velocity. 
We interpret the drop in the total turbulent
velocity at small radii as the flow settling into more ordered motion in an accretion disk.

\section{The Initial Mass Function}

\begin{figure}
\plotone{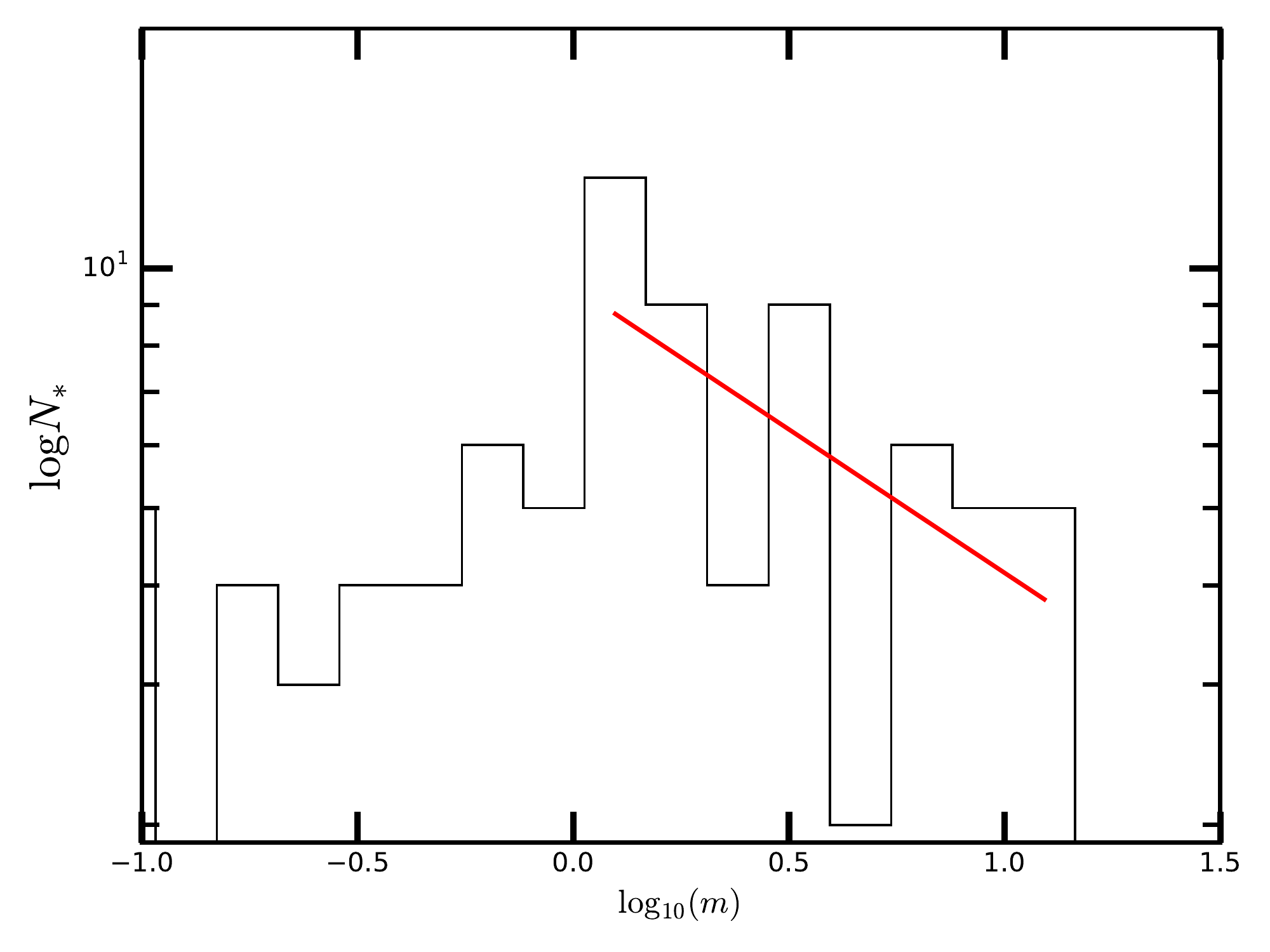} 
\caption{Initial mass function for the Ramses run with $N_J = 32$. The peak of the IMF is at $2 M_\odot$ and the high mass slope is $\Gamma = 1.36$, where the Saltpeter slope is $\Gamma = 2.35$. The red line shows the least-squares fit to the mass function for $M \geq 1.0 M_\odot$. Our IMF varies with time with the peak mass moving to lower mass and the value of $\Gamma$ increasing with time.}
\label{fig:imf}
\end{figure}

\begin{figure}
\plotone{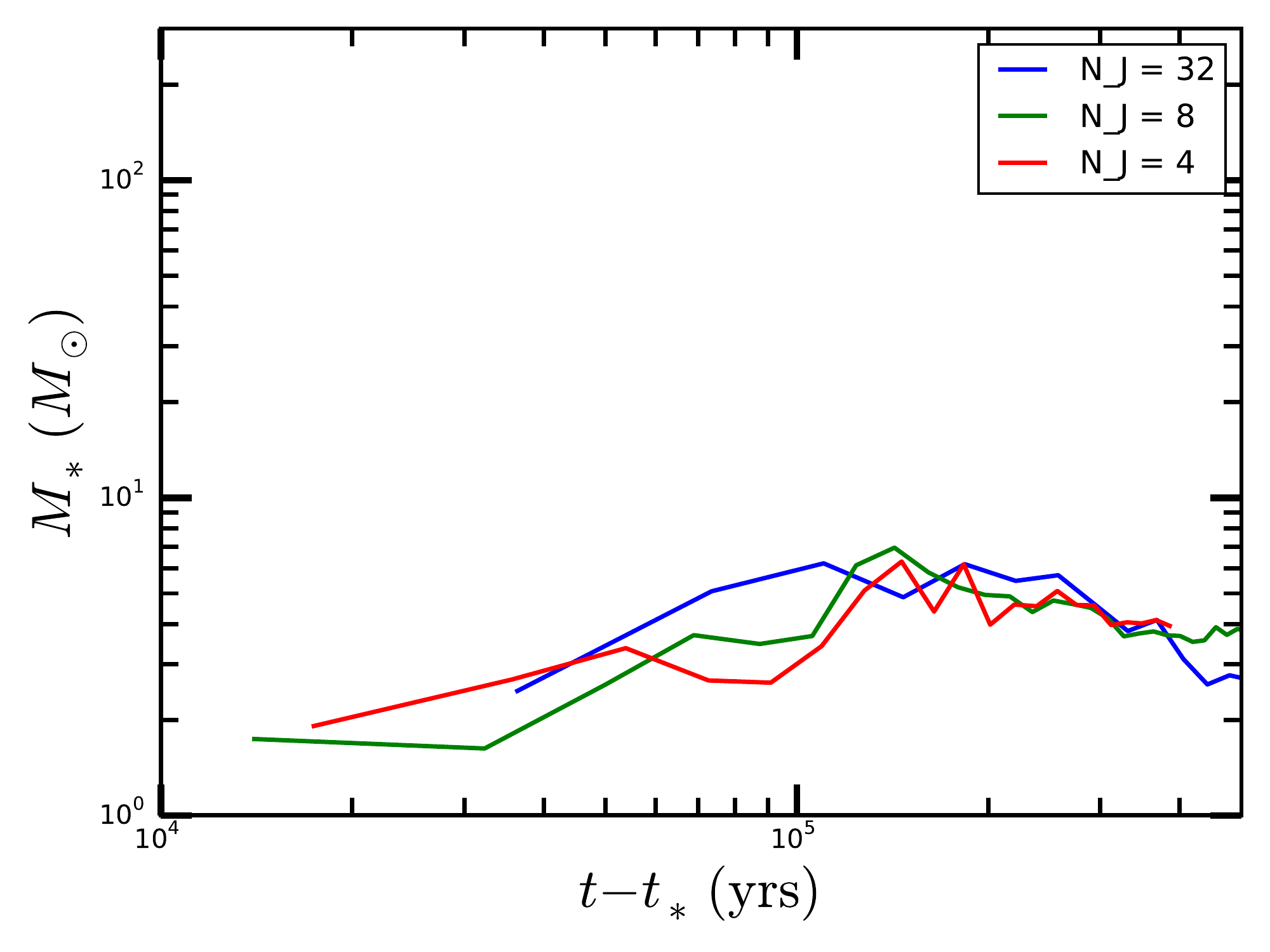} 
\caption{Average stellar mass as a function of time for the Ramses run with $N_J = 32$ in green. Average stellar mass starts at $\approx 1 M_\odot$ increasing to $\approx 6 M_\odot$ and then decreasing as low mass star formation begins. We also show the average stellar mass for $N_J = 8$ (blue) and $N_J = 4$ (red). This demonstrates that $N_J = 4$ runs are converged.}
\label{fig:avg_mass}
\end{figure}

For completeness we report the IMF in this subsection, though we caution the reader again that, because we do not handle the thermal physics properly, the location of the break in the IMF is unlikely to be correct; however it is commonly believed that the slope at the high mass end is set by the turbulence so that the thermal properties will not have much effect there.
Figure \ref{fig:imf} shows the IMF at the end of our Ramses run with $N_J = 32$. 
The plot includes a total of 90 stellar particles, with a total mass of $240 M_\odot$, or about  $\approx 1 \%$ of the total mass in the simulation. The time that Figure \ref{fig:imf} is plotted corresponds roughly to the right edge ($\approx 600,000$ years after the first star forms) in figure \ref{fig:avg_mass}. 
It shows a form that is roughly consistent with observed IMFs, in that it has a power law at high masses, a peak around a solar mass, and a fall off at lower mass. The peak however, is at $2 M_\odot$ which is about a factor of four higher than observed IMFs, and the fall off at high mass is too flat, indicating that we are top heavy. If the Saltpeter slope is denoted by $\Gamma = 2.35$, our slope is $\Gamma = 1.36$. 

Figure \ref{fig:avg_mass} shows the average stellar mass as a function of time. We see that the average mass is significantly higher than that of observed IMFs, where it is in the range of $0.3 - 1.0$. In addition, we see that this average mass rises initially as the massive stars grow and then decreases as low mass star formation kicks in.

\section{Convergence with $N_J$}

In this appendix, we examine how our results for the mass accretion rate depend on the resolution of the Jeans length, as quantified by $N_J$. 

\begin{figure}
\plotone{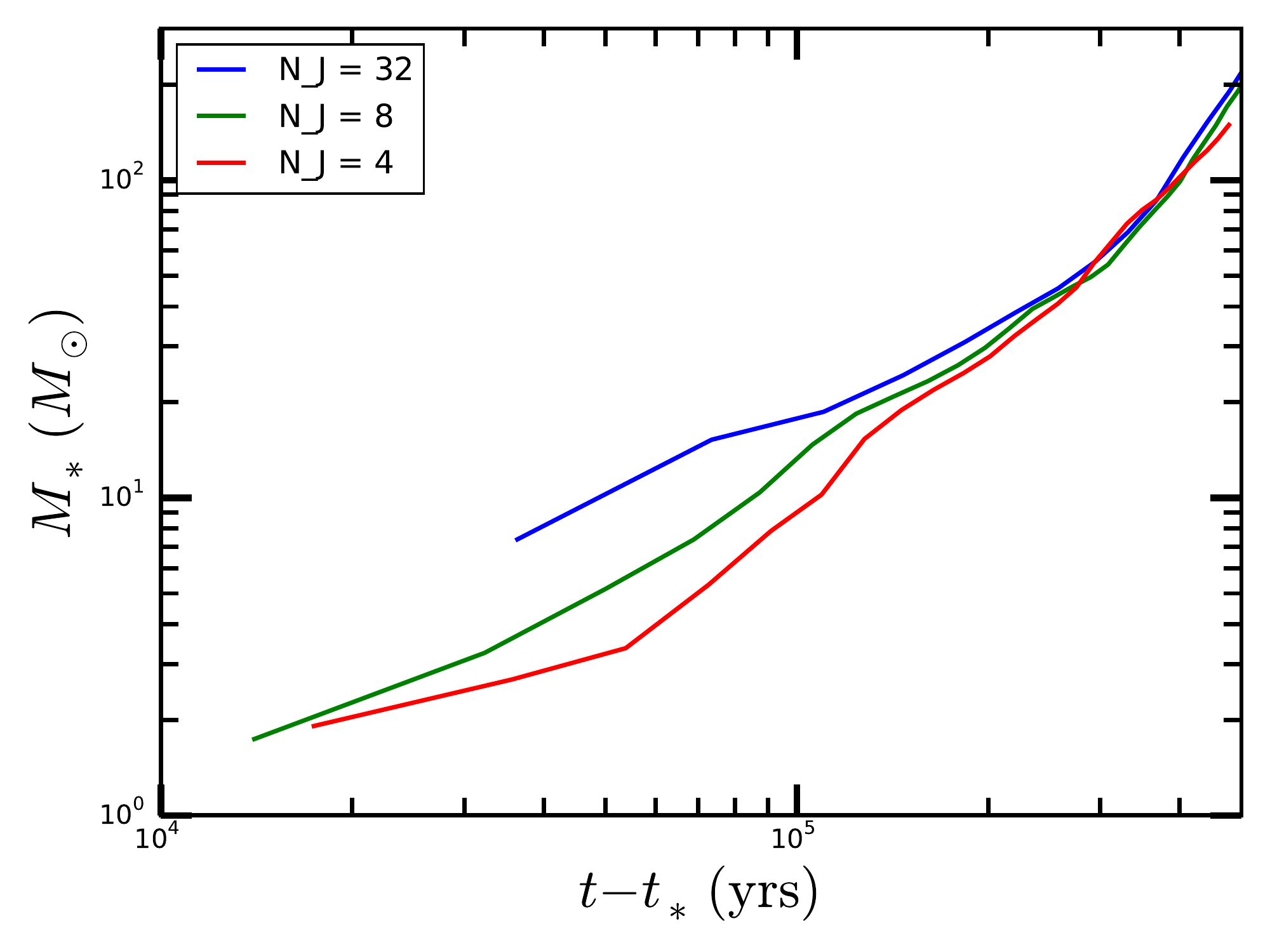} 
\caption{Plot of $M_*(t-t_*)$ for Ramses runs with $N_J = 32$ (blue), $N_J = 8$ (green), $N_J = 4$ (red). At the end of the $N_J = 32$ run, the total stellar mass was $M_*\approx 240M_\odot$.}
\label{fig:Convergence}
\end{figure}

Figure \ref{fig:Convergence} shows the total mass in stars plotted as a function of time since the time $t_*$ at which the first star particle formed, for $N_J=4,\,8$, and $32$.
For the $N_J=32$ run, the final value of the total stellar mass $M_*\approx 240M_\odot$, for $N_J =32$) 
is about $1\%$ of the total gas mass in the box, while the final $t-t_*\approx 0.6\Myrs$, 
about 15\% of the free fall time $3.8\Myrs$ for the mean density of the box. 
The green line shows the total stellar mass for $N_J = 8$, while the red line shows the same quantity for $N_J = 4$.

The figure shows that for $t - t_* > 200,000$ years the stellar mass as a function of $t-t_*$ is converged to within $10 \%$, and to even better accuracy at late times.

We have also done a convergence study for the average mass, see figure \ref{fig:avg_mass}, showing that the mean stellar mass is converged for $N_J=4$. This is consistent with the IMF being converged, albeit to a form that is not in good agreement with observations. We remind the reader that because of our use of an isothermal equation of state, we do not expect the IMF to match measured IMFs.

\begin{figure}
\plotone{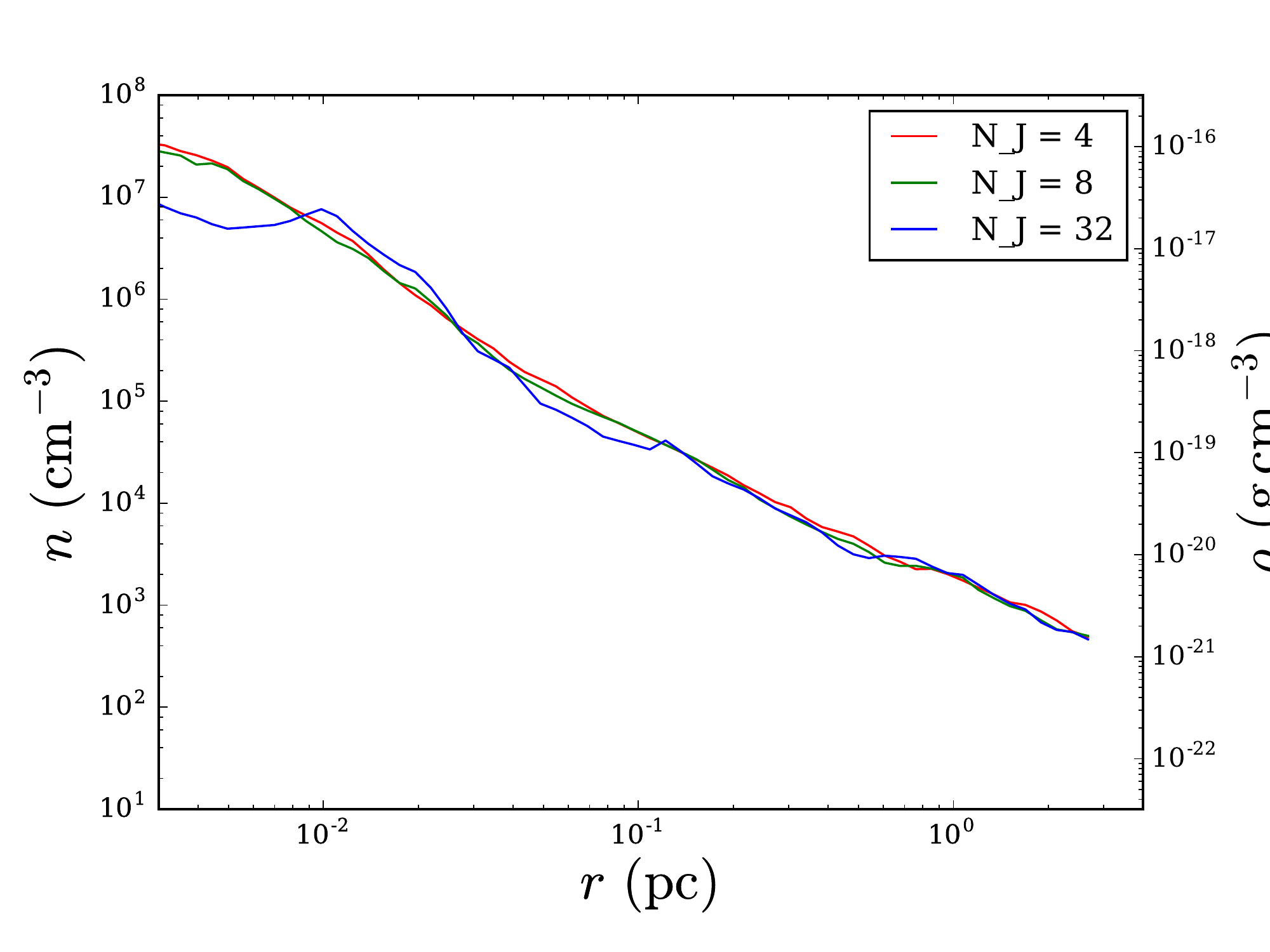} 
\caption{Plot of $\rho(r)$ for Ramses runs with $N_J = 32$ (blue), $N_J = 8$ (green), $N_J = 4$ (red). These are the averaged profiles of the density for star particles with $0.5M_\odot\leq M_*\leq 3 M_\odot$. Each simulation had $\approx 88 M_\odot$ worth of gas in star particles. The density for $r>r_d$ is the same all three runs, showing that the $N_J=4$ run is converged for $r>r_d$.}
\label{fig:density_convergence}
\end{figure}

As a further convergence check, Figure \ref{fig:density_convergence} shows the run of density as a function of radius for three different Ramses simulations. The $N_J = 32$ (blue) run had $3$ star particles at $0.5M_\odot<M_*<3 M_\odot$, while both the $N_J = 8$ (green line) run and the $N_J = 4$ (red line) had $9$ star particles.
We see convergence for all radii larger than the disk radius, $r_d$.  This illustrates that the density approaches an attractor solution that is robust against the underlying numerical technique. 

\bsp	
\label{lastpage}

\end{document}